\documentclass{article}
\usepackage{amsfonts,latexsym,eucal,amsmath,amsthm,amssymb}
\usepackage{pgf,tikz}
\usetikzlibrary{arrows}
\usepackage{fullpage}
\usepackage{enumerate}
\usepackage{comment}

\newcommand{\bea}{\begin{eqnarray}}
\newcommand{\eea}{\end{eqnarray}}
\theoremstyle{definition}

\usepackage{hyperref}
\usepackage{bm}
\usepackage{subfig}
\usepackage{float}
\usepackage{xcolor}
\usepackage{epstopdf}
\usepackage{multirow}
\DeclareMathOperator*{\argmin}{argmin}
\usepackage[section]{placeins}

\title{Mathematical models of drug delivery \\via a contact lens during wear}

\author{D.M. Anderson and R.A. Luke\\
Department of Mathematical Sciences\\
George Mason University\\
Fairfax, Virginia 22030
}

\date{\today}

\begin{document}

\maketitle

\begin{abstract}

In this work we develop and investigate mathematical and computational models
that describe drug delivery from a contact lens during wear.  Our models are
designed to predict the dynamics of drug release from the contact lens and subsequent transport into the adjacent pre-lens tear film and post-lens tear film as well as into the ocular tissue (e.g.~cornea), into the eyelid, and out of these regions.  These processes are modeled by one dimensional diffusion
out of the lens coupled to compartment-type models for drug concentrations
in the various accompanying regions.   In addition to numerical solutions that
are compared with experimental data on drug release in an {\it in vitro} eye model,
we also identify a large diffusion limit model for which analytical solutions 
can be written down for all quantities of interest, such as cumulative release of the drug from the contact lens.  We use our models to make assessments about possible mechanisms and drug transport pathways through the pre-lens and post-lens tear films and provide interpretation of experimental observations.
We discuss successes and limitations of our models as well as their potential to guide further research to help understand the dynamics of ophthalmic drug delivery via
drug-eluting contact lenses.
    
\end{abstract}

\section{Introduction}

%{\color{red}{Tasks:

%\begin{itemize}
%\item check force, pressure and $\Delta h_{\rm post}$ values in Table 1
%\item double check $k_{lid}$ in Table 1 
%\item is this 'model' or 'models'? (update title or abstract accordingly)
%\end{itemize}

%}}

Ophthalmic drugs are commonly delivered via eye drops.  This approach has at least two significant limitations.  First, the residence time of the drug in the eye is relatively short due
 to drainage out of the eye during a blink and the resupply of fresh tears.  
Thus, eye drops must be
given frequently for sustained drug delivery.    Second,
relatively large doses of drug must be delivered in each eye drop, as much of the drug goes elsewhere in the body 
-- up to 95\% of the drug delivered to the eye via a drop is lost through the canaliculi, lacrymal sac, nasolacrymal duct and bloodstream
\cite{LC2006,Bengani_etal_2013,Carvalho_etal_2015} -- reducing the efficacy of the treatment and leading to the possibility of drug side effects \cite{FM1987}.  Combined with the difficulty of a patient administering the correct dosage to the eye, the bioavailability of drugs delivered by drops intended to reach the cornea, for example, is quite low.

There are a number of alternative methods to deliver ophthalmic drugs to the eye, including drug/particle injections and implants into various 
compartments of the eye \cite{Kompella_etal2021}.
Another option uses drug-eluting contact lenses.
The possibility of using contact lenses to deliver drugs to the eye is appealing for treatment of ocular diseases/conditions (e.g.~glaucoma, fungal keratitis, 
antibacterial considerations, and myopia) \cite{Ciolino_etal_2009,Ciolino_etal_2009b, Bengani_etal_2013,Carvalho_etal_2015,Hatzav_etal_2016,Prausnitz_etal_1998,Fea_etal2022,Kang_etal2022,Wong_etal2022,Mun_etal2022}.  
Indeed, much work has appeared in the medical literature on drug-eluting contact lenses and several reviews have been written (e.g.~\cite{Ciolino_etal_2009,Ciolino_etal_2009b, Bengani_etal_2013,Carvalho_etal_2015,Hatzav_etal_2016}).  
The close proximity of the contact lens to the eye allows for much better
efficiency in terms of drug delivery to the cornea via a drug-soaked contact lens  (predictions range from 20\%--95\%) \cite{LC2006}.  
Carvalho {\it et al.} \cite{Carvalho_etal_2015} point out the difficulty in control of diseases
such as glaucoma, whose treatment requires a constant level of therapeutic drug in the eye, and the possible use of drug-infused contact lenses to supply a more predictable dose in a manageable way.
Understandably, there has been significant efforts towards the development of drug-eluting contact lens technology
(\cite{PdM_etal2022a,Peng_etal2022,Nguyen_etal2021,Rykowska_etal2021,FdM2021,Jones_etal2021}),
including `smart' contact lenses \cite{Mun_etal2022,Du_etal2022} and the use of 3D printing technologies \cite{phan2021development,Larochelle_etal2021}.
Drug transport to other parts of the eye besides the cornea may also be desirable \cite{Prausnitz_etal_1998}.

Although contact lens drug delivery
%with its potential for vast increases in drug bioavailability, 
has been studied as a method of ophthalmic drug delivery since at least the 1970s \cite{li2006modeling}, therapeutic options have only been proposed recently and remain limited \cite{franco2021contact}. The first drug-eluting contact lens to receive regulatory approval, for allergic eye itch, was authorized in Japan in 2021 and the United States in 2022; the latter approval came after at least 18 years of development \cite{novack2023us}. Barriers to commercial implementation include lack of \textit{in vivo} and even \textit{in vitro} studies \cite{phan2021development}, and lack of a full understanding and control of drug delivery rates.  These rates depend not only on the diffusion and transport mechanisms in the contact lens, but also, critically, on surrounding fluid dynamics in the tear film and related reservoirs that act as the conduit for the drug to reach a target tissue. 
Some situations, such as glaucoma treatment, call for slower rates where a drug can be released over the course of 
one day to one month \cite{Bengani_etal_2013}, while other situations, such as applications involving drugs used to dilate eyes, require shorter time scales \cite{Melki_etal_ARVO2022}.  
Molecular diffusion is the main release mechanism, but soft contact lenses are hydrogels \cite{Liu_etal_2013,PSP2015} and even the diffusion of water through a lens (in the absence of drugs)
 requires a nonlinear diffusion model that accounts for swelling and the possibility of glass formation \cite{FPR2006}.  
For contact lens drug delivery, the most common way to load the drug into the lens is to soak the lens in a solution, which is a diffusion up-take problem \cite{PC_2011}.  However, attempts to meet the broad range of
desired release rates of the drug have involved novel contact lens design and drug-infusing techniques \cite{PKC2010,Carvalho_etal_2015,Zhu_etal2018},
including inkjet printing of nanoparticles onto the contact lens \cite{Tetyczka_etal2022}.
%including vitamin E pretreatment \cite{PKC2010}, molecular imprinting and the use of nanoparticles \cite{Carvalho_etal_2015}.   
Spatially-dependent diffusion properties may also be introduced, for example with the use of a drug-polymer film coated by a HEMA hydrogel (e.g.~\cite{Ciolino_etal_2009, Ciolino_etal_2009b}) as a possible means of slowing the drug-delivery release rate from the contact lens.  
%However, drug transport within and drug release from the contact lens is only the beginning of the story.
Ultimately, in order to assess the effectiveness of wearing a drug-eluting contact lens, one must also track the drug after it leaves the contact lens -- into the pre- and post-lens tear films, 
out underneath the eyelids, and to the desired target.  The influence of blinking and evaporation during daytime contact
lens wear as well as the impact of the closed eye configuration and eye motion (e.g.~REM sleep) during night-time wear must also be assessed for a complete understanding of drug delivery via contact lenses.  

Theoretical approaches have modeled drug transport in hydrogels (e.g., \cite{huang2001importance,pimenta2016controlled}); most are diffusion-based. 
Many studies have examined drug uptake and release in well-controlled settings such as vials 
(e.g.~\cite{PC_2011,Lanier_etal2021,Dixon_etal2018a,Dixon_etal2018b,Liu_etal2022,Nguyen_etal2022}).
Recently, Pereira-da-Mota {\it et al.} \cite{PdM_etal2022b} (but see also \cite{PdM_etal2022a}) reported data on {\it ex vivo} and {\it in vivo} (rabbit) drug-release from contact lenses. 
Navarro-Gil {\it et al.} \cite{Navarro-Gil_etal2022} also studied {\it in vivo} drug release from contact lenses in rabbits over 2 hours of wear.
Another recent study reports drug-delivery via contact lens in an {\it in vitro}, 3D printed, realistic eye model that include effects such as blinking \cite{phan2021development} during the 
drug release process.  This study was notable in that it reported a significant
and measurable changes in the rate of drug loss from the lens in the vial setting versus when it was `worn' in the {\it in vitro} eye model.  

Models  have been examined for contact lens drug delivery \cite{LC2006} and transport across a layer separating a drug supply from a sink 
\cite{Prausnitz_etal_1998} that implement similar approaches.
Prausnitz {\it et al.} \cite{Prausnitz_etal_1998} studied transient diffusion across the sclera (a pathway to the retina) by employing one dimensional diffusive transport across the sclera separating a donor compartment with a finite amount of a drug and a receiver compartment.  The donor and receiver drug concentrations were modeled with ordinary differential equations and coupled by diffusion across the sclera.   Models focusing on other regions of the eye (e.g.~anterior chamber) have also been examined \cite{Bhandari2021}.
Li \& Chauhan \cite{LC2006} studied contact lens drug delivery and predicted the fractions of
drug delivered to the pre-lens, post-lens, and cornea.  Drug transport in the contact lens
was treated by one-dimensional diffusion across the lens.  The post-lens tear film was modeled as a squeeze layer with an upper boundary (the contact lens) that moves laterally (superior/inferior) and orthogonally (anterior/posterior) to the eye during a blink.  They did not resolve pre-lens tear film dynamics but rather modeled the drug loss out of the contact lens
on the pre-lens side using two limiting boundary conditions. Based on the assumption that the pre-lens tear film breaks up rapidly due to evaporation, the first assumed that no drug reached the pre-lens tear film, thus preventing drug transport into the pre-lens film 
(i.e.~no flux boundary condition). The second boundary condition assumed the drug concentration in the pre-lens tear film was zero. 
The second represents the limit in which drug diffuses from the contact lens into the pre-lens film but is quickly and completely removed by mixing and drainage.
%While this greatly simplified the problem, predictions for the percentage of drug that reached the cornea in these two cases varied from as little as 20\% to as much as 95\%. 

%{\color{red}{check also compartment models (e.g.~\cite{ZC2005a,ZC2005b})}}

%\subsection{Prior work on absorption mechanisms}

To the best of our knowledge, there is little experimental evidence of the action of the upper eyelid on the tear film during a blink, but many studies have modeled its effect in various ways. Doane %(1980) 
\cite{doane1980interactions} took high-speed images of the human tear film to measure upper and lower lid dynamics, including blink time, velocity, and  lid positioning. More recent imaging studies include Wu \textit{et al.} %(2014) 
\cite{wu2014} and Awisi-Gyau \textit{et al.} %(2020) 
\cite{awisigyau2020}. Modeling examples include Anderson \textit{et al.} %(2021)
\cite{anderson2021tear} and Zubkov \textit{et al.} %(2013) 
\cite{zubkov2013meniscal}, which both include Couette flow of the tear film. The latter specifically assumes that tear film fluid is supplied from under the eyelid by this flow. The model in Braun \textit{et al.} %(2015) 
\cite{braun2015} assumes a rectangular eyelid that moves across the tear film domain while maintaining a constant distance from the cornea. This model in particular, in contrast to ours, allows for the tear film height to vary spatially across the cornea as a result of the eyelid motion. Recently, Ramasubramanian \textit{et al.} \cite{ramasubramanian2022finite} created a finite element method 3D model of the ``lid wiper'', the part of the eyelid that comes in contact with the tear film, and looked at the stresses imposed on the cornea and eyelid as a result of a blink. Some models (see, e.g., \cite{jones2005dynamics}) incorporate the draining of fluid as part of the blink cycle.  Braun \textit{et al.} %(2015) 
\cite{braun2015}  note the difficulty in understanding tear film dynamics surrounding a blink due to imaging challenges %. Several of these %theoretical 
%studies %(\cite{ramasubrama
%note 
and that further experimental evidence is needed.
%to better understand tear film dynamics during the blink cycle.

Drug absorption into the eyelid is also possible.
Farkouh \textit{et al.} \cite{farkouh2016systemic} confirm this and discuss potential absorption pathways of the drug from the tear film, and include the eyelids as a minor route contributing to systemic absorption, which overall comprises more than half of the dose. See {\it et al.} \cite{see2017eyelid} note that the thinner the skin layer, the greater the drug permeability and, at less than 1 mm, the eyelid is the thinnest skin layer on the human body. %They also note that the conjunctiva is more permeable to drugs than the cornea (2 - 30 times), and so in reverse, it may deliver more drug to the eyelid.
The authors found the permeability of the eyelid in rats to be on the order of $(10^{-9} - 10^{-7})$ m/s for several drugs.
For the experimental setting used by Phan {\it et al.}~\cite{phan2021development}, which we examine in detail below,
they also report absorption into the eyelid.
To the best of our knowledge, no mathematical models of contact lens drug delivery have been designed specifically to include eyelid absorption; in fact, some  (i.e., \cite{habibi2022drug}) explicitly assume zero absorption of drug by the eyelid.
A WHO report by Kielhorn {\it et al.} \cite{kielhorn2006dermal} on dermal absorption includes a section on mathematical models; one such model treats the skin as a single pseudo-homogeneous membrane and models absorption via Fick's first law. Models such as  Kr{\"u}se {\it et al.} \cite{kruse2007analysis} describe the absorption equilibrium by partition and diffusion coefficients in the cornea and tear film. Selzer {\it et al.} \cite{selzer2015mathematical}  review different types of mathematical models for adsorption, including pharmacokinetic ordinary differential equations (ODEs), quantitative structure-activity relationship (QSAR) models, and diffusion and partitioning. They include models that allow for concentration variation by skin depth. The authors note that the ODE models often do not fit experimental data as well as diffusion models with spatial heterogeneity, but the latter data is not always available. %Selzer et al. \cite{selzer2015mathematical} also point to papers regarding transient drug exposure (e.g., Frasch and Barbero (2008) \cite{frasch2008transient}).
%, which is addressed in our Discussion section.

In the present work, to join theoretical and experimental information, we extend a mathematical model of contact lens drug delivery by \cite{liu2016diffusion} to align with the experimental \textit{in vitro} model eye system designed by Phan {\it et al.}~\cite{phan2021development}. The model couples a partial differential equation for diffusion of drug inside the contact lens with ordinary differential equations governing the dynamics of the pre- and post-lens tear films and also the eyelid. 
%These films protect the ocular surface and promote clear vision. 
The model simulates tear film dynamics during blinking over a 24 hour period. We compute the total drug lost from the contact lens and compare our results to experimental data from Phan {\it et al.}~\cite{phan2021development}. By isolating certain mechanisms or combinations of mechanisms in our models, we attempt to better understand the mechanics and dynamics of contact lens drug delivery.

Our paper is organized as follows.  In section 2 we describe the basic eye model configuration of interest along with key concepts and variables.  In Section 3 we briefly visit the vial model configuration to identify diffusion and partition coefficients for the two different lens types examined in the Phan {\it et al.}~\cite{phan2021development} study.
In Section 4 we outline the various components of our eye model and present its nondimensionalized version and some related quantities in Section 5.  Section 6 presents details of a simplified model based on a large diffusion limit in which relatively simple analytical formulas can be obtained for key quantities of interest.  In Section 7 we give the main results of our eye model in comparison to the Phan {\it et al.}~\cite{phan2021development} data.  Section 8 includes further discussion and avenues for additional research.  Section 9 contains our conclusions.

\section{Configuration, etc.}

We use data from a study conducted by Phan \textit{et al.} \cite{phan2021development}. Those authors designed a 3D-printed model eye system with blinking action to measure the cumulative release of ``drug'' (mimicked by a water-soluble red dye) from two types of contact lenses: a conventional hydrogel (etafilcon A) and a silicone hydrogel (senofilcon A). The contact lens is situated in an upright position to mimic \textit{in vivo} wear, and an upper eyelid ``blinks'' every 10 seconds. Tubing delivers fluid to the eye from under the upper eyelid during a blink. The system is mounted on a collection plate so that the amount of dye released can be measured. The model eye and collection unit are printed from hydrophobic material, but the eyelid is not. See the authors' Figure 1 for a schematic \cite{phan2021development} and our Figures \ref{fig:est_eye1} and \ref{fig:est_eye2} for example video stills. We will compare the predictions of our models with the cumulative drug release data reported by Phan {\it et al.} \cite{phan2021development} in their Figure 3.   Some aspects of our models are not directly applicable for the Phan {\it et al.} data, such as drug permeability into the cornea, but are included into our models as our longer-term goals include prediction of drug delivery to tissues such as the cornea.

\subsection{Parameters}
 \label{sec:calc_params}

As much as possible, our models use parameters taken directly from Phan \textit{et al.} \cite{phan2021development}.
 The authors state that each contact lens was soaked in red dye ($1.0048$ mg / $\mu $L) for 24 hours.   They report an initial amount of drug in the
contact lens as $22.4 \pm 2.0$ mg (etafilcon A) and $47.8 \pm 2.7$ mg (senofilcon A).\footnote{We believe Phan {\it et al.} listed these values reversed with respect to the lens type -- see their Figures 3 and 4 and text on the same page.  That is, etafilcon A seems to release close to 100\% of its dye in Figure 4 and reaches a maximum released value of around 22 mg in Figure 3 and so it seems corresponds to an initial amount of drug of $22.4$ mg.}  We find corresponding contact lens volumes for each lens using $V_{\rm cl} = A_{\rm cl} h_{\rm cl}$ where $V_{\rm cl}$ is contact lens volume, $A_{\rm cl}$ is contact lens area, and $h_{\rm cl}$ is contact lens thickness, given in Table~\ref{table-compA}.
 
%\begin{align}
%V_{\rm cl}^{\rm eta} & =  10.78 \times 10^{-9} \; \mbox{m}^3 = 10.78 \; \mu \mbox{L}, \\
%V_{\rm cl}^{\rm seno} & =  13.08 \times 10^{-9} \; \mbox{m}^3 = 13.08 \; \mu \mbox{L}.
%\end{align}
%$V_{cl} = 0.0154$ cm$^3$ and $C_0 = 1.45$ mg mm$^{-3}$ (etafilcon A) and $C_0 = 3.10$ mg mm$^{-3}$ (senofilcon A).

As a basis for comparison for their eye model results, Phan {\it et al.} first examined drug release from the contact lens submerged in a large fluid volume in a vial
configuration.
The volume of fluid in the vial was reported to be $V_{\rm vial} = 2$ mL. Assuming spatial uniformity, the initial drug concentration in the lens  is computed as $C^{\rm init} = M^{\rm init}/V_{\rm cl}$
and is reported in Table~\ref{table-compA} for each lens.
%\begin{align}
%C^{\rm init}_{\rm eta} = \frac{22.4 \; \mbox{mg} }{10.78 \; \mu \mbox{L}} = 2.078 \; \mbox{mg}/ \mu \mbox{L}, \\
%C^{\rm init}_{\rm seno} = \frac{47.8 \; \mbox{mg} }{13.08 \; \mu %\mbox{L}} = 3.654 \; \mbox{mg}/ \mu \mbox{L}.
%\end{align}
We assume that these concentrations have reached equilibrium, and thus predict the partition coefficient as
$K = C^{\rm init}/C^{\rm load}_{\rm vial}$, which is 
reported in Table~\ref{table-compA} for each lens
(see also Liu {\it et al.} \cite{liu2022transport}, equation 2).

%\begin{align}
%K_{\rm eta} = \frac{2.078}{1.0048} = 2.07, \\
%K_{\rm seno} = \frac{3.654}{1.0048} = 3.64,
%\end{align}

\begin{table}[h!]
\begin{center}
\begin{tabular}{llllll}
      Parameter  &  \multicolumn{2}{c}{etafilcon A}            & \multicolumn{2}{c}{senofilcon A} & Source \\ 
         (units)                           &  vial   & eye model &  vial  & eye model & \\ \hline
 %--------------
$V_{\rm vial}$ (mL) &  2 &  -- & 2     &  -- & \dag \\
 %--------------
$C_{\rm vial}^{\rm load}$ (mg $\mu$L$^{-1}$) &  1.0048 &  -- & 1.0048     &  -- & \dag \\
 %--------------
$M^{\rm init}$ (mg) &  $22.4 \pm 2$ &  $22.4 \pm 2$ &  $47.8 \pm 2.7$   & $47.8 \pm 2.7$ & \dag \\   
 %--------------
$h_{\rm cl}$ ($\mu$m) & 70     & 70 & 85    &  85 & \dag \\
 %--------------
$R_{\rm cl}$ (mm) &  7   &  7 & 7   & 7 & \dag \\
 %--------------
 $ A_{\rm cl} = \pi R_{\rm cl}^2$ (mm$^2$) & 153.9 & 153.9 & 153.9 & 153.9 & \dag \\
 %-------------------
$V_{\rm cl}=\pi R_{\rm cl}^2 h_{\rm cl}$ ($\mu$L) &  10.78   &  10.78 & 13.08   & 13.08 & \dag \\
 %--------------
$C^{\rm init}$ (mg $\mu$L$^{-1}$) &  2.078 &  2.078 & 3.654  & 3.654 & \dag \\
%--------------
$D$ (m$^2$ s$^{-1}$) & $2.24 \times 10^{-13}$ 
%&  {\color{red}{$4 \times 10^{-13}$}}
& $2.24 \times 10^{-13}$ 
%&  {\color{red}{$4 \times 10^{-13}$}}
& $7.50 \times 10^{-15}$
%&  {\color{red}{$1 \times 10^{-14}$ }}  
& $7.50 \times 10^{-15}$ &  $\ominus$\\
%& {\color{red}{$1 \times 10^{-14}$}}  \\
 %--------------
$K$ &  2.07  &  2.07 &  3.64 &  3.64 & $\ominus$, $\diamond$ \\
%--------------
 $h_{\rm lid}$ ($\mu$m) & -- & 500 & -- & 500 & $\diamond$ \\
$A_{\rm lid}$ (mm$^2$) & -- & 290 & -- & 290 & $\diamond$ \\
 $V_{\rm lid}$ (mm$^3$) & -- & 145 & -- & 145 & $\diamond$ \\
   %--------------
% $k_{\rm lid}$ (m/s) & -- & $(0.00189 - 572) \times 10^{-8}$ & -- & ? & $\bowtie$ \\
 $k_{\rm lid}$ (m/s) & -- & $(3 - 700) \times 10^{-9}$ & -- & $(3 - 700) \times 10^{-9}$ & $\bowtie$ \\
%--------------
 %--------------
$h_{\rm pre}^{\rm init}$ ($\mu$m) & -- &  1.16 -- 5.39 & -- & 1.16 -- 5.39 & $\star$,$\diamond$\\
  %--------------
$h_{\rm post}^{\rm init}$ ($\mu$m) & -- &  5 & -- & 5 & $\star$\\
%--------------
 $\Delta X_{\rm cl}$ (mm) & -- & \begin{tabular}[c]{@{}l@{}}$(0.1-4)^{\star}$;\\ $(0.03-0.06)^{\bowtie}$ \end{tabular} & -- & \begin{tabular}[c]{@{}l@{}} $(0.1-4)^{\star}$; \\ $(0.04-0.35)^{\bowtie}$ \end{tabular} & $\star$;  $\bowtie$  \\
%--------------
 $\Delta h_{\rm post}$ ($\mu$m) & -- & $0.01 - 0.02$  & -- & $0.014 - 0.1$ & $\bowtie$ \\
%--------------
%$J_{T}^{\rm out}$ ($\mu$L min$^{-1}$) & -- &  5 & -- & 5 & \dag \\
$\tau$ (s) &  --   &  10 &  -- &  10 & \dag \\
 %--------------
 $\Delta t_{\rm blink}$ (s) & -- &  0.2579 & -- & 0.2579 & $\star$\\
$P_{\rm blink}$ (Pa) & -- &  $29.7-59.6$ & -- & $41.7-306$ & $\bowtie$\\
%--------------
$F_{\rm blink} = \pi R^2 P_{\rm blink} $ (N) & -- &  $0.0046-0.0092$ & -- & $0.0064-0.047$ & $\bowtie$\\
 %--------------
$\bar{J}_{E}$ & -- &  0 & -- & 0 & $\bowtie$\\
 %--------------
%%$\bar{k}_r$ & -- &  0 & -- & 0 & \\
 %--------------
$\bar{k}_c$ & -- &  0 & -- & 0 & $\star$ \\
 %--------------
$\bar{J}_{\rm osmotic}$ & -- &  0 & -- & 0 & $\star$ \\
%--------------
%$\bar{F}$ & -- & $117.7$ (\mbox{or} $0$) & -- & $151.2$ (\mbox{or} $0$) & $\star$\\
%--------------
%%$\bar{k}_r$ & -- & $10000$ & -- & $10000$ & \\
 \hline
\end{tabular}
\end{center}
\caption{
Parameter values used for the two different contact lenses - etafilcon A and senofilcon A. Source column: \dag: Phan \textit{et al.} (2021) \cite{phan2021development}, $\ominus$: found via optimization, $\diamond$: estimated from experimental  data \cite{phan2021development}, $\star$: estimated from literature, $\bowtie$: hand-tuned to match experimental data \cite{phan2021development}. Dimensionless parameters are grouped at the end of the table and denoted with bars.}
\label{table-compA}
\end{table}
%{\color{red}{DMA: When measurements were made by removing $2\mu$L of fluid an equal volume of 'fresh PBS' was replaced to 'maintain sink conditions' -- try to confirm what this means if we need to adjust our value of $C_{pre}$ and $C_{post}$ in some way at these timepoints.}}

Table \ref{table:Phan21_Fig3} reports the cumulative drug release data extracted from Phan \textit{et al.}'s Figure 3. Values listed here are obtained via {\tt Matlab}'s image analysis software {\tt grabit.m} (MathWorks, Natick, MA, USA).
The time values are those reported by Phan {\it et al.}; our approximate time values extracted from their figure were within $1$\% of these 
reported values.  Their reported final time values are listed in parenthesis; our approximate values are well within their reported error bars. Table \ref{table:drug_exp} shows the drug release values recorded by Phan \textit{et al.} \cite{phan2021development} as well as the calculated eyelid absorption values using their percentages.
  The authors state that 31.2 \% and 17.5 \% of released drug are absorbed by the eyelid for the etafilcon A and senofilcon A lenses. %Note here again that we believe Phan {\it et al.} listed some values reversed with respect to the lens type.

 \begin{table}[h]
\begin{center}
\begin{tabular}{lllll}
      Time  &  \multicolumn{2}{c}{etafilcon A (mg)}            & \multicolumn{2}{c}{senofilcon A (mg)} \\ 
      (hours)                       &  vial   & eye model &  vial  & eye model \\ \hline
 %--------------
$0.5$ 
&  14.76
&  0.2044
&  7.651   
& 0.7461  \\
 %--------------
$1$ 
& 18.28     
& 1.492
& 10.97    
&  2.372 \\
 %--------------
$2$ 
&  21.13   
&  3.120
& 11.65   
& 3.797  \\
 %--------------
$4$ 
&  22.42
&  6.036
& 14.57     
&  5.359 \\
 %--------------
$8$ 
&  22.30  
&  8.823 
&  19.25 
&  8.146 \\
 %--------------
$12$ 
&  22.24   
&  10.59 
&  22.17 
&  11.07 \\
 %--------------
$24$ 
& 21.93 ($21.8 \pm 4.0$) 
&  13.07 ($13.2 \pm 2.9$)
& 28.16  ($28.0 \pm 1.0$) 
& 14.83  ($14.7 \pm 3.9$) \\
 \hline
\end{tabular}
\end{center}
\caption{
Cumulative drug release data (in mg) as a function of time from Phan {\it et al.}\cite{phan2021development} Figure 3 for two different contact lens types (etafilcon A) and (senofilcon A) 
and two configurations (in a vial and in an {\it in vitro} eye model). 
}
\label{table:Phan21_Fig3}
\end{table}

%Phan {\it et al.} \cite{phan2021development} state that each contact lens was soaked in red dye ($1.0048$ mg / $\mu $L) for 24 hours.  
%The  uptake of drug was reported as $22.4 \pm 2.0$ mg (etafilcon A) and $47.8 \pm 2.7$ mg (senofilcon A) (see earlier footnote).  
%If we estimate the contact lens volume to be $V_{cl} = A H$ where $A = \pi R^2$ with contact lens radius $R=7$mm and contact lens thickness $H$ given in  Table~\ref{table-compA} we find that this gives lens volume

 \begin{table}[h]
 \centering
\begin{tabular}{|c|r|r|r|}
\hline
\textbf{Lens type} & \multicolumn{1}{c|}{\textbf{\begin{tabular}[c]{@{}c@{}}Total drug \\ absorbed (mg)\end{tabular}}} & \multicolumn{1}{c|}{\textbf{\begin{tabular}[c]{@{}c@{}}Drug released \\ from contact lens (mg)\end{tabular}}} & \multicolumn{1}{c|}{\textbf{\begin{tabular}[c]{@{}c@{}}Drug absorbed \\ by eyelid (mg)\end{tabular}}} \\ \hline
\textbf{Etafilcon A} & 22.4 & 13.2 & 4.12 \\ \hline
\textbf{Senofilcon A} & 47.8 & 14.7 & 2.57 \\ \hline
\end{tabular}
\caption{Reported drug absorbed and released by contact lens from Phan {\it et al.} (2021) \cite{phan2021development} and calculated drug absorbed by eyelid.}
\label{table:drug_exp}
\end{table}

\begin{figure}[h]
 \centering
 \includegraphics[scale=0.45]{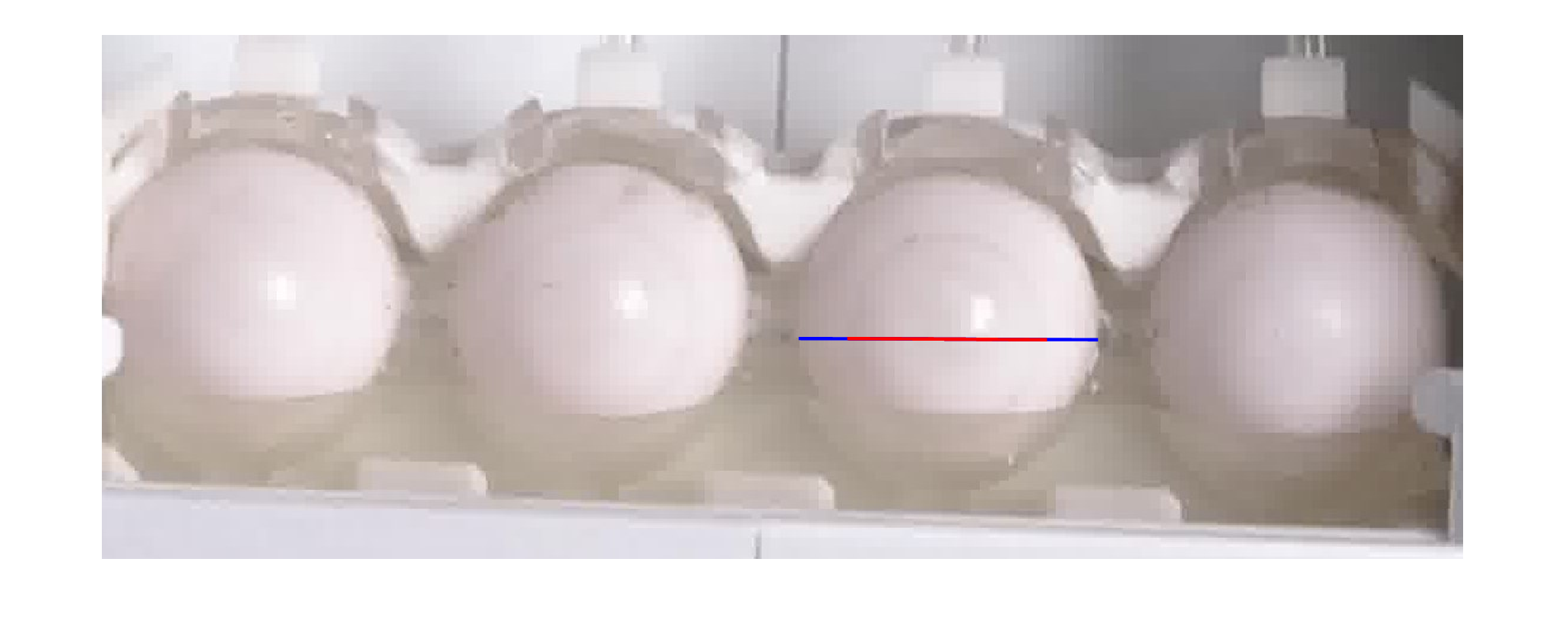} 
 \caption{Estimating the model eye diameter from a video still
 of Phan {\it et al.} \cite{phan2021development}. The red line estimates the contact lens diameter in pixels (known in mm) and the blue line estimates the model eye diameter. }
 \label{fig:est_eye1}
 \end{figure}

\begin{figure}[h]
 \centering
 \subfloat[][contact lens diameter]{\includegraphics[scale=0.2]{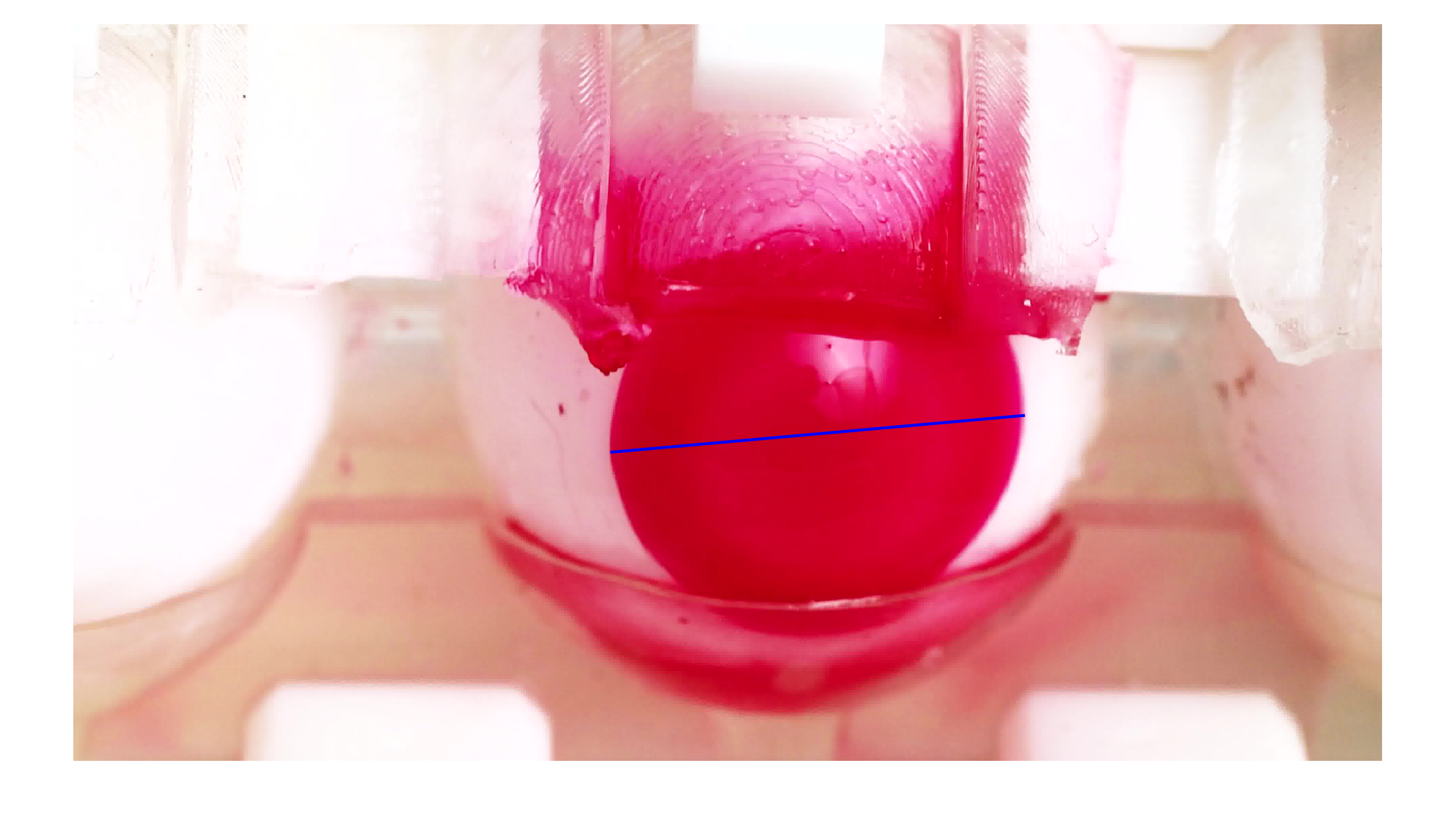}} 
 \subfloat[][Eyelid area]{\includegraphics[scale=0.2]{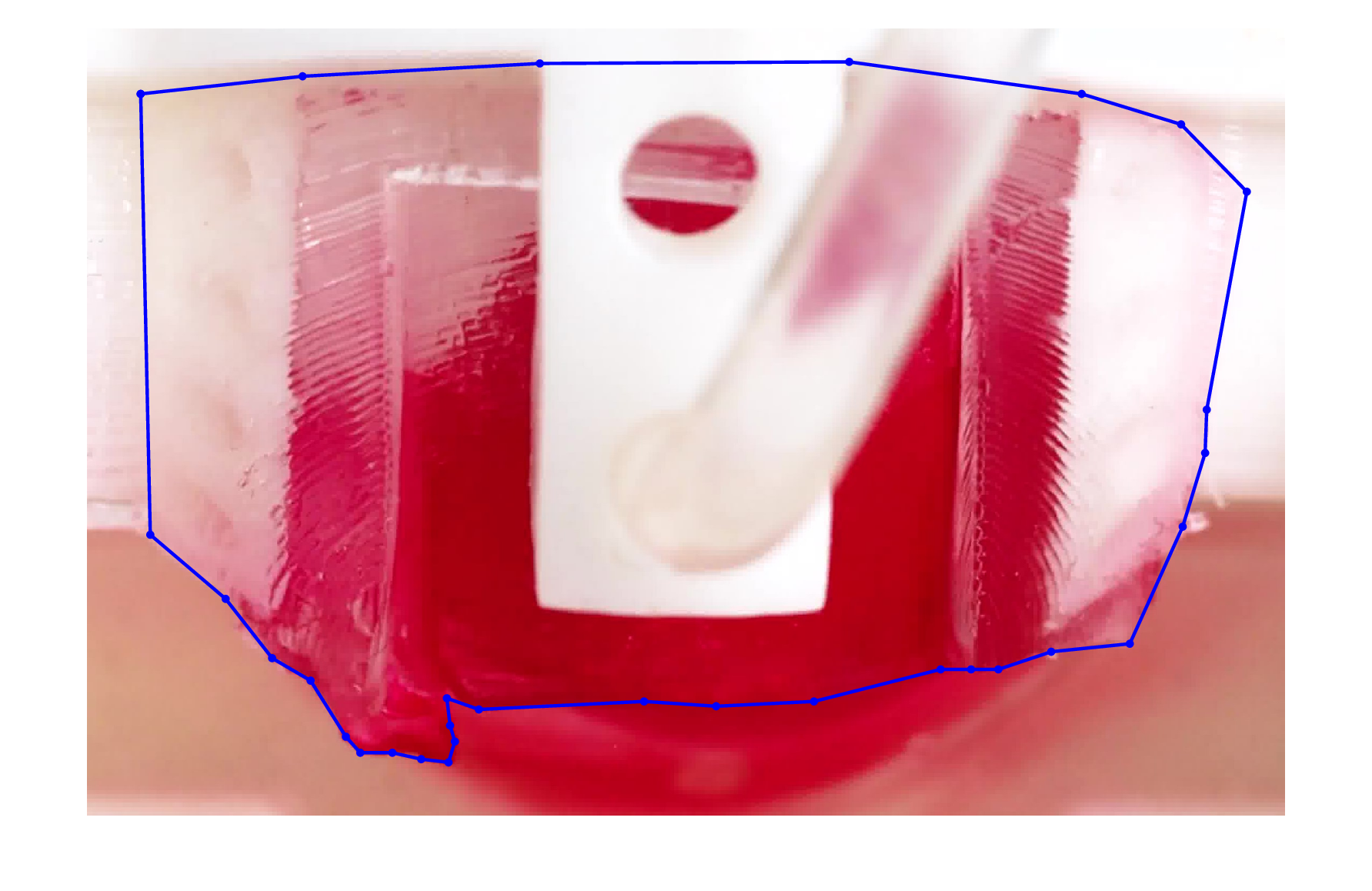}}
 \caption{Estimating the eyelid area from video stills. The blue line in (a) estimates the contact lens diameter in pixels (known in mm) and the blue box in (b) estimates the eyelid area. Image from Phan {\it et al.} \cite{phan2021development}. }
 \label{fig:est_eye2}
 \end{figure}

We estimate the model eye diameter and eyelid area via simple image analysis as shown in Figures \ref{fig:est_eye1} and \ref{fig:est_eye2}. Repeating the procedure three times, the average eye diameter estimate is 21.3 mm and the eyelid area estimate is 290 mm$^2$; estimates for the size of the eyeball in the transverse diameter are 21--27 mm \cite{bekerman2014variations}.  
 
Phan \textit{et al.} \cite{phan2021development} use their tear flow rate of 5 $\mu$L/min (0.083 $\mu$L/s) and blink rate of 1 blink/10 s to find that ``approximately 0.83 $\mu$L of tear fluid that is spread on top of the contact lens with each blink.'' If we assume that the volume of tear fluid (0.83 $\mu$L) is evenly spread only over the contact lens area (see Table \ref{table-compA}), we can compute the pre-lens thickness at the start of an interblink as $h_{\rm pre}(t_{\rm blink}^+) = V_{\rm pre}/A_{\rm cl} = 
%(0.83 \times 10^{-9} \text{ m}^3)/(153.9 \times 10^{-6} \text{ m}^2) = 
5.39 \ \mu \text{m}.$
Realistically, the fluid spreads over more than just the contact lens.  The contact lens area and the surface area of the front of the eyeball provide reasonable lower and upper bounds for the area over which the tear fluid spreads. Using our estimated eyeball diameter 
%estimated from the Phan {\it et al.} video stills 
to find half the surface area of the eye sphere, $2 \pi R_{\rm eye}^2$, the same calculation yields $h_{\rm pre}(t_{\rm blink}^+) = 1.16 \ \mu$m. This may provide a realistic range of initial pre-lens tear film thicknesses of 1.16 -- 5.39 $\mu$m.   While estimates for human tear flow rates are smaller than that used by Phan and coauthors %(e.g., (0.5 - 2.2) $\mu$L/min 
\cite{mishima1966determination},  the calculated tear film thickness fall within or near experimental ranges 
% ($(3.98 \pm 1.06) \ \mu$m)
\cite{nichols2005}.

%Using the reported update of drug by the contact lens, $M_{\rm dye}^{\rm init}$, and the lens volume $V_{\rm cl} = \pi R^2 H$, the initial concentration of the dye in the contact lens, $C_{\rm cl}^{\rm init}$, can be computed (assuming spatial uniformity).  Assuming these concentrations have reached equilibrium, we have the prediction that the partition coefficient for each lens is given by $K = C_{\rm cl}^{\rm init}/C_{\rm vial}^{\rm init}$ (see Table \ref{table-compA} and also Liu {\it et al.} \cite{Liu_etal2022}, equation (2)).
%\begin{eqnarray}
%K^{eta A} = \frac{2.078}{1.0048} = 2.07, \\
%K^{seno A} = \frac{3.654}{1.0048} = 3.64,
%\end{eqnarray}
%(see also Liu {\it et al.} \cite{Liu_etal2022}, equation (2)).

%From Kimura and Tojo (2007) \cite{kimura2007development} we find a rough estimate of the diffusion coefficient of the eyelid (in rabbits) to be $3.75 \times 10^{-11}$ m$^2$/s. While this may become a parameter for optimization, for now we use this value as an estimate in order to match $k_p$, the permeability coefficient, to the experimental data from Phan et al. (2021) \cite{phan2021development} in order to compare to the max flux situation.

\section{Vial drug release model}

%\subsection{Dimensional model}

In the vial release configuration, we assume that the contact lens is contained in fluid of constant volume $V_{\rm vial}$.
We assume that the transport of drug concentration $C(z,t)$ within the contact lens is dominated by diffusion across the lens (anterior/posterior)
\begin{eqnarray}
\label{eq:Ceq_vial}
\frac{\partial C}{\partial t} & = & D \frac{\partial^2 C}{\partial z^2},
\end{eqnarray}
where $D$ is the diffusion coefficient.
This equation applies on $0< z < h_{\rm cl}$, where $h_{\rm cl}$ is a constant thickness of the contact lens, and is subject to boundary conditions 
\begin{eqnarray}
\label{eq:Cbc12_vial}
C(z=0,t) = K C_{\rm vial}(t), \quad
C(z=h_{\rm cl},t)  =  K C_{\rm vial}(t),
\end{eqnarray}
where $C_{\rm vial}(t)$ is the drug concentrations in the vial, and $K$ is the partition coefficient.  The drug in the vial is assumed to be well-mixed
so that the concentration $C_{\rm vial}(t)$ is uniform in space and the same on both sides of the lens.
An initial condition corresponding to a drug-soaked contact lens is given by $C(z,t=0)=C^{\rm init}(z)$.  Estimates for the size of the diffusion coefficient can be obtained from recent work by Liu {\it et al.} \cite{Liu_etal_2013}, who have characterized the diffusive processes that can occur for macromolecules in hydrogels, such as those used in contact lens manufacture.    We can also estimate it here using data reported by Phan {\it et al.} \cite{phan2021development}. We assume the diffusion coefficient is constant but note that a spatially-dependent diffusion coefficient may also be of clinical interest, as would be the case for the drug-polymer film coated by a HEMA hydrogel \cite{Ciolino_etal_2009, Ciolino_etal_2009b} investigated as a possible means to reduce the contact lens drug-delivery release rate.

%\begin{eqnarray}
%\label{eq:Cbc12}
%C(z=0,t) = K C_{\rm post}(t), \quad
%C(z=H,t)  =  K C_{\rm pre}(t),
%\end{eqnarray}
%where $C_{\rm post}(t)$ and $C_{\rm pre}(t)$ are the drug concentrations in the post-lens tear film and pre-lens tear film whose values are determined by the models described below, and $K$ is the partition coefficient. 
%An initial condition corresponding to a drug-soaked contact lens is given by $C(z,t=0)=C_0(z)$.   

The flux of drug out of the contact lens is driven by concentration
gradients in the lens. We assume the
fluid in the vial remains well-mixed and that the mass flux out of either
side of the lens (the `pre-lens' and `post-lens' sides) contributes to a 
uniformly mixed vial fluid volume.  It follows that
\begin{equation}
\frac{d \left( C_{\rm vial} V_{\rm vial} \right) }{dt} =  
\left( D \left. \frac{\partial C}{\partial z} \right|_{z=0} 
- D \left. \frac{\partial C}{\partial z} \right|_{z=h_{\rm cl}} \right) A_{\rm cl}.
\end{equation}

%Conservation of drug mass implies 
%\begin{equation}
%\frac{d}{dt} \left( C_{\rm pre} V_{\rm pre} \right)  =  - D \left. \frac{\partial C}{\partial z} \right|_{z=H} A,
%\end{equation}
%where $A$ is the contact lens area and $V_{\rm pre}$ is the volume associated with the `pre-lens half' of the vial,
%which we interpret as $V_{\rm pre} = \frac{1}{2} V_0$.
%Similarly,
%\begin{equation}
%\frac{d}{dt} \left( C_{\rm post} V_{\rm post} \right)  =  D \left. \frac{\partial C}{\partial z} \right|_{z=0} A,
%\end{equation}
%where $V_{\rm post} = \frac{1}{2} V_0$.

Below we will discuss solutions to this model and identify values for parameters such as the diffusion coefficient $D$ and partition coefficient $K$.
In related studies (e.g.~\cite{PC_2011,liu2016diffusion,Liu_etal2022}) a common assumption is that $V_{\rm vial}$ is sufficiently large in comparison to the contact lens volume $V_{\rm cl}$ so that $C_{\rm vial} \approx 0$ (also called `perfect sink conditions') and the drug diffusion problem is independent of the
partition coefficient, $K$.  
This leaves the diffusion coefficient $D$ and geometric properties such as thickness $h_{\rm cl}$ and surface area $A_{\rm cl}$
along with initial drug mass $M^{\rm init}$ as the key parameters.

\subsection{Vial model solutions and optimization strategy}

\begin{figure}[h]
\centering
\includegraphics[scale=.45]{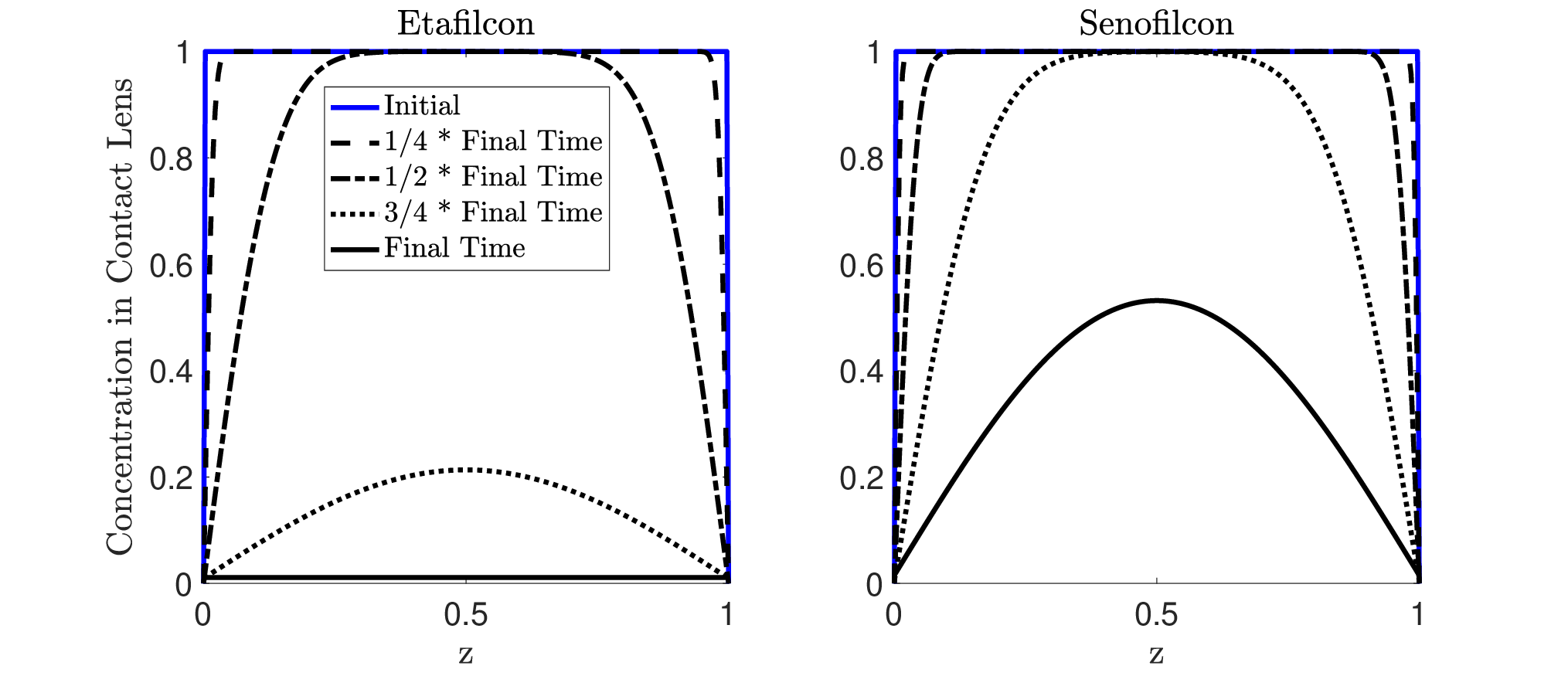}
\caption{Snapshots at different time points of the concentration in each lens in a vial over time.} 
\label{fig:vial_conc}
 \end{figure}

 We compute sample solutions with representative parameters. In Figure \ref{fig:vial_conc} we show the contact lens concentration at different times throughout the experiment. Note that the etafilcon lens has lost essentially all of its drug concentration by the final time (24 hours).

%\subsection{Vial optimization}

Of interest is the amount of drug lost from the contact lens over time. We keep track of the mass of drug in the contact lens at time $t$, given by
\begin{equation}
\label{eq:Mass_CL}
    M(t) = A_{\rm cl} \int_0^{h_{\rm cl}} C(z,t) dz.
\end{equation}
From this we determine the cumulative mass of drug lost at a given time $t$:
\begin{equation}
\label{eq:Mass_CL_Lost}
    M^{\rm lost}(t) = M^{\rm init} - A_{\rm cl} \int_0^{h_{\rm cl}} C(z,t) dz.
\end{equation}
 
 In the vial setting we can set up straightforward optimizations over the few model parameters not determined from the experimental settings of Phan \textit{et al.} \cite{phan2021development}.

 Expressed in continuous variables, we seek to minimize $||M^{\rm lost}(t) - M^{\rm lost}_{\rm exp}(t)||_2^2$ over the diffusion coefficient, $D$. Here, $M^{\rm lost}_{\rm exp}$ denotes the drug mass lost from the contact lens reported by Phan \textit{et al.} \cite{phan2021development}. The variable $t$ corresponds to the time after insertion of the contact lens into the model eye system. We use the partition coefficient $K$  found from the equilibrium balance above. 
 %All variables have been nondimensionalized with the scalings given above. 
 The norm is over all $t \in [0, T]$, where $T$ corresponds to the total length of the experiment (24 hours). The optimization problem may be written
 \begin{equation}
     \argmin_{D} ||M^{\rm lost}(t; D) - M^{\rm lost}_{\rm exp}(t)||_2^2.
 \end{equation}
 Both derivative-free and steepest descent-based algorithms are used to test consistency of the solution. These are implemented in MATLAB via \verb|fminsearch| and \verb|lsqnonlin|. Default step-size and objective function value tolerances are set.  In another case, we 
 %drop the assumption that the partition coefficient $K$ can be determined separately, and 
 let $K$ vary as a second optimization parameter. We also conduct the optimization by setting the partition coefficient equal to zero,  which corresponds to a max flux (perfect
 sink) situation, to test for an effect on the optimal diffusion coefficient $D$.
 
 \subsection{Fitting}
 \label{sec:vial_fitting}
 
We fit drug lost from the contact lens computed from our model to the experimental drug lost recorded by Phan \textit{et al.} \cite{phan2021development}. The optimal values are $D_{\rm eta} = 2.25 \times 10^{-13}$ m$^2/s$ and $D_{\rm seno} = 7.47 \times 10^{-15}$ m$^2/s$.
 The optimization converged numerically by meeting the step-size and objective function tolerances for both the etafilcon A and senofilcon A lenses.  The optimal values from the two algorithms implemented are consistent.

 \begin{figure}[h]
\centering
\includegraphics[scale=.5]{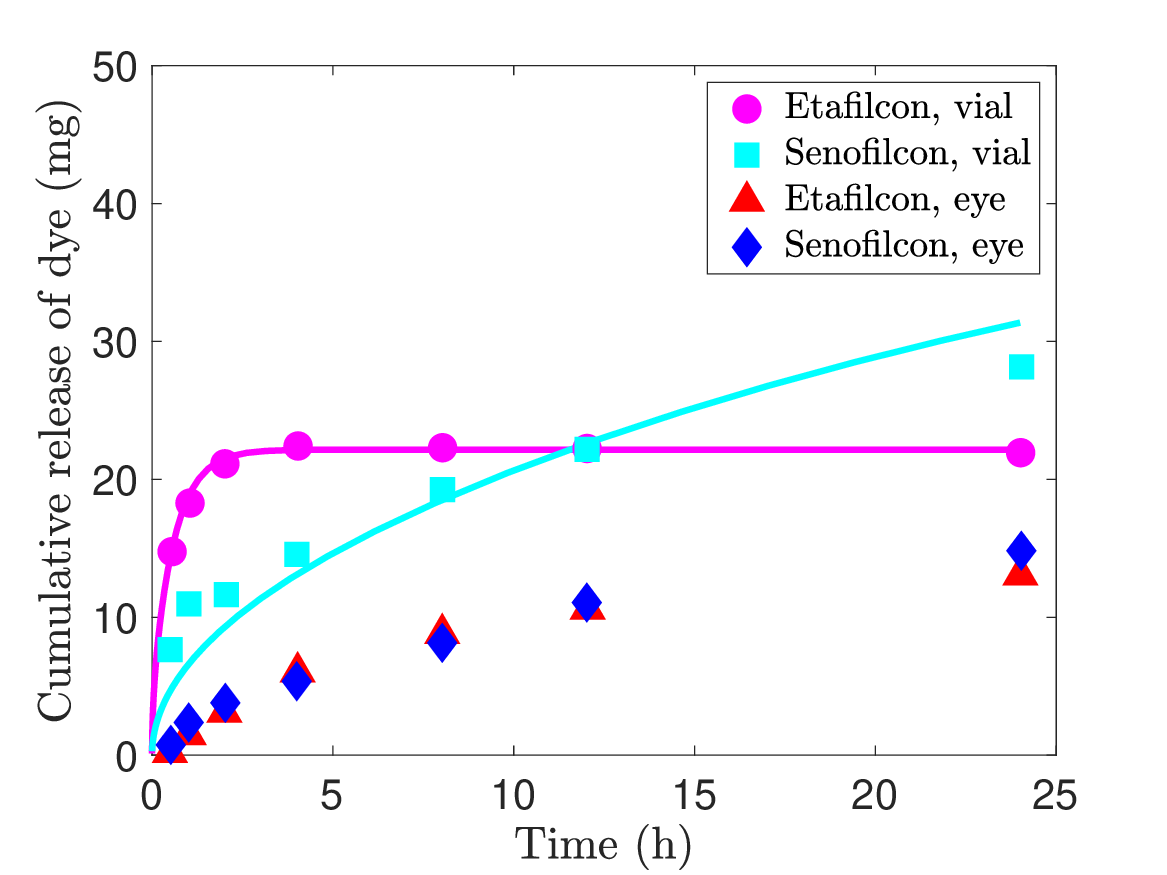}
\caption{Optimized single-variable vial model solutions with experimental data.} 
\label{fig-Phan_Vial_Fitting}
 \end{figure}

 We also optimize over both the diffusion coefficient $D$ and partition coefficient $K$. The optimal values are $D_{\rm eta} = 2.24 \times 10^{-13}$ m$^2/s$, $K_{\rm eta} = 2.07$, $D_{\rm seno} = 7.50 \times 10^{-15}$ m$^2/s$, and $K_{\rm seno} = 4.31$.  Both optimal diffusion coefficients agree with the values from the single-variable optimization to within 0.5\%. The etafilcon A partition coefficient agrees extremely well with the equilibrium balance-calculated value from above, while the senofilcon A value is about 18\% larger.  However, by varying the initial parameter guesses for the optimization, the fit seems highly insensitive to changes in $K$. In fact, by setting $K = 0$, which corresponds to a max flux (perfect sink) situation, we achieve identical-looking solutions. This is likely due to the product $K C_{\rm vial}$ being quite small and as 
 this only influences  the concentration at the lens/fluid boundaries, it may have
 a relatively small influence on the diffusive flux out of 
 the lens.
 
%  \begin{figure}[h]
%\centering
%\includegraphics[scale=.6]{Figs/4_27_23_opt_Dk.eps}
%\caption{Optimized two-variable vial model solutions with experimental data.} 
% \end{figure}
 
% \begin{figure}[H]
%\centering
%\includegraphics[scale=.4]{Figs/4_27_23_opt_Dk_conc.eps}
%\caption{Snapshots at different time points of the concentration in each lens in a vial over time. Note that the etafilcon lens has lost essentially all of its drug concentration by the final time (24 hours).} 
% \end{figure}

A notable observation from the Phan {\it et al.} experiments (e.g.~see Figure~\ref{fig-Phan_Vial_Fitting}) is that the etafilcon A lens loses essentially all of its drug mass in 24 hours whereas
the senofilcon A retains a large amount of the initial drug. This is supported by our finding that the two optimized diffusion coefficients are of different orders of magnitude, and the significantly smaller value corresponds to the senofilcon A lens.  It seems that diffusion within the contact lens is the rate-limiting  mechanism in the vial setting. In contrast, in the Phan {\it et al.}~{\it in vitro} experiments, both lenses retain a significant
portion of the original amount of drug in the lens. This must seemingly be explained by some limiting rate associated with the post-lens and/or pre-lens film.    That is, there appears to be some other rate-limiting mechanism(s), in addition to or instead of diffusion in the contact lens, 
in the {\it in vitro} eye model.  We investigate several possible mechanisms
in the framework of the eye model outlined in the next section.  %We use the parameter values as listed in Table~\ref{table-compA} to make comparison with the Phan {\it et al.} {\it in vitro} data across various versions of our model.

%\begin{equation}
%\frac{d}{dt} \left( C_{\rm pre} V_{\rm pre} \right)  =  - D \left. %\frac{\partial C}{\partial z} \right|_{z=H} A,
%\end{equation}
%where $A$ is the contact lens area ($=\pi R_{\rm CL}^2$ where $R_{\rm CL}$ is %the radius of the contact lens) and $V_{\rm pre}$ is the volume associated %with the `pre-lens half' of the vial,
%which we interpret as $V_{\rm pre} = \frac{1}{2} V_0$.
%Similarly,
%\begin{equation}
%\frac{d}{dt} \left( C_{\rm post} V_{\rm post} \right)  =  D \left. %\frac{\partial C}{\partial z} \right|_{z=0} A,
%\end{equation}
%where $V_{\rm post} = \frac{1}{2} V_0$.

%\subsection{Nondimensional model}
%
%This is a dimensionless model for vial drug release.
%\begin{equation}
%\dct = D \dczz, \quad 0 < z < 1.
%\label{eq:diff_eqn}
%\end{equation}
%
%Initial condition:
%\begin{equation}
%C(z, t = 0) = 1.
%\end{equation}
%
%Boundary conditions:
%\begin{subequations}
%\begin{eqnarray}
%C(z = 0, t) & = & K C_{\rm post}(t), \\
%C(z = 1, t) & = & K C_{\rm pre}(t),
%\end{eqnarray}
%\end{subequations}
%
%where
%\begin{subequations}
%\begin{eqnarray}
%\frac{1}{2} \frac{d C_{\rm pre}}{dt} & = & - D A \left. \dcz \right|_{z = 1}, \\
%\frac{1}{2} \frac{d C_{\rm post}}{dt} & = &  D A \left. \dcz \right|_{z = 0}.
%\end{eqnarray}
%\end{subequations}

\section{Dimensional eye model}

In the eye model, the drug-soaked contact lens is present during normal lens wear.  In this configuration, a pre-lens tear film
(between the lens and the air) and a post-lens tear film (between the lens and the cornea) surround the contact lens.  The fluid volumes of the pre- and post-lens films are both several orders of magnitude smaller than a typical vial volume.  The drug transport problem is also complicated by blinking.  As described in more detail below, we assume that when the eyelid moves the contact lens may also move,
and that between blinks (the interblink) when the eyelid is assumed stationary, the contact lens is also stationary.  
Eyelid motion and contact lens motion both can generate further, non-diffusive, transport of the drug.  Our model outlined in this section
attempts to describe the transport of drug from the contact lens into the pre- and post-lens films and further transport either lost during blinking or absorbed into surrounding tissue such as the cornea and/or eyelid.

The following model takes the form of a compartment-type model coupled to a diffusion problem in the contact lens.  We assume that the 
contact lens concentration is a function of space and time as in the vial setting, but that the pre- and post-lens thicknesses and drug concentrations are functions of time only.

\begin{figure}[h]
\centering
\includegraphics[scale=.4]{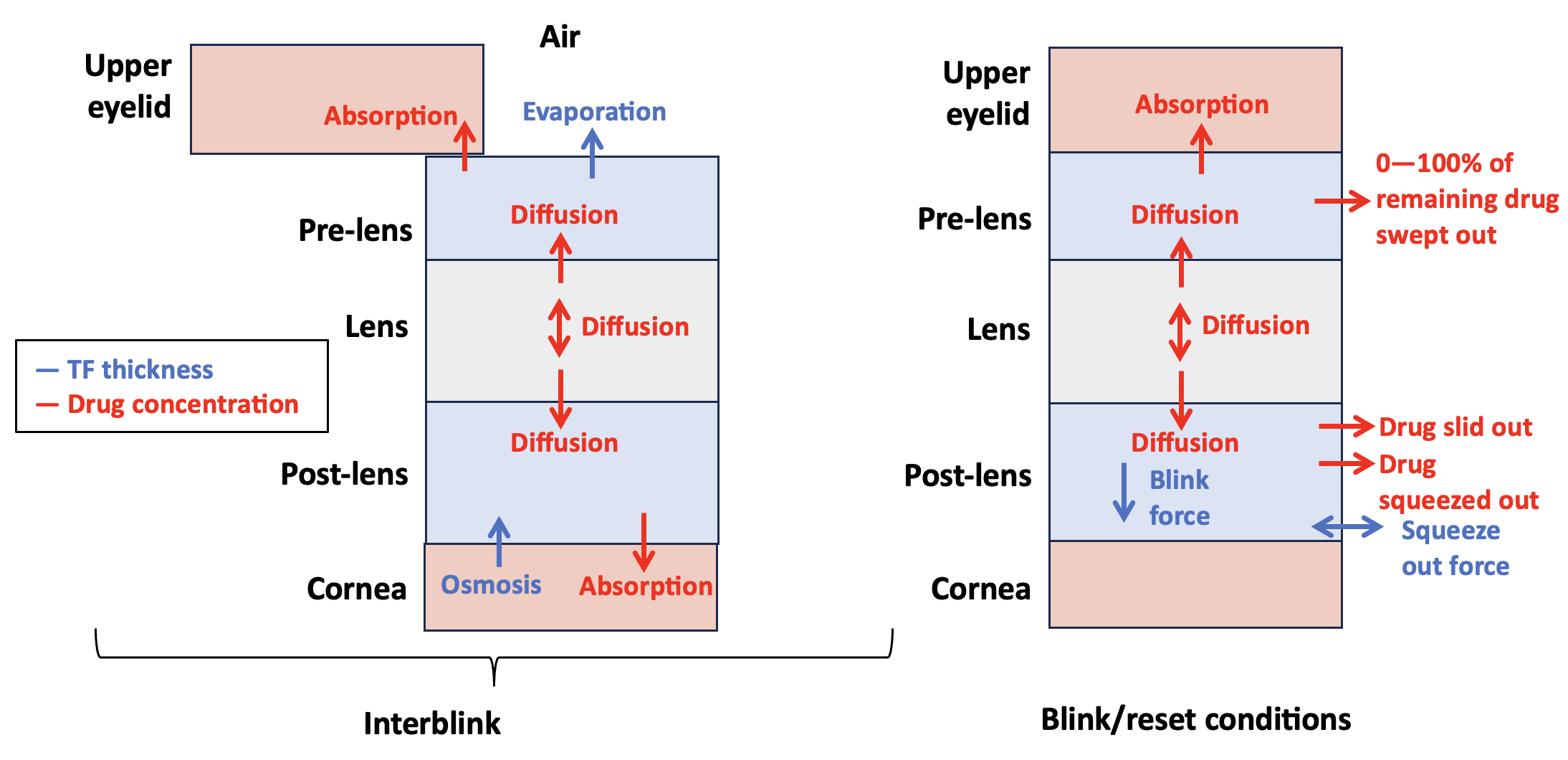}
%Figs/CLDD_model_noflow.png
%Figs/schematic_10_11_23.png}
\caption{Schematic that describes the mechanisms/processes occurring between and during blinks. Blue and red ink denote processes affecting tear film thickness and drug concentration, respectively.}
\label{fig:schematic}
\end{figure}

In Fig. \ref{fig:schematic} we represent all the mechanisms/processes occurring during the interblink and during a blink. Vertical arrows indicate processes with $z$-direction relevance, either across a compartment boundary or within the contact lens. Horizontal arrows represent rates in or out of a compartment and have no meaningful left-to-right directionality. Motivated by the experimental data from \cite{phan2021development}, we assume that the upper eyelid partially covers the region of interest affecting the contact lens during the interblink and completely covers the region of interest during a blink.  

Phan {\it et al.} \cite{phan2021development} note that  ``tear fluid first pools underneath the eyelid and is only delivered to the eyeball by the blink.'' This suggests that there is no flow in/out of the pre-lens during the interblink, and that no drug is swept out during the interblink. One possible assumption is that the blink sweeps all drug out of the pre-lens, as fresh fluid is delivered with the blink, pushing out the drug-filled fluid. As a more general assumption, outlined in more detail below, we allow for partial loss of drug mass in the pre-lens as  a consequence of a blink.

\subsection{Contact lens}

Drug transport in the contact lens is governed by the same equation as in 
the vial configuration
\begin{eqnarray}
\label{eq:Ceq}
\frac{\partial C}{\partial t} & = & D \frac{\partial^2 C}{\partial z^2},
\end{eqnarray}
for $0< z < h_{\rm cl}$.   The boundary conditions take a similar form 
\begin{eqnarray}
\label{eq:Cbc12}
C(z=0,t) = K C_{\rm post}(t), \quad
C(z=h_{\rm cl},t)  =  K C_{\rm pre}(t),
\end{eqnarray}
where now $C_{\rm post}(t)$ and $C_{\rm pre}(t)$ are the drug concentrations in the post-lens tear film and pre-lens tear film whose values are determined by the models described below.
An alternate boundary condition on the pre-lens side that we consider in some cases
is a no-flux boundary condition, $\partial C/\partial z (z=h_{\rm cl}) =0$.
Similar boundary conditions have been considered by Li \& Chauhan \cite{LC2006}.
%\cite{LC2006,PSP2015,Pimenta_etal_2016}.   
The initial condition is $C(z,t=0)=C^{\rm init}(z)$.

\subsection{Upper eyelid}

The mechanisms of eyelid absorption in a blinking tear film system are not well understood, and minimal experimental evidence exists to support theories. Phan {\it et al.} \cite{phan2021development} refer to the phenomenon as ``nonspecific absorption'' and report uptake of approximately 25 \% of released drug by the upper eyelid, therefore warranting modeling attention. To this end, we postulate a reasonable model for upper eyelid absorption of drug that is motivated by their experimental videos \cite{phan2021development}. We assume that some fraction of the eyelid remains in contact with the tear film just above the lens, and that absorption occurs during the interblink. 
%The concentration of drug in the upper eyelid is governed by a concentration difference. The corresponding term in the equation for pre-lens drug enforces conservation of mass.
%\begin{equation}
    %\frac{d(A_{\rm lid} h_{\rm lid} C_{\rm lid})}{dt} = k_{\rm lid} (C_{\rm pre} - C_{\rm lid}) \times A_{\rm overlap}.
%\end{equation}
Further, we model the upper eyelid as a sink due to its large volume relative to that of the pre-lens. The concentration of drug in the upper eyelid is governed by
\begin{equation}
    \frac{d(A_{\rm lid} h_{\rm lid} C_{\rm lid})}{dt} = k_{\rm lid} C_{\rm pre} A_{\rm overlap}.
\end{equation}
The corresponding term in the equation for pre-lens drug enforces conservation of mass.

\subsection{Pre-lens}

The pre-lens thickness $h_{\rm pre}$ evolves during the interblink according to
\begin{equation}
\label{eq:hpre}
\frac{d (A_{\rm cl} h_{\rm pre})}{dt}  =  - J_E 
%- J_T^{\rm out} + J_T^{\rm in}
,
\end{equation}
where $J_E$ is a fluid volume loss per unit time due to evaporation of the pre-lens tear film. %$J_T^{\rm out}$ represents a volumetric flow rate out of the pre-lens tear
%film and $J_T^{\rm in}$ represents a volumetric flow rate into the pre-lens tear film.   
The pre-lens drug concentration is governed during the interblink by
%\begin{equation}
%\label{eq:Cpre}
%\frac{d (A_{\rm cl} h_{\rm pre} C_{\rm pre})}{dt}  =  - D A_{\rm cl} \left. \frac{\partial C}{\partial z}\right|_{z=h_{\rm cl}} - J_T^{\rm out} C_{\rm pre} - k_{\rm lid} (C_{\rm pre} - C_{\rm lid}) \times A_{\rm overlap}.
%\end{equation}
\begin{equation}
\label{eq:Cpre}
\frac{d (A_{\rm cl} h_{\rm pre} C_{\rm pre})}{dt}  =  - D A_{\rm cl} \left. \frac{\partial C}{\partial z}\right|_{z=h_{\rm cl}} 
%- J_T^{\rm out} C_{\rm pre} 
- k_{\rm lid} C_{\rm pre} A_{\rm overlap}.
\end{equation}
This equation represents diffusive mass transport from the contact lens and absorption of drug in the eyelid.

\subsection{Post-lens}

The post-lens tear film thickness, $h_{\rm post}$, evolves during the interblink according to 
\begin{equation}
\label{eq:hpost}
\frac{d (A_{\rm cl} h_{\rm post})}{dt}  =  Q_{\rm in} - Q_{\rm out} + J_{\rm osmotic}.
\end{equation}
Here, $Q_{\rm in}$ and $Q_{\rm out}$ are non-negative flux terms describing fluid that may enter or leave the post-lens film during the interblink.  
%The forms of $Q_{\rm in}$ and $Q_{\rm out}$ are yet to be described.
The quantity $J_{\rm osmotic}$ represents an osmotic flow that may be present from the cornea into the post-lens tear film.

The post-lens drug concentration is governed during the interblink by
\begin{equation}
\label{eq:Cpost}
\frac{d (A_{\rm cl} h_{\rm post} C_{\rm post})}{dt}  =   D A_{\rm cl} \left. \frac{\partial C}{\partial z}\right|_{z=0} - k_c C_{\rm post} A_{\rm cl}  -  C_{\rm post} Q_{\rm out}.
\end{equation}
This equation represents diffusive mass transport from the contact lens, absorption of drug into the cornea, and advective transport of drug mass out of the lens.  We assume that there is no advective transport of drug mass into the post-lens during the interblink associated with $Q_{\rm in}$.

When we compare with the Phan {\it et al.} \cite{phan2021development} 
experimental data we 
note that their ``cornea'' is made out of hydrophobic material and hence
$k_c=0$.  In those comparisons we also assume that
$Q_{\rm in}=0$, $Q_{\rm out}=0$, and $J_{\rm osmotic}=0$, so that the post-lens thickness remains constant during an interblink.

\subsection{Reset conditions}
\label{sec-reset_conditions}

In general, a blink is modeled by stopping the evolution defined by equations~(\ref{eq:Ceq}), (\ref{eq:hpre}), (\ref{eq:Cpre}), (\ref{eq:hpost}), 
and (\ref{eq:Cpost}) at time $t=t_{\rm blink}$, and restarting them with new initial conditions
\begin{subequations}
\begin{eqnarray}
C(z,t_{\rm blink}^+)  & = &  C(z,t_{\rm blink}^{-}), \\
C_{\rm pre}(t_{\rm blink}^+) & = & (1-p) C_{\rm pre}(t_{\rm blink}^-), \\
h_{\rm pre}(t_{\rm blink}^+) & = &  h_{\rm pre}^{\rm init},\\
C_{\rm post}(t_{\rm blink}^+)  & = & \begin{cases}
    C_{\rm post}^{\rm squeeze} \\
    C_{\rm post}^{\rm slide}
\end{cases}, \\
h_{\rm post}(t_{\rm blink}^+) & = & h_{\rm post}^{\rm init}, \\
C_{\rm lid}(t_{\rm blink}^+) & = & C_{\rm lid}(t_{\rm blink}^-),
\end{eqnarray}
\end{subequations}
where $t_{\rm blink}^{-}$ and $t_{\rm blink}^{+}$ indicate the times just before and just after a blink.
The interpretations, respectively, are 

\begin{enumerate}
\item[(21a)] the concentration in the lens is assumed to be the same immediately before and after a blink,
\item[(21b)] the blink washes out a certain proportion of the drug from the pre-lens where $p \in [0,1]$.  For
example, $p=0$ means no pre-lens drug mass is lost during a blink (the pre-lens concentration resets to its pre-blink value) and $p=1$ means that all pre-lens drug mass is lost as the result of a blink (the pre-lens concentration resets to zero),
\item[(21c)] the pre-lens  thickness resets to the initial value after a blink,
\item[(21d)] the concentration in the post-lens after the blink is the determined by motion of the contact lens for either squeezing out or sliding out mechanisms (outlined in more detail below), 
\item[(21e)] the post-lens thickness resets to its initial value after a blink,
\item[(21f)]  and the concentration in the upper eyelid is assumed to be the same immediately before and after a blink.
\end{enumerate} 
The cases for $C_{\rm post}(t_{\rm blink}^+)$ warrant further explanation. In general, the eyelid motion that occurs during a blink generates forces that
can result in motion of the contact lens in the posterior/anterior directions (in and out), superior/inferior directions (up and down), and lateral/medial directions (side to side) (e.g.~see Gilman \cite{Gilman1982} and  Hayashi \cite{Hayashi1977}).  Lateral/medial motion is
expected to be small and is neglected in our study.  Posterior/anterior motion is typically characterized by some type of lubrication/squeeze flow type model (e.g.~Chauhan \& Radke \cite{ChauhanRadke2002} and Maki \& Ross \cite{MR2014a,MR2014b}).  Other work has explored sliding-type motion described by
Couette flow (e.g.~Chauhan \& Radke \cite{ChauhanRadke2001} and Anderson {\it et al.} \cite{anderson2021tear}) or combined squeezing--sliding effects (e.g.~Creech {\it et al.} \cite{Creech_etal2001}, Li \& Chauhan \cite{LC2006}, and Dunn {\it et al.} \cite{Dunn_etal2013}).
Although posterior/anterior and superior/inferior motions of the contact lens can occur simultaneously 
during blinking and can be included together in 
fluid dynamics models as in earlier studies noted above,  we choose to examine these two types of motion independently. This treatment will allow us to examine how each of these types of motion may or may not play a role in the experimental observations of drug release in the 
eye model of Phan {\it et al.} \cite{phan2021development}.
Therefore, we outline below the reset conditions for our model that include posterior/anterior motion (squeeze out of drug) and superior/inferior motion (slide out of drug).

%\begin{figure}[H]
%\centering
%\includegraphics[scale=.225]{Figs/interblink_diagram.pdf}\\ \vspace{-0.25in}
%\includegraphics[scale=.225]{Figs/squeeze_out_downstroke_diagram.pdf}
%\includegraphics[scale=.225]{Figs/slide_out_downstroke_diagram.pdf}\\ %\vspace{-0.25in}
%\includegraphics[scale=.225]{Figs/fill_in_upstroke_diagram.pdf}
%\includegraphics[scale=.225]{Figs/slide_out_upstroke_diagram.pdf}\\
%\vspace{-0.25in}
%\caption{Schematics for interblink (top), squeeze out/fill in (left, middle and bottom) and slide out (right, middle and bottom) in the post-lens.  We assume that any drug mass that escapes from under the contact lens by either sliding or squeezing mechanisms does not return in the upstroke.}
%\label{fig:squeeze_and_slide}
%\end{figure}

\begin{figure}[h]
\centering
\includegraphics[scale=.32]{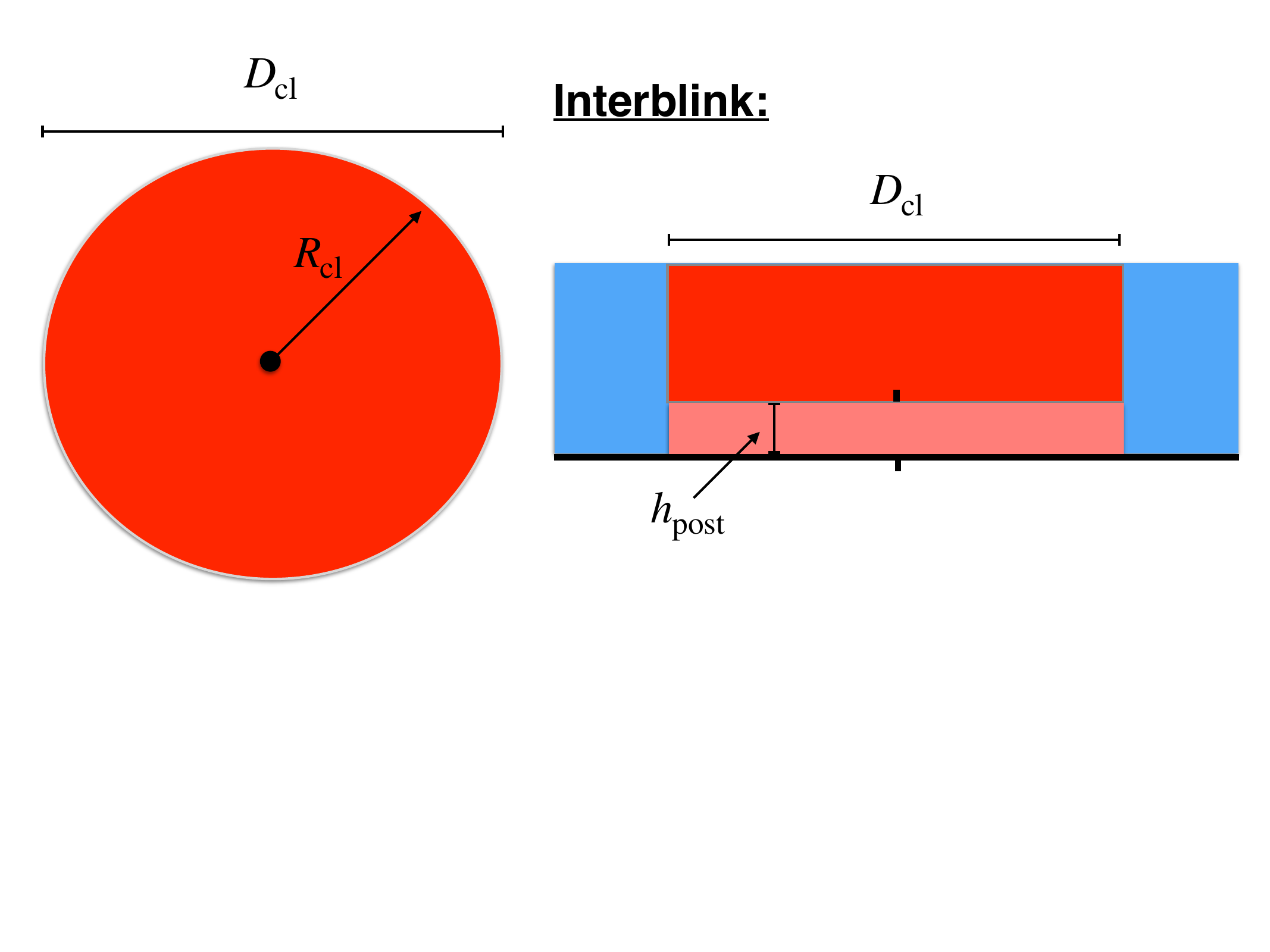}\\ \vspace{-1.25in}
\includegraphics[scale=.32]{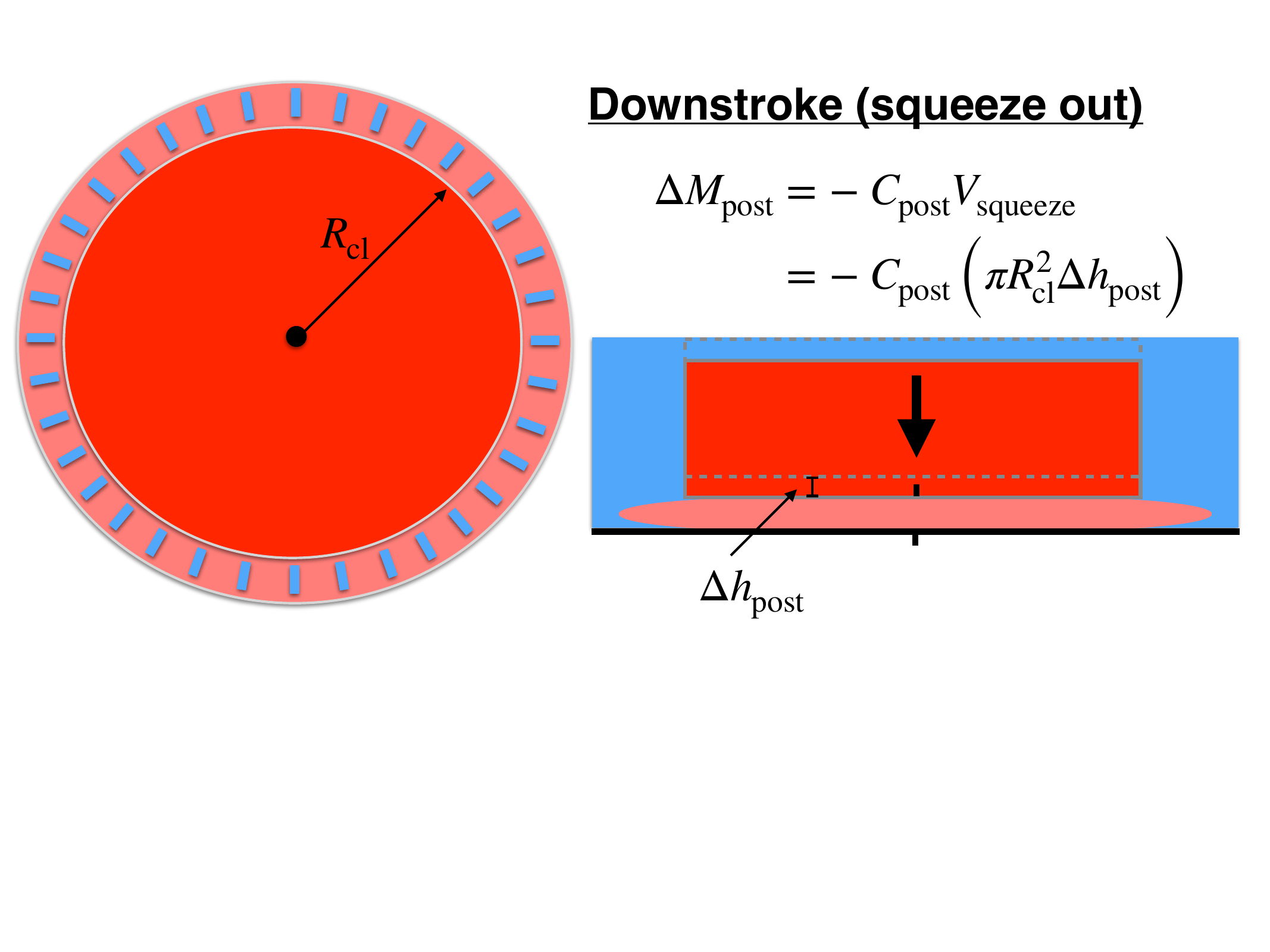}\\
\vspace{-1.15in}
\includegraphics[scale=.32]{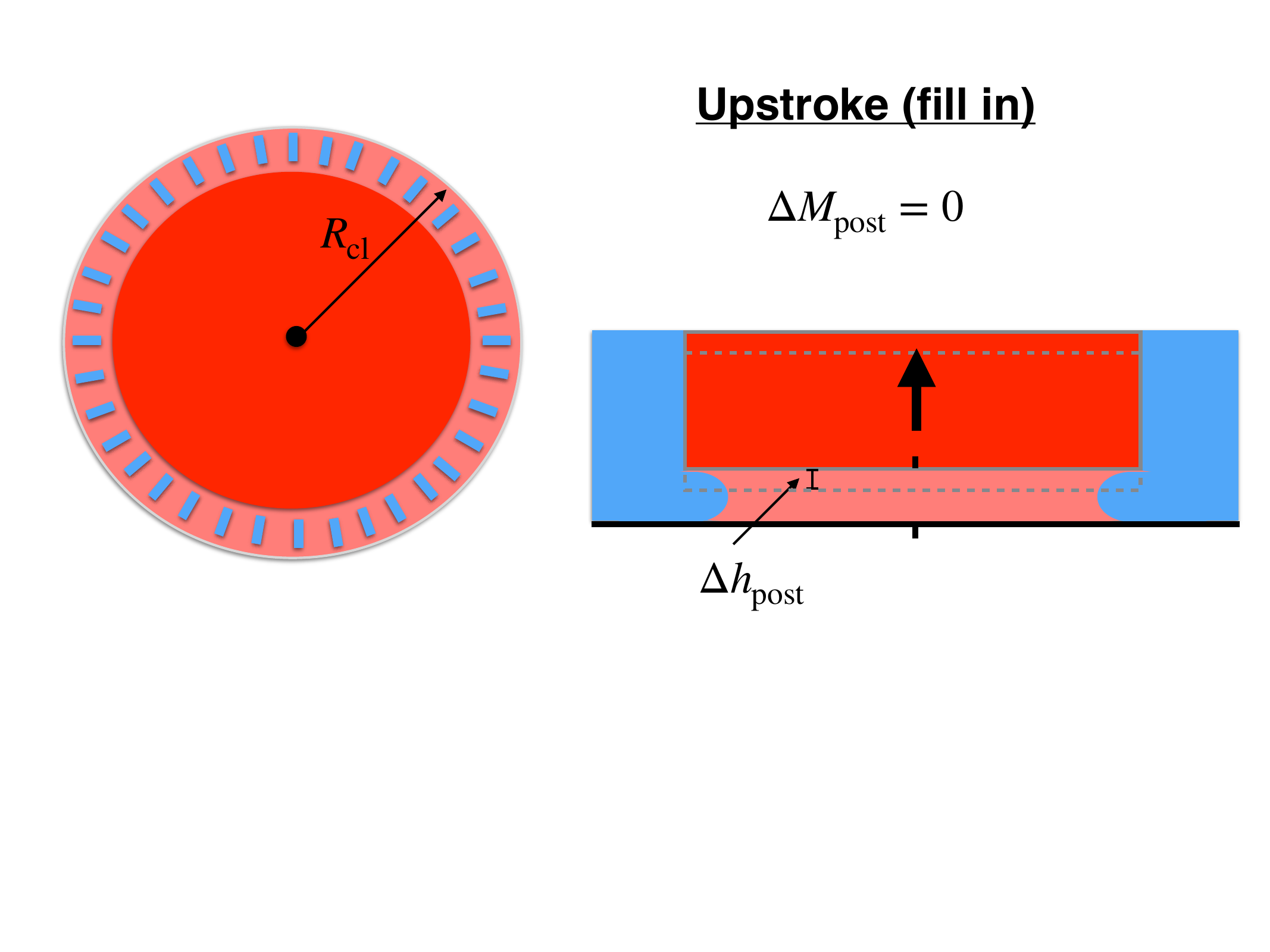}
\vspace{-1.25in}
\caption{Schematics for interblink (top), squeeze out (middle), and fill in (bottom) in the post-lens.  We assume that any drug mass that squeezed out
during the downstroke is permanently lost from the post-lens and does not return during the upstroke.}
\label{fig:squeeze_out}
\end{figure}

\paragraph{Squeeze out of drug.} %We represent the effect of contact lens movement as a reset condition. 
In the case of squeeze out of drug, we assume 
that the blink generates posterior/anterior contact lens
motion (normal to the ocular surface with no tangential motion of the lens along the ocular surface) such that 
the post-lens depresses and then returns to its original thickness by the end of the blink duration.   Our assumption that all of the contact lens motion occurs during the blink -- the downward depression as well as the upward relaxation -- is a model convenience that allows us to assign drug mass
lost from the post-lens tear film as a result of a blink directly as a blink/reset condition.
We note that more sophisticated models of post-lens tear film suggest that the relaxation stage may actually occur over the entire interblink period (e.g.~see Figures 2, 4, and 6 of Maki \& Ross \cite{MR2014b}).
We use a classical squeeze film theory result in radial coordinates to find the depressed post-lens thickness value in order to compute the mass of drug lost.
The dynamics of squeezing the film are modeled by
\begin{equation}
\frac{d h_{\rm post}^{\rm squeeze} }{dt}  =  - \frac{2F}{3\pi \mu R_{cl}^4} (h_{\rm post}^{\rm squeeze})^3,
\end{equation}
where $R_{\rm cl}$ is the radius of the contact lens, $F$ is the force applied ($F > 0$ for a downward force that generates $dh_{\rm post}^{\rm squeeze}/dt <0$), and $\mu$ is the viscosity of the fluid.
Rearranging, integrating, and applying $h_{\rm post}^{\rm squeeze}(t=0) = h_{\rm post}^{\rm init}$ gives
\begin{equation}
\label{eq:h_new_value}
 h_{\rm post}^{\rm squeeze}(t=\Delta t_{\rm blink})  =  \left[ \frac{4F}{3\pi \mu R_{\rm cl}^4} \Delta t_{\rm blink} + \left( h_{\rm post}^{\rm init} \right)^{-2} \right]^{-1/2}.
\end{equation}
This gives a formula for the depressed post-lens thickness as a function of its pre-blink thickness, the time duration associated with the blink, $\Delta t_{\rm blink}$, and the blink force $F >0$.  Note that $h_{\rm post}^{\rm squeeze}(\Delta t_{\rm blink}) < h_{\rm post}^{\rm init}$ when
$\Delta t_{\rm blink}>0$. We can either use the blink force $F$ as a parameter or consider a depression amount $\Delta h_{\rm post} = h_{\rm post}^{\rm init} - h_{\rm post}^{\rm squeeze}(\Delta t_{\rm blink})$, and then express the required force $F$ via
\begin{equation}
F = \frac{3 \pi \mu R_{\rm cl}^4}{4 \Delta t_{\rm blink}} \left[ \left( h_{\rm post}^{\rm init} - \Delta h_{\rm post}\right)^{-2} - \left(h_{\rm post}^{\rm init} \right)^{-2} \right].
\end{equation}
The mass of drug lost from the post-lens as a result of this squeeze out process can be found via
\begin{equation}
    \Delta M_{\rm post} = -C_{\rm post} V_{\rm squeeze} = -C_{\rm post} A_{\rm cl} \Delta h_{\rm post}.
\end{equation}
For a schematic see Figure~\ref{fig:squeeze_out}.
To compute the concentration at the end of a blink that results from the squeeze out of drug, we determine $C_{\rm post}(t_{\rm blink}^+) = C_{\rm post}^{\rm squeeze}$ via
\begin{equation}
 C_{\rm post}^{\rm squeeze} = \frac{M_{\rm post}(t_{\rm blink}^-) + \Delta M_{\rm post}}{V_{\rm post}(t_{\rm blink}^+)}  = \frac{C_{\rm post}(t_{\rm blink}^-) A_{\rm cl} [ h_{\rm post}^{\rm init} - \Delta h_{\rm post} ]}{h_{\rm post}^{\rm init}A_{\rm cl}} = C_{\rm post}(t_{\rm blink}^-) \frac{h_{\rm post}^{\rm squeeze}}{h_{\rm post}^{\rm init}}. 
 \end{equation}

 \paragraph{Slide out of drug.} From the video from 
 Phan {\it et al.}~\cite{phan2021development}, it is evident that there is (superior/inferior) sliding motion of the contact lens as the result of a blink (it is difficult to confirm or rule out the squeeze out mechanism from the video). The geometry under consideration is approximated to be a circular disk (the contact lens) of radius $R_{\rm cl}$ separated from a flat surface (the eye) by a thin fluid layer of thickness $h_{\rm post}$. In this context we assume the post-lens thickness to be constant.
The superior/inferior motion of the lens is characterized by a velocity component $U_{\rm cl}(t)$.  A very simple model
for the unidirectional fluid velocity in this post-lens region is Couette flow of the form
\begin{equation}
u(z,t) = \frac{z}{h_{\rm post}} U_{\rm cl}(t),
\end{equation}
where $z \in [0,h_{\rm post}]$.  Our main objective is to estimate the net volume flux out of the post-lens region, i.e., out from under the contact lens at the lens periphery.

 \begin{figure}[h]
\centering
\includegraphics[scale=.32]{Figs/interblink_diagram.pdf}\\ \vspace{-1.05in}
\includegraphics[scale=.32]{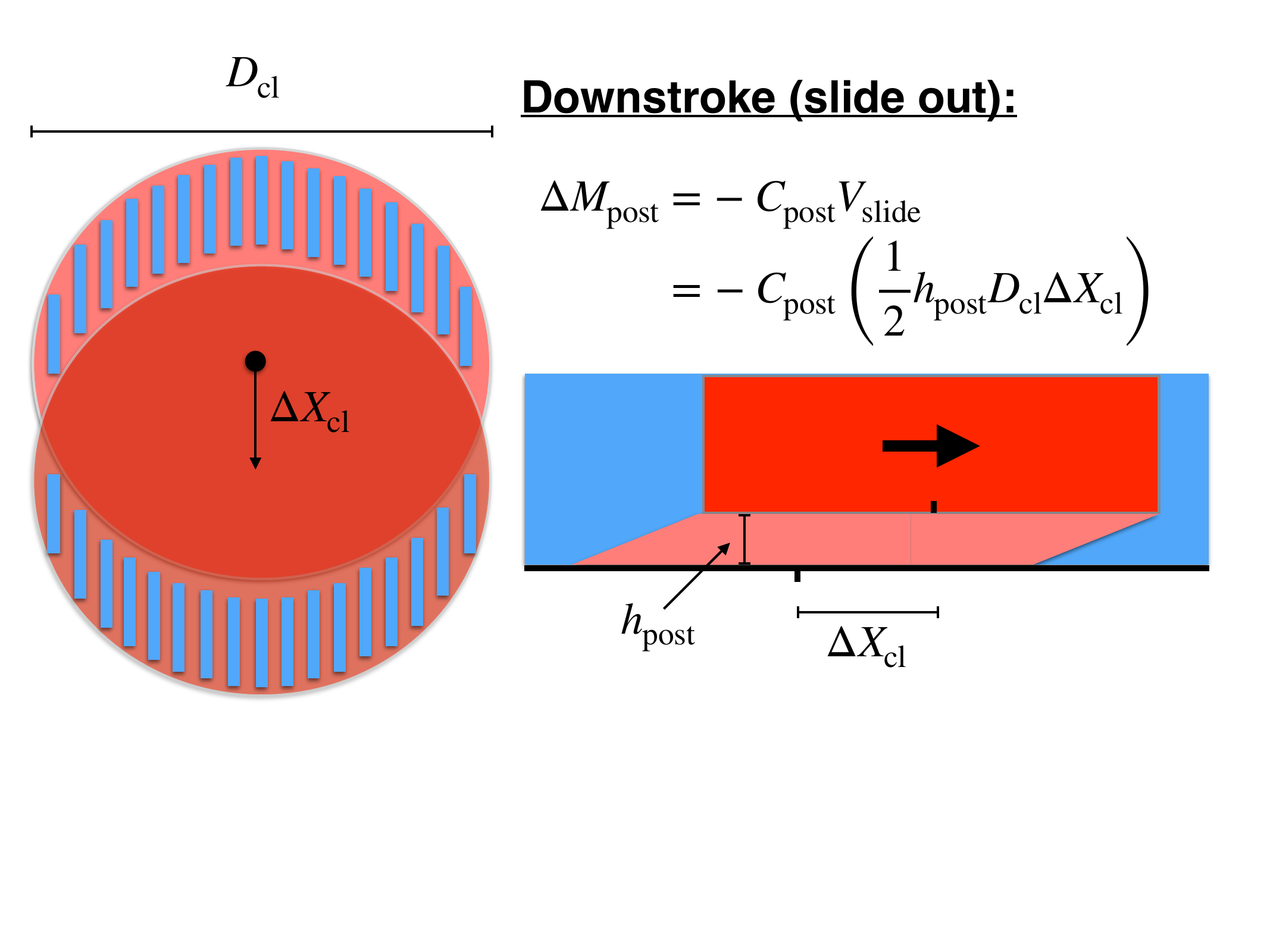}\\
\vspace{-0.75in}
\includegraphics[scale=.32]{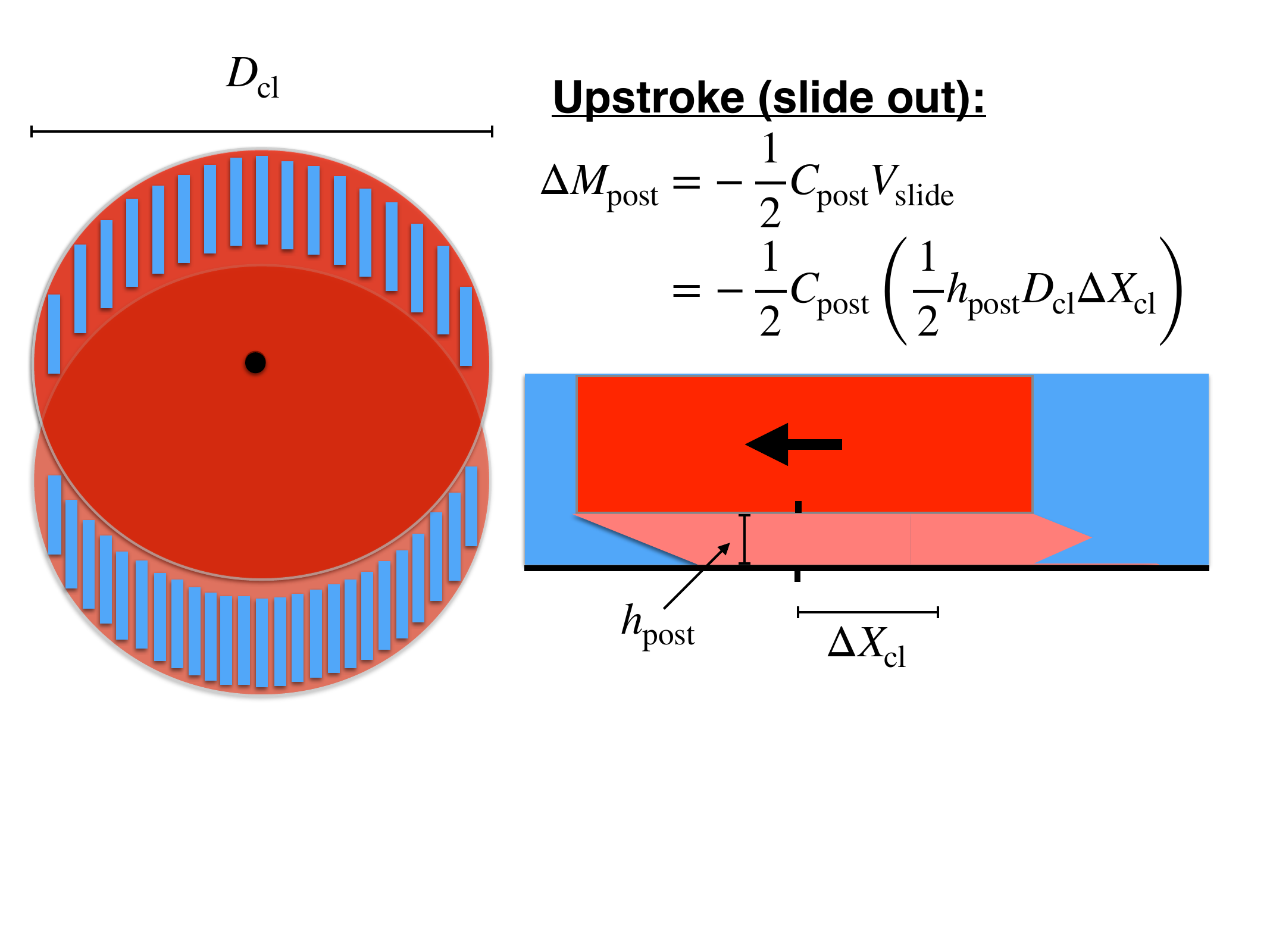}
\vspace{-0.75in}
\caption{Schematics for interblink (top), slide out downstroke (middle), and slide out upstroke (bottom) in the post-lens.  We assume that any drug mass that escapes from under the contact lens (e.g. on the upper lid/left side
in the downstroke schematic) does not return.}
\label{fig:slide_out}
\end{figure}

The volumetric flow in/out of the upper half of the circular disk from time $t=0$ to time $t=T$, say during downward motion of the lens, is
\begin{equation}
V_{\rm slide}  =  \int_0^T \int_0^\pi \int_0^{h_{\rm post}} u(z,t) \hat{j} \cdot \hat{n}_{\rm disk} \ R_{\rm cl} dz \ d\theta \ dt.
\end{equation}
Here $\hat{j}$ is a unit vector in the vertical direction (i.e.~up for a standing contact lens wearer), and $\hat{n}_{\rm disk} = (\cos \theta, \sin \theta,0)$
is the unit outward normal of the circular disk in the radial direction.  Our assumption is that $u(z,t)$ is independent of $\theta$ and so the space portion
of these integrals can be evaluated to give
\begin{align}
V_{\rm slide} & =  \int_0^T \int_0^\pi \int_0^{h_{\rm post}} u(z,t) \hat{j} \cdot \hat{n}_{\rm disk} R_{\rm cl} \  dz \  d\theta \  dt, \nonumber \\
& =  R_{\rm cl} \int_0^T U_{\rm cl}(t) \left( \int_0^\pi \sin \theta d\theta \right) \left( \int_0^{h_{\rm post}} \frac{z}{h_{\rm post}} dz \right) dt, \nonumber \\
& =  R_{\rm cl} \int_0^T U_{\rm cl}(t) \left(2 \right) \left( \frac{1}{2} h_{\rm post} \right) dt, \nonumber \\
& =  R_{\rm cl} h_{\rm post} \int_0^T  U_{\rm cl} dt, \nonumber \\
& =  R_{\rm cl} h_{\rm post} \Delta X_{\rm cl} = \frac{1}{2} D_{\rm cl} h_{\rm post} \Delta X_{\rm cl},
\end{align}
%
%If we now assume that $h_{\rm post}$ is a constant, i.e., no squeeze out motion occurs simultaneously with slide out motion,  then
%\begin{align}
%V_{\rm slide} & =  R_{\rm cl} h_{\rm post}  \int_0^T U_{\rm cl} dt, \nonumber \\
%& =  R_{\rm cl} h_{\rm post} \Delta X_{\rm cl} = \frac{1}{2} D_{\rm cl} h_{\rm post} \Delta X_{\rm cl},
%\end{align}
where $D_{cl}$ is the diameter of the contact lens and $\Delta X_{\rm cl}$ is the transverse distance that the contact lens travels in time $T$.  
Some experimental studies report $\Delta X_{\rm cl} = 1$--$4$ mm (see Hayashi \cite{Hayashi1977},
Gilman \cite{Gilman1982}) while others observe
smaller values $\Delta X_{\rm cl} = 0.1$--$0.3$mm
(see Cui {\it et al.} \cite{Cui_etal2012} and Wolffsohn {\it et al.} \cite{Wolffsohn_etal2009}).
Theoretical models show that even without the complications of elastic deformation of the contact lens and curvature of the lens and ocular surface, prediction of transverse contact lens motion is
sensitive to post-lens thickness \cite{anderson2021tear}.

In terms of mass transfer of drug in/out of the post-lens we need to distinguish portions of the lens that are `exposing' the 
previous post-lens from those that are `creating' new post-lens. A blink comprises two components:
a downstroke (eye closing) where $U_{\rm cl} <0$ and an upstroke (eye opening) where $U_{\rm cl} >0$ and drug
mass can be lost in both of these stages. 
For a schematic see Figure~\ref{fig:slide_out}.
%If we assume that any fluid entering the postlens region has concentration, $C_{\rm upper lid}$ and $C_{\rm lower lid}$ for the two lid regions, (possibly we just take $C_{\rm upper lid}=C_{\rm lower lid}  = 0$)
In particular, we assume the following situations hold for post-lens drug mass lost as a result of lens sliding motion:
\begin{itemize}
%\item Interblink:
%\begin{equation}
%L_{\rm post}  =  0.
%\end{equation}
\item Downstroke:
\begin{equation}
\Delta M_{\rm post}^{\rm down}  =  - C_{\rm post} V_{\rm slide} = - C_{\rm post} \left( \frac{1}{2} h_{\rm post} D_{\rm cl} \Delta X_{\rm cl} \right).
\end{equation}
\item Upstroke:
\begin{equation}
\Delta M_{\rm post}^{\rm up}  =   - \frac{1}{2} C_{\rm post} V_{\rm slide} = -\frac{1}{2} C_{\rm post} \left( \frac{1}{2} h_{\rm post} D_{\rm cl} \Delta X_{\rm cl} \right).
\end{equation}
\end{itemize}
We assume Couette flow occurs during the upstroke as well; this dragging flow results in an additional factor of one-half to calculate the amount of drug lost (again
see Figure~\ref{fig:slide_out}).
Then, over one blink (downstroke plus upstroke), we have net loss of drug mass in the post-lens given by 
\begin{equation}
\Delta M_{\rm post}  =  -\frac{3}{4}  C_{\rm post} h_{\rm post} D_{\rm cl}\Delta X_{\rm cl} .
\end{equation}
To compute the concentration at the end of a blink that results from the slide out of drug, we determine $C_{\rm post}(t_{\rm blink}^+) = C_{\rm post}^{\rm slide}$ via
\begin{equation}
 C_{\rm post}^{\rm slide} = \frac{M_{\rm post}(t_{\rm blink}^-) + \Delta M_{\rm post}}{V_{\rm post}(t_{\rm blink}^+)}  
 %= \frac{C_{\rm post}(t_{\rm blink}^-) h_{\rm post}^{\rm init}\left[ A_{\rm CL} - \frac{3}{4}  D_{\rm CL} \Delta X_{\rm CL} \right]}{h_{\rm post}^{\rm init}A_{\rm CL}} 
 = C_{\rm post}(t_{\rm blink}^-) \frac{\left[ A_{\rm cl} - \frac{3}{4}  D_{\rm cl} \Delta X_{\rm cl} \right]}{A_{\rm cl}} = C_{\rm post}(t_{\rm blink}^-)  \left(1 - \frac{3}{2} \frac{\Delta X_{\rm cl}}{\pi R_{\rm cl}} \right). 
 \end{equation}
%Here $V_{\rm slide}$ is interpreted to be the positive volume
%\begin{equation}
%V_{\rm slide} = R_{\rm CL} h_{\rm post} \Delta X_{\rm CL}.
%\end{equation}

\section{Nondimensional eye model}
\label{sec:nondim_eye_model}

We introduce dimensionless variables
\begin{equation}
\bar{C} = \frac{C}{C^{\rm init}},\quad
\bar{z} = \frac{z}{h_{\rm cl}},\quad
\bar{t} = \frac{t}{\tau},\quad
\bar{h}_{\rm pre} = \frac{h_{\rm pre}}{h_{\rm cl}},\quad
\bar{h}_{\rm post} = \frac{h_{\rm post}}{h_{\rm cl}},
\end{equation}
where $\tau$ is a typical time scale associated with blinking (e.g.~$10$ seconds in Phan {\it et al.} \cite{phan2021development}). 

{\bf Interblink:} The corresponding dimensionless model is 
\begin{eqnarray}
\label{eq:CL_dimensionless}
\frac{\partial \bar{C}}{\partial \bar{t}} & = & \bar{D} \frac{\partial^2 \bar{C}}{\partial \bar{z}^2},
\end{eqnarray}
in the contact lens $0< \bar{z} < 1$, subject to initial condtion $\bar{C}(\bar{z},\bar{t}=0) = 1$ and the boundary conditions 
\begin{eqnarray}
\label{eq:Cbc12_dimensionless}
\bar{C}(\bar{z}=0,\bar{t}) = K \bar{C}_{\rm post}(\bar{t}), \quad
\bar{C}(\bar{z}=1,\bar{t})  =  K \bar{C}_{\rm pre}(\bar{t}),
\end{eqnarray}
(or $\partial \bar{C}/\partial \bar{z} =0$ at $\bar{z}=1$ 
instead for the no-flux condition on the pre-lens side).
In the pre-lens tear film during an interblink we have
\begin{eqnarray}
\label{eq:pre_dimensionless}
\frac{d \bar{h}_{\rm pre}}{d\bar{t}}  & = &  - \bar{J}_E %+ \bar{J}_T^{\rm in} - \bar{J}_T^{\rm out}
,\\
\label{eq:pre_dimensionless_Ceq}
\frac{d ( \bar{h}_{\rm pre} \bar{C}_{\rm pre})}{d\bar{t}} & = & - \bar{D}  \left. \frac{\partial \bar{C}}{\partial \bar{z}}\right|_{\bar{z}=1} 
%- \bar{J}_T^{\rm out} \bar{C}_{\rm pre} 
- \bar{k}_{\rm lid} \bar{C}_{\rm pre} ,
\end{eqnarray}
where dimensionless diffusion coefficient, evaporation parameter, 
%tear flow parameters, 
and lid permeability constant are given by 
\begin{equation}
\bar{D} = \frac{D \tau}{h_{\rm cl}^2}, \quad \bar{J}_E = \frac{J_E \tau}{h_{\rm cl} A_{\rm cl}},\quad
%\bar{J}_T^{\rm out} = \frac{J_T^{\rm out} \tau}{h_{\rm CL} A_{\rm CL}}, \quad 
%\bar{J}_T^{\rm in} = \frac{J_T^{\rm in} \tau}{h_{\rm CL} A_{\rm CL}}, \quad 
\bar{k}_{\rm lid} = \frac{k_{\rm lid} \tau}{h_{\rm cl}}\frac{A_{\rm overlap}}{A_{\rm cl}}.
\end{equation}
In the post-lens film during an interblink we have
\begin{eqnarray}
\label{eq:post_dimensionless}
\frac{d \bar{h}_{\rm post}}{d\bar{t}}  & = &  \bar{Q}_{\rm in} - \bar{Q}_{\rm out} + \bar{J}_{\rm osmotic},\\
\label{eq:post_dimensionless_Ceq}
\frac{d  (\bar{h}_{\rm post} \bar{C}_{\rm post})}{d\bar{t}}  & = &   \bar{D}  \left. \frac{\partial \bar{C}}{\partial \bar{z}}\right|_{\bar{z}=0} - \bar{k}_c \bar{C}_{\rm post}   -  \bar{C}_{\rm post} \bar{Q}_{\rm out},
\end{eqnarray}
where
$\bar{Q}_{\rm in} = Q_{\rm in} \tau/(A_{\rm cl} h_{\rm cl})$,
$\bar{Q}_{\rm out} = Q_{\rm out} \tau/(A_{\rm cl} h_{\rm cl})$,
$\bar{J}_{\rm osmotic} = Q_{\rm osmotic} \tau/(A_{\rm cl} h_{\rm cl})$,
and
$\bar{k}_c = k_c \tau/h_{\rm cl}$. 

In the upper eyelid during the interblink we have
\begin{equation}
    \frac{d \bar{C}_{\rm lid}}{d\bar{t}} = \bar{k}_{\rm lid} V_{\rm ratio} \bar{C}_{\rm pre},
\end{equation}
where $V_{\rm ratio} = V_{\rm cl}/V_{\rm lid} = (h_{\rm cl} A_{\rm cl})/(h_{\rm lid} A_{\rm lid})$. Thus, we expect the nondimensional permeability constant for the lid, $\bar{k} V_{\rm ratio}$, to be much smaller than that for the pre-lens due to their relative volumes ($V_{\rm ratio}$ is about 0.08 or 0.09 depending on which lens is selected).

{\bf Blink/reset:} Equations~(\ref{eq:CL_dimensionless}), (\ref{eq:pre_dimensionless}), (\ref{eq:pre_dimensionless_Ceq}), (\ref{eq:post_dimensionless}), and (\ref{eq:post_dimensionless_Ceq}) are solved
from $\bar{t} = 0$ to $\bar{t}= \bar{t}_{\rm blink}^-$.   Then, to account for the effects of a blink, we define the following reset conditions
and restart the evolution with initial conditions at $\bar{t}_{\rm blink}^+$ defined by: 
\begin{subequations}
\begin{eqnarray}
\bar{C}(z,\bar{t}_{\rm blink}^+)  & = & \bar{C}(z,\bar{t}_{\rm blink}^{-}), \\
\bar{C}_{\rm pre}(\bar{t}_{\rm blink}^+) & = & (1-p) \bar{C}_{\rm pre}(\bar{t}_{\rm blink}^-), \\
\bar{h}_{\rm pre}(\bar{t}_{\rm blink}^+) & = &\bar{h}_{\rm pre}^{\rm init},\\
\bar{C}_{\rm post}(\bar{t}_{\rm blink}^+)  & = & \begin{cases} \bar{C}_{\rm post}^{\rm squeeze} \\
\bar{C}_{\rm post}^{\rm slide} \end{cases}\\
\bar{h}_{\rm post}(\bar{t}_{\rm blink}^+) & =  &\bar{h}_{\rm post}^{\rm init}, \\
\bar{C}_{\rm lid}(\bar{t}_{\rm blink}^+) & = & \bar{C}_{\rm lid}(\bar{t}_{\rm blink}^-).
\end{eqnarray}
\end{subequations}
In these expressions
\begin{equation}
 \bar{h}_{\rm pre}^{\rm init} = \frac{h_{\rm pre}^{\rm init}}{h_{\rm cl}}, \quad \bar{h}_{\rm post}^{\rm init} = \frac{h_{\rm post}^{\rm init}}{h_{\rm cl}}.
 \end{equation}

 %Note that in Eq. \ref{eq:Cpost} we include the term $-k_c C_{\rm post}$ to model the absorption of drug into the cornea with rate $k_c$, but for the Phan et al. model, $k_c = 0$ because the ``cornea'' is made out of hydrophobic material. Thus, in the versions of the model that follow, $k_c$ is set to zero.

\subsection{Quantities of interest}

As in the vial setting, of interest is the amount of drug in each model compartment over time. Specifically, at any 
given time $t$, we keep track of the mass of drug in the contact lens, $M(t)$, given by
equation~(\ref{eq:Mass_CL}) as well as the cumulative
mass of drug lost, $M^{\rm lost}(t)$, given by
equation~(\ref{eq:Mass_CL_Lost}).
%\begin{equation}
%    M(t) = A_{\rm cl} \int_0^{h_{\rm cl}} C(z,t) dz.
%\end{equation}
%From this we determine the cumulative mass of drug lost at a given time $t$:
%\begin{equation}
%    M^{\rm lost}(t) = M^{\rm init} - A_{cl} \int_0^{h_{\rm cl}} C(z,t) dz.
%\end{equation}
The quantity $M^{\rm lost}(t)$ is the specific one reported in the experimental work 
of Phan \textit{et al.} \cite{phan2021development} for both the vial setting and the eye model setting.
This drug lost from the contact lens can be decomposed into the amounts in the other compartments:
\begin{equation}
\begin{split}
    M^{\rm lost}(t) & = \sum (\text{Collected from pre-lens}) + \sum (\text{Collected from post-lens}) \\
    & + \sum (\text{Absorbed by lid})  + \sum ( \text{In post-lens})+ \sum (\text{In pre-lens}).
    \end{split}
 \end{equation}
 Before presenting solutions of this model, 
 in the next section we outline a simplification of this model that admits a closed form solution for these quantities of interest and provides some reference and context for parameter selection in the full eye model.

%%%%%%%%%%%%%%%%%%%%%%%%%%%%%%%%%%%%%%%%

%\subsection{Model solutions}

\section{Large diffusion limiting case}
\label{sec:large_diff}
A mathematically-appealing limit of the dimensionless model outlined in Section~\ref{sec:nondim_eye_model}
has $\bar{D} \gg 1$, with a consequence being that the drug concentration in the contact lens is
approximately uniform in space (time dependent only).  Technically speaking, for the known parameter values associated with the two contact lenses of interest in this study, $\bar{D}$ does not have this property (for the etafilcon A lens $\bar{D} \approx 5 \times 10^{-4}$ and for the senofilcon A lens $\bar{D} \approx 1 \times 10^{-5}$).  Despite this, we pursue this special limit for two reasons: (1) approximately spatially-uniform concentrations in the contact lens are possible even when $\bar{D}$ is not
large, and (2) this approximation allows analytical expressions for all quantities of interest and, as such, provides 
predictions that can be used to validate the full model and further guide analysis/interpretation.  In order to frame these predictions as closely as possible to the experimental
quantities of interest, we present the details of this approximate solution in dimensional form.

Consider a limit in which diffusion in the contact lens is sufficiently fast -- or the lens sufficiently thin or the interblink sufficiently long, i.e.~$\bar{D}$ is sufficiently large -- to allow the approximation that the contact lens drug concentration becomes spatially uniform -- i.e.~$C = C(t)$ -- by the end of each interblink.  As this particular limit
puts emphasis on the partition coefficient concentration balance, we introduce a partition coefficient for each of the pre-lens and post-lens boundaries of the contact lens, $K_{\rm pre}$ and $K_{\rm post}$.  
That is, the ratio of the spatially-uniform
contact lens concentration with the pre-lens concentration is fixed by the partition coefficient, $K_{\rm pre}$, and
similarly 
%its ratio with the post-lens concentration is fixed by the corresponding partition coefficient,
for the post-lens by $K_{\rm post}$.  Thus, the simplified model is based on the assumption that by the end of the interblink the pre-lens, contact-lens, and post-lens concentrations have together reached equilibrium levels consistent with the partition coefficients.
For simplicity, we ignore drug loss into the eyelid or  the cornea.
The only loss of mass out of the whole system (pre-lens, contact-lens, and post-lens) happens at a blink
(i.e.~between interblinks) and as in our earlier framework is treated as a reset type condition.
Therefore, we can frame this simplified model as a discrete dynamical system and keep track of time in terms of interblink number, $j = 1,2,3,\ldots$.
Since it is convenient to count blinks, we refer to blink $j$ as the blink that occurs after interblink $j$.  Our variables are then $C_{\rm pre}^j$, $C^j$, $C_{\rm post}^j$
for the concentrations in the pre-lens, contact-lens, and post-lens at the end of interblink $j$. Similarly we introduce $M^j_{\rm tot}$ as the total drug mass in the system (including pre-lens, contact-lens,
and post-lens compartments) at the end of interblink $j$.  We reserve $j=0$ to denote the initial conditions (i.e.~before the first interblink) where all the drug mass is in the contact lens so that $C_{\rm pre}^0=0$, 
$C_{\rm post}^0=0$ and $C^0 = C^{\rm init}$ and $M^0_{\rm tot} = C^{\rm init} V_{\rm cl}$.

We apply boundary conditions associated with the partition coefficient, although below we define a switching parameter $f$ that allows us to also capture the case in which a no-flux condition is imposed at the contact-lens/pre-lens boundary (more details are given below).  In this situation, by the end of each interblink $j$ we assume that concentrations in the pre-lens,
contact-lens, and post-lens regions have equilibrated so that
\bea
\label{eq:concentrations}
K_{\rm post} C_{\rm post}^j = C^j = K_{\rm pre} C_{\rm pre}^j,
\eea
for $j=1,2,3,\ldots$,
which indicates that the concentrations in the pre-lens, post-lens, and contact lens are in proportion with one another.
As a result of equations~(\ref{eq:concentrations}) at the end of interblink $j$, the total drug mass, $M^j_{\rm tot}$, in the entire pre-lens, contact-lens, post-lens system can be expressed as
\bea
\label{eq:interblink_mass}
M^j_{\rm tot} & = & C_{\rm post}^j V_{\rm post} + C^j V_{\rm cl} + f C_{\rm pre}^j V_{\rm pre}, \nonumber \\
 & = & C_{\rm post}^j \left( V_{\rm post} + K_{\rm post} V_{\rm cl} + f \frac{K_{\rm post}}{K_{\rm pre}} V_{\rm pre}\right),
\eea
where $V_{\rm post}$, $V_{\rm cl}$, and $V_{\rm pre}$ are the volumes of the post-lens, contact-lens, and pre-lens regions.  The parameter
$f$ allows consideration of the case in which a no-flux boundary condition is used between the contact lens and the pre-lens.  In particular, we set $f=0$
to impose a no-flux condition into the pre-lens region (i.e.~zero drug mass escapes into the pre-lens), and $f=1$ otherwise.
Since we assume that the volumes and partition coefficients are constants, this relation reveals that the post-lens
concentration at the end of interblink $j$ is proportional to the total drug mass in the system at that moment.

Note that $M^1_{\rm tot}$ represents the drug mass in the system at the end of interblink 1 and since no blink has yet happened, $M^1_{\rm tot} = M^0_{\rm tot}$.  However, since the drug
has redistributed itself according to~(\ref{eq:concentrations}) during interblink 1, in general $C_{\rm post}^1 \neq C_{\rm post}^0=0$, $C_{\rm pre}^1 \neq C_{\rm pre}^0=0$, 
and $C^1 \neq C^0$.

In this setting, drug mass leaves the system only during a blink.  Losses of mass from the post-lens and/or pre-lens that occur as a consequence of blink $j$ imply that
\bea
M^{j+1}_{\rm tot} & = & M^j_{\rm tot} - \Delta M_{\rm post}^j - \Delta M_{\rm pre}^j,
\eea
where $M^j_{\rm tot}$ is the drug mass in the system right before blink $j$ (i.e.~at the end of interblink $j$), $M^{j+1}_{\rm tot}$ is the drug mass in the system
right after blink $j$ (i.e.~beginning of interblink $j+1$), and $\Delta M_{\rm post}^j$ and $\Delta M_{\rm pre}^j$ are the amount of drug mass lost from the 
post-lens and pre-lens regions due to blink $j$. Note that the total mass does not change during an interblink so $M^{j+1}_{\rm tot}$ also represents the
drug mass at the end of interblink $j+1$.
We have already argued that the drug mass lost during a blink from the pre-lens and post-lens regions is proportional to the drug concentration at the moment
of the blink, with the proportionality factor being the total volume exiting the system, which we denote
here by $\Delta V_{\rm pre}$ and $\Delta V_{\rm post}$.  That is,
\bea
\label{eq:Delta_Mj}
\Delta M_{\rm pre}^j = f \Delta V_{\rm pre} C_{\rm pre}^j, \quad
\Delta M_{\rm post}^j = \Delta V_{\rm post} C_{\rm post}^j ,
\eea
where we have again inserted the factor $f$ so that drug loss to the pre-lens can be shut off ($f=0$) in the case of a no-flux boundary condition at the contact-lens/pre-lens boundary.  Further details for $\Delta V_{\rm post}$ and $\Delta V_{\rm pre}$ will be given below.
Then, the drug lost due to blink $j$ generates the updated total mass $M^{j+1}_{\rm tot}$
\bea
\label{eq:mass_update}
M^{j+1}_{\rm tot} & = & M^{j}_{\rm tot} - \Delta V_{\rm post} C_{\rm post}^j - f \Delta V_{\rm pre} C_{\rm pre}^j,
\eea
for $j=1,2,3,\ldots$.

Using~(\ref{eq:interblink_mass}) to substitute for $M^j_{\rm tot}$ and $M^{j+1}_{\rm tot}$ in equation~(\ref{eq:mass_update}) and also using~(\ref{eq:concentrations}) gives
\bea
C_{\rm post}^{j+1}  & = & C_{\rm post}^j - 
\frac{ \Delta V_{\rm post} + f \frac{K_{\rm post} }{K_{\rm pre} } \Delta V_{\rm pre}  }
{ \left( V_{\rm post} + K_{\rm post} V_{\rm cl} + f \frac{K_{\rm post}}{K_{\rm pre}} V_{\rm pre}\right) } C_{\rm post}^j .
\eea
Therefore, the post-lens concentration at the end of interblink $j+1$ can be written as
\bea
C_{\rm post}^{j+1}  & = & ( 1 - s) C_{\rm post}^j,
\eea
for $j=1,2,3,\ldots$ where $s$ is a constant given by
\bea
\label{eq:s_factor}
s & = & \frac{ \Delta V_{\rm post} + f \frac{K_{\rm post} }{K_{\rm pre} } \Delta V_{\rm pre}  }
{ \left( V_{\rm post} + K_{\rm post} V_{\rm cl} + f \frac{K_{\rm post}}{K_{\rm pre}} V_{\rm pre}\right) }.
\eea
An explicit solution follows: $C_{\rm post}^{j} = (1-s)^{j-1} C_{\rm post}^1$ for $j=1,2,3,\ldots$.  Expressions for $C_{\rm pre}^j$, $C^j$, and $M^j_{\rm tot}$ also follow.

\begin{figure}[h]
\centering
\vspace{-0.9in}
\includegraphics[scale=.37]{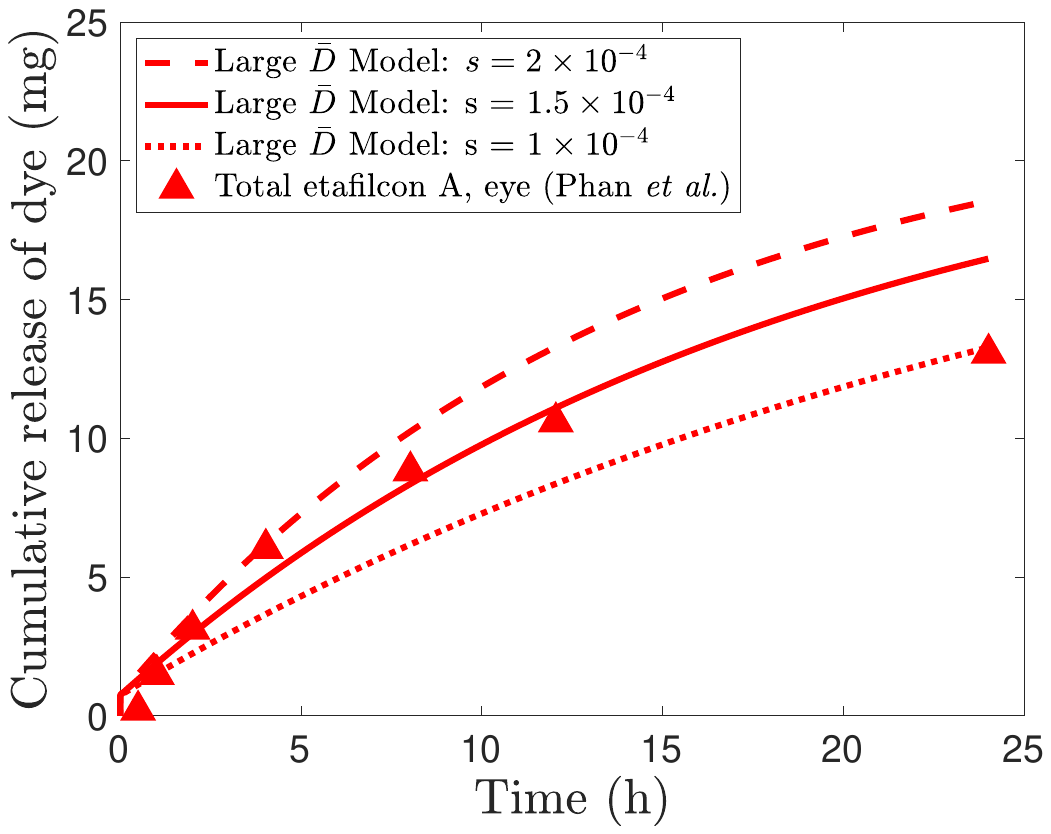}
\includegraphics[scale=.37]{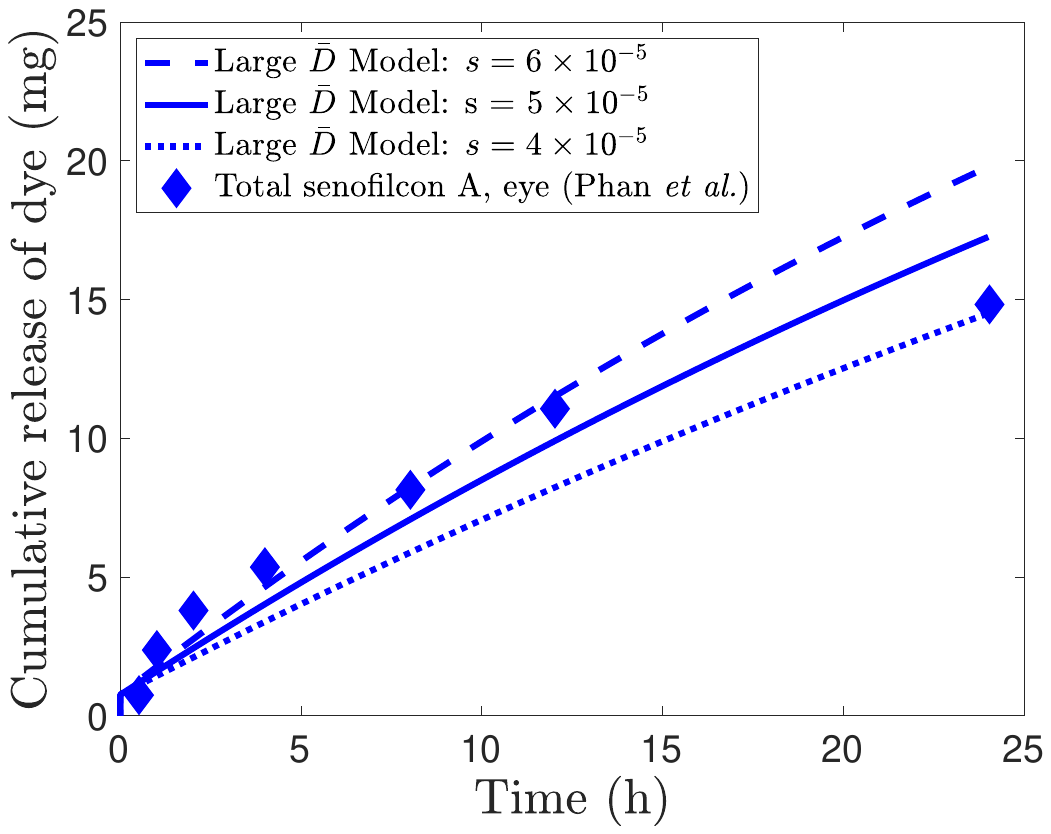}\\
\vspace{-0.75in}
\caption{The large diffusion limit predictions are shown for the two contact lens types,
etafilcon A on the left and senofilcon A on the right, each with different values for the parameter $s$ as indicated.    
The parameter values for $h_{\rm pre}$ (5 $\mu$m), $h_{\rm cl}$, $h_{\rm post}$, $\tau$, and $K_{\rm pre}= K_{\rm post} =K$ are as in Table~\ref{table-compA}.  The corresponding Phan {\it et al.} experimental eye model data are also shown.}
\label{fig_CDR_LargeD}
\end{figure}

It will be convenient to express these solutions in terms of a discrete time variable $t_j$.  In particular, for interblink time $\tau$, we have the analytical solution
\bea
t_j & = & j \tau, \\
C_{\rm post}(t_j) & = & (1 - s)^{j-1} C_{\rm post}^1, \\
C_{\rm pre}(t_j) & = & f \frac{K_{\rm post} }{K_{\rm pre} } C_{\rm post}(t_j), \\
C(t_j) & = & K_{\rm post}  C_{\rm post}(t_j), \\
M_{\rm tot}(t_j) & = & C_{\rm post}(t_j) \left( V_{\rm post} + K_{\rm post} V_{\rm cl} + f \frac{K_{\rm post}}{K_{\rm pre}} V_{\rm pre}\right),
\eea
for $j=1,2,3, \ldots$ where
\bea
C_{\rm post}^1 & = & \frac{M^{\rm init}}{ \left( V_{\rm post} + K_{\rm post} V_{\rm cl} + f \frac{K_{\rm post}}{K_{\rm pre}} V_{\rm pre}\right) } ,
\eea
along with $t_0=0$, $C_{\rm post}(0) = 0$, $C_{\rm pre}(0) = 0$, $C(0)=M^{\rm init}/V_{\rm cl}$.   
These formulas constitute closed-form solutions to the multiple blink drug delivery problem in this special large diffusion limit.
It is also convenient to identify the total drug mass in the contact lens at time $t=t_j$ as
\bea
M(t_j) & = & V_{\rm cl} C(t_j),
\eea 
so that the cumulative drug released from the contact lens, $M^{\rm lost}(t_j) = M^{\rm init} - M(t_j)$, is 
\bea
\label{eq:CDR_formula_LargeD}
M^{\rm lost}(t_j) & = &  M^{\rm init} - K_{\rm post} V_{\rm cl}   C_{\rm post}(t_j), \nonumber \\
  & = & M^{\rm init} - K_{\rm post} V_{\rm cl}   (1 - s)^{j-1} C_{\rm post}^1, \nonumber \\
  & = & M^{\rm init} \left[ 1 -   
  \frac{ K_{\rm post} V_{\rm cl} }{ \left( V_{\rm post} + K_{\rm post} V_{\rm cl} + f \frac{K_{\rm post}}{K_{\rm pre}} V_{\rm pre}\right) }
  (1 - s)^{j-1} \right],
\eea
for $j=1,2,3,\ldots$ with $M^{\rm lost}(0)=0$.
Note that the factor
\bea
\frac{ K_{\rm post}V_{\rm cl} }{ \left( V_{\rm post} + K_{\rm post} V_{\rm cl} + f \frac{K_{\rm post}}{K_{\rm pre}} V_{\rm pre}\right) } & = &
\frac{1}{1 + \frac{V_{\rm post}}{ K _{\rm post} V_{\rm cl}} + f \frac{V_{\rm pre}}{ K _{\rm pre} V_{\rm cl}} } \approx 1,
\eea
since, typically, partition coefficients are ${\cal O}(1)$ and $V_{\rm post}$ and $V_{\rm pre}$ are expected to be much smaller than $V_{\rm cl}$.
This tells us that, at least under the assumptions of the present reduced model, the single constant $s$ defined in equation~(\ref{eq:s_factor}) controls nearly completely the dynamics
of the cumulative drug release.  That is, it appears a good approximation that may allow us to gain intuition is
\bea
\frac{M^{\rm lost}(t_j)}{M^{\rm init}} & \approx & 1 -   (1 - s)^{j-1}.
\eea
This suggests that comparison to data on cumulative drug release in the large diffusion limit amounts to the optimal selection of a single
parameter combination $s$ through which all pre-lens and post-lens drug mass loss mechanisms are represented.    
We show solutions below but for now we shall suppose that a particular choice $s= s^*$ achieves
this goal for a given contact lens.  With respect to the Phan {\it et al.}~\cite{phan2021development} data we recall that the `vial' drug release differs significantly
between the two lens types and appears to be reasonably-well described by linear diffusion in the contact lens
with each lens characterized by a different diffusion coefficient.  On the other hand, the `eye model' drug release data is effectively indistinguishable between the two contact lens types, suggesting that 
a non-diffusive mechanism exerts some control on the drug release dynamics.  
 Figure~\ref{fig_CDR_LargeD} plots our prediction in equation~(\ref{eq:CDR_formula_LargeD})  for representative ranges of the parameter $s$
 for each of the two contact lenses along with the
data from Phan {\it et al.}~\cite{phan2021development}.

%%%%%%%%%%%%%%%%%%%%%%%%%%%%%%%%%%%%%%%%%%%%%%%%%%%%%%%%%%%%%%%%%%
%%%%%%%%%%%%%%%%%%%%%%%%%%%%%%%%%%%%%%%%%%%%%%%%%%%%%%%%%%%%%%%%%%

Two critical contributions to the parameter $s$ defined in equation~(\ref{eq:s_factor}) come from $\Delta V_{\rm post}$ and $\Delta V_{\rm pre}$.  
For the post-lens volume lost, using the results discussed in Section~\ref{sec-reset_conditions} we have
\bea
\Delta V_{\rm post} & = & 
\left\{
\begin{array}{ll}
\frac{3}{2} R_{\rm cl} h_{\rm post} \Delta X_{\rm cl}, & \mbox{Slide Out} \\
\pi R_{\rm cl}^2 \Delta h_{\rm post}, & \mbox{Squeeze Out}
\end{array}
\right. ,
\eea
as the post-lens tear volume that moves out from under the contact lens during a blink.  These values
of $\Delta V_{\rm post}$ are used in $\Delta M_{\rm post}^j$ in~(\ref{eq:Delta_Mj}) with the assumption that this lost volume
transports drug mass out of the post-lens tear film and none of this drug returns.
For the pre-lens volume lost we write
\bea
\Delta V_{\rm pre} & = & p V_{\rm pre}, 
\eea
where $p \in [0,1]$ is an adjustable parameter that represents the proportion of pre-lens mass lost as a result of a blink. Here, 
$p=0$ means no pre-lens drug mass is lost and $p=1$ means all pre-lens drug mass is lost as the result of a blink.  
So in~(\ref{eq:Delta_Mj}) this means $\Delta M_{\rm pre}^j = f p V_{\rm pre} C_{\rm pre}^j$, which involves
two model parameters $f$ and $p$.  The case $f=0$ corresponds to no-flux into the pre-lens and thus here the value of $p$ has no impact on the predictions because there is never any drug mass in the pre-lens tear film.
The case $f=1$ allows the standard diffusive flux from
the contact lens into the pre-lens and here the parameter $p$ establishes how much of the pre-lens drug mass is swept away as the result of a blink.
Below, we explore more closely the constant $s$ and its dependence on the model parameters in the pre-lens and post-lens including the cases of slide out and squeeze out 
mechanisms of drug loss in the post-lens region.

\subsection{Slide-out cases:}
In this case the constant $s$ in equation~(\ref{eq:s_factor}) is
\bea
\label{eq:s_factor_slideA}
s & = & \frac{ \frac{3}{2} R_{\rm cl} h_{\rm post} \Delta X_{\rm cl} + f p V_{\rm pre} \frac{K_{\rm post} }{K_{\rm pre} } }
{ \left( V_{\rm post} + K_{\rm post} V_{\rm cl} + f \frac{K_{\rm post}}{K_{\rm pre}} V_{\rm pre}\right) }.
\eea
Using $V_{\rm post} = \pi R_{\rm cl}^2 h_{\rm post}$, $V_{\rm pre} = \pi R_{\rm cl}^2 h_{\rm pre}$, and $V_{\rm cl} = \pi R_{\rm cl}^2 h_{\rm cl}$, we find that
\bea
\label{eq:s_factor_slideB}
s & = & \frac{ \frac{3}{2\pi} h_{\rm post} \frac{\Delta X_{\rm cl}}{R_{\rm cl}} + f p  h_{\rm pre} \frac{K_{\rm post} }{K_{\rm pre} } }
{  \left(  h_{\rm post} + K_{\rm post}  h_{\rm cl} + f \frac{K_{\rm post}}{K_{\rm pre}} h_{\rm pre}\right) } 
= \frac{ \frac{3}{2\pi} \frac{h_{\rm post}}{h_{\rm cl}} \frac{\Delta X_{\rm cl}}{R_{\rm cl}} + f p  \frac{h_{\rm pre}}{h_{\rm cl}} \frac{K_{\rm post} }{K_{\rm pre} } }
{  \left(  K_{\rm post}  + \frac{h_{\rm post}}{h_{\rm cl}} + f \frac{K_{\rm post}}{K_{\rm pre}} \frac{h_{\rm pre}}{h_{\rm cl}} \right) }.
\eea
If we define
\bea
{\cal H} & = & K_{\rm post}  + \frac{h_{\rm post}}{h_{\rm cl}} + f \frac{K_{\rm post}}{K_{\rm pre}} \frac{h_{\rm pre}}{h_{\rm cl}},
\eea
then
\bea
\label{eq:s_factor_slideC}
s & = & \frac{1}{ {\cal H} } \left( \frac{3}{2\pi} \frac{h_{\rm post}}{h_{\rm cl}} \frac{\Delta X_{\rm cl}}{R_{\rm cl}} + f p  \frac{h_{\rm pre}}{h_{\rm cl}} \frac{K_{\rm post} }{K_{\rm pre} }\right) .
\eea
If $s=s^*$ is the value of $s$ for which the large diffusion model best fits the cumulative drug release data, then we can identify a closed-form expression for the required 
contact lens displacement, $\Delta X_{\rm cl}$, that would be necessary to achieve such a fit, given the other parameters:
\bea
\frac{\Delta X_{\rm cl}}{R_{\rm cl}} & = & \frac{2\pi}{3} \frac{h_{\rm cl}}{h_{\rm post}} \left( s^* {\cal H} - f p  \frac{h_{\rm pre}}{h_{\rm cl}} \frac{K_{\rm post} }{K_{\rm pre} } \right), \nonumber \\
& = & \frac{2\pi}{3} \left[ s^* \left( 1 + \frac{K_{\rm post} h_{\rm cl}}{h_{\rm post}} \right) + (s^* - p) f \frac{K_{\rm post} }{K_{\rm pre} }  \frac{h_{\rm pre}}{h_{\rm post}} \right].
\eea

It remains to explore these predictions numerically to assess if the requirements for $\Delta X_{\rm cl} / R_{\rm cl}$ are feasible for known or estimated values
for the other parameters.  Several particular cases are outlined below.  We assume in these cases that $K_{\rm pre} = K_{\rm post} =K$.  Since
we know $K$, $h_{\rm cl}$, and $R_{\rm cl}$ for the experimental data and we assume $s^*$ is fixed by the fit to the cumulative drug release curve, 
we fix those parameters and explore how the predictions vary with $h_{\rm post}$, $h_{\rm pre}$, $f$, and $p$.

\begin{figure}[h]
\centering
\vspace{-0.9in}
\includegraphics[scale=.35]{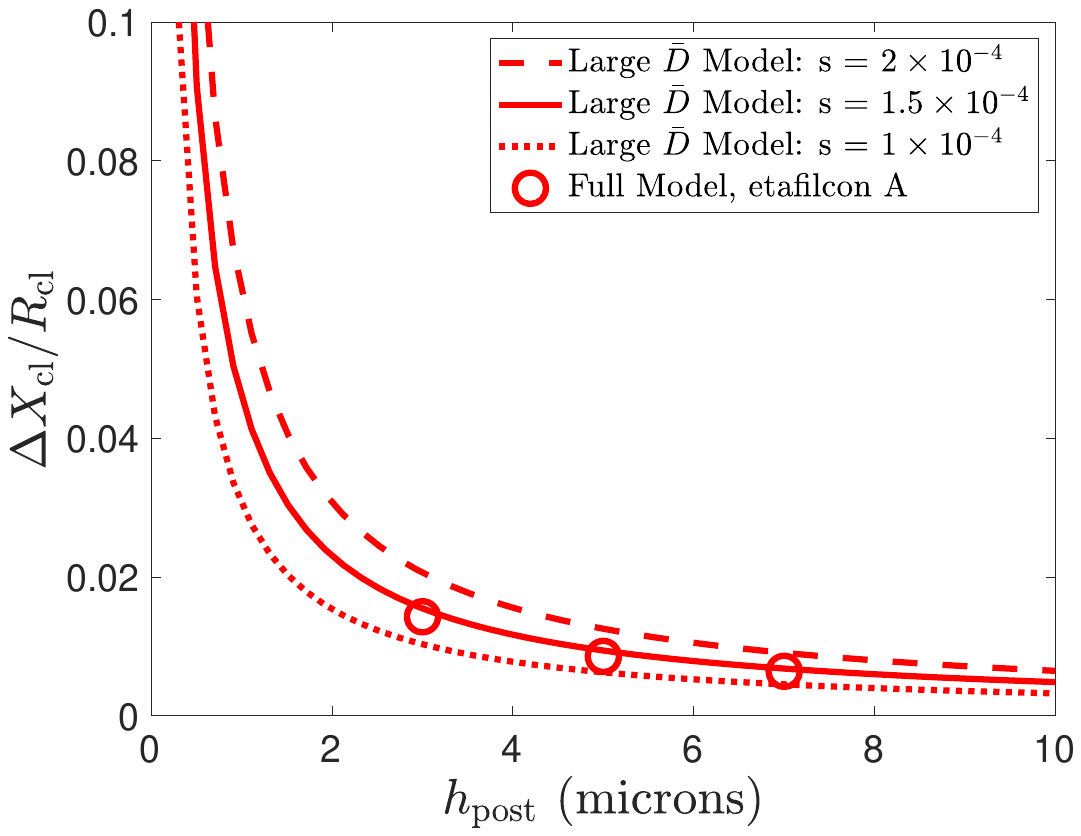}
\includegraphics[scale=.35]{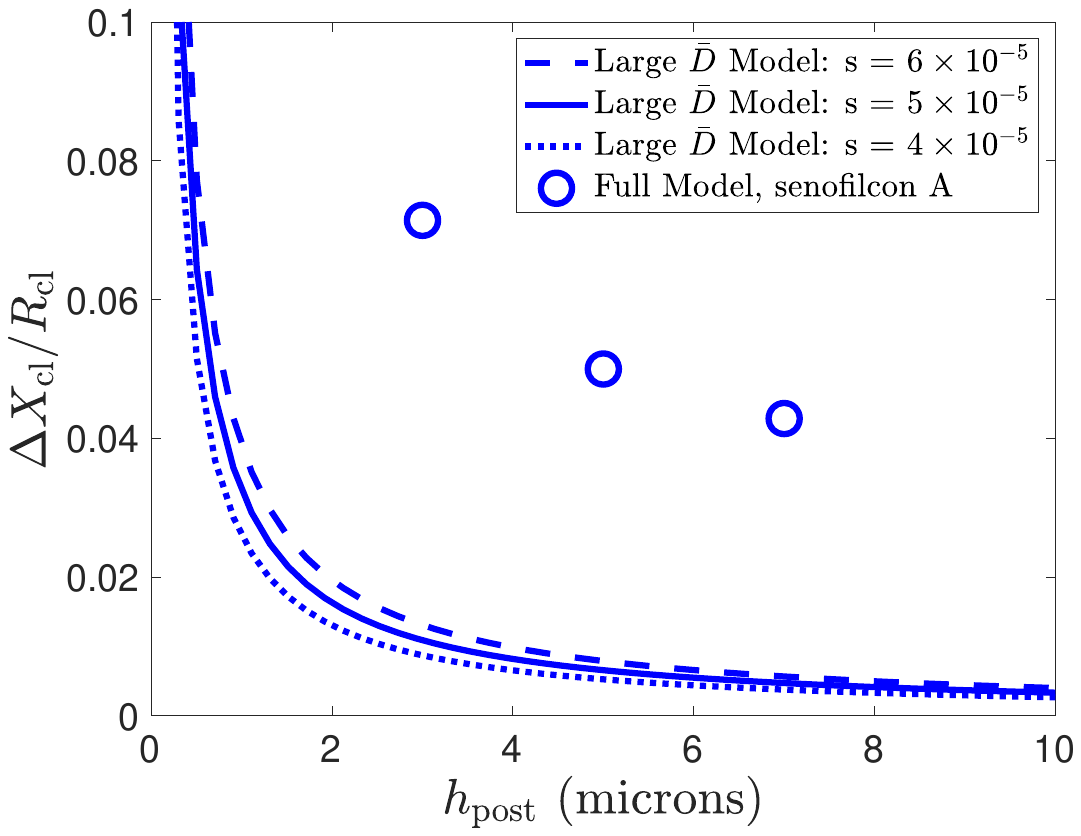} \vspace{-1.75in} \\
\includegraphics[scale=.35]{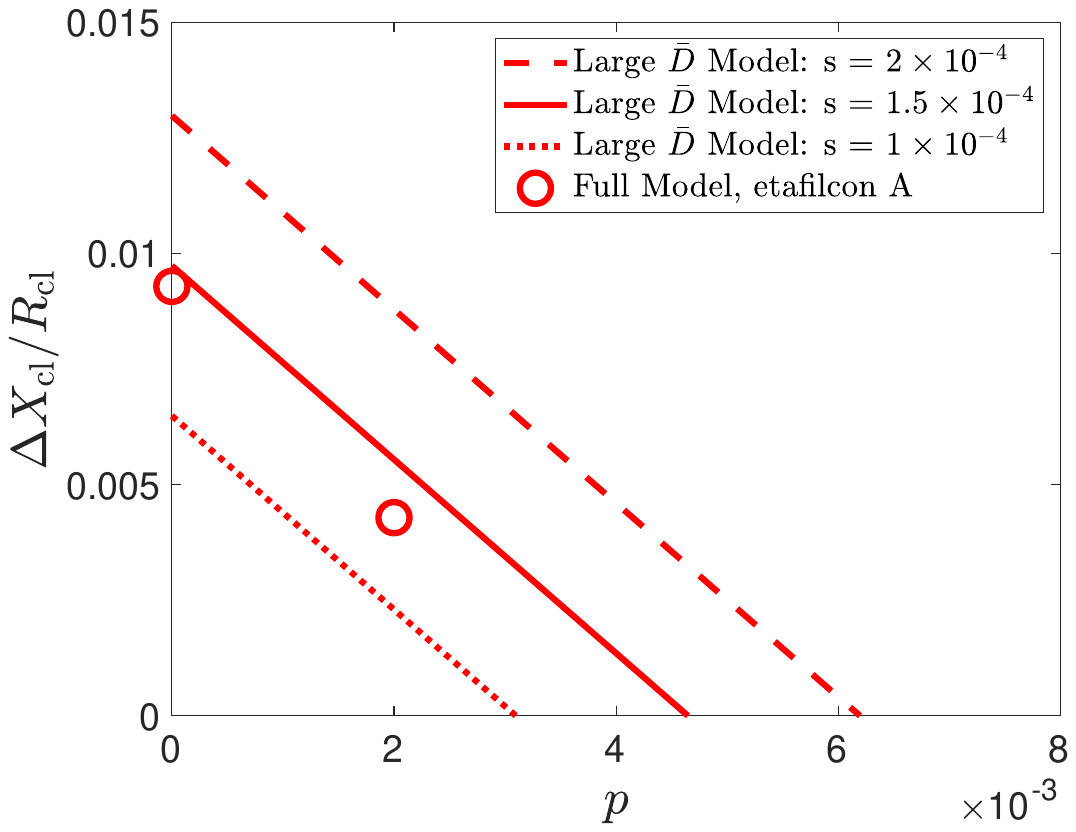}
\includegraphics[scale=.35]{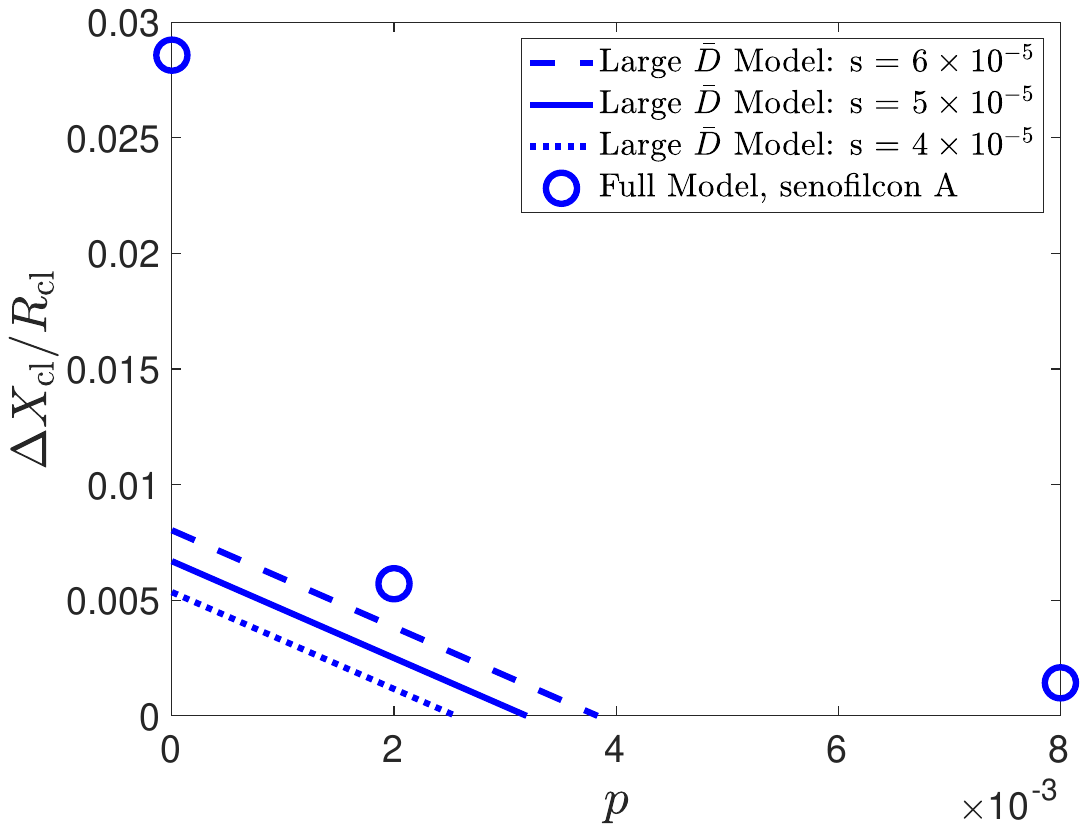}
\vspace{-0.95in}
\caption{Slide-Out Cases.  The upper plots shows the value of $\Delta X_{\rm cl}/R_{\rm cl}$ 
required to achieve the specified value of $s$ (and fit the Phan {\it et al.} eye model data) for the case with no-flux into the pre-lens (upper left: etafilcon A lens; upper right: senofilcon A lens).
The lower plots show the value of $\Delta X_{\rm cl}/R_{\rm cl}$ required to achieve the 
specified value of $s$ with partial mass loss from the pre-lens specified by the value $p$
(lower left: etafilcon A lens; lower right: senofilcon A lens). In these cases with
$p \neq 0$ we use the partition coefficient boundary condition at the contact lens/pre-lens boundary.  
In the lower plots we have used $h_{\rm pre} = h_{\rm post} = 5$ $\mu$m.
Also shown in the plots are predictions from the full model fits (red and blue open circles).
%Further details are discussed in the text.
}
\label{fig_SlideOut_LargeD}
\end{figure}

\begin{itemize}
\item {\bf Post-lens slide-out, pre-lens no flux:} Here we examine the no-flux boundary condition on the contact-lens/pre-lens boundary, $f=0$.  We find
\bea
\label{eq:slide_out_DeltaX_v_hpost}
\frac{\Delta X_{\rm cl}}{R_{\rm cl}} & = & \frac{2\pi}{3} s^* \left( 1 + \frac{K  h_{\rm cl}}{h_{\rm post}} \right) .
\eea
The result in equation~(\ref{eq:slide_out_DeltaX_v_hpost}) is shown in the upper two plots of Figure~\ref{fig_SlideOut_LargeD}.  In these plots, in addition to the large diffusion limit solution we also show several
computed solutions from the full model (open circles; see Section \ref{sec:full_model_results}).  These predictions are in close agreement for the etafilcon A lens.  For the senofilcon A lens, which has a much smaller value of $\bar{D}$  for the full model, the large diffusion limit predictions require a noticeably reduced post-lens motion for the same cumulative drug release.

\item {\bf Post-lens slide-out, partial pre-lens loss:} Here we assume $f=1$ and $p \in [0,1]$.  We find
\bea
\label{eq:slide_out_DeltaX_v_p}
\frac{\Delta X_{\rm cl}}{R_{\rm cl}} & = & \frac{2\pi}{3} \left[ s^* \left( 1 + \frac{K h_{\rm cl}}{h_{\rm post}} \right) + (s^* - p)  \frac{h_{\rm pre}}{h_{\rm post}} \right].
\eea
For $\Delta X_{\rm cl} \ge 0$ we must limit $p \in [0,p_{\rm max}]$ where $\Delta X_{\rm cl}=0$ gives $p = p_{\rm max}$ and 
\bea
\label{eq:pmax}
p_{\rm max} & = & s^* \left( 1 + \frac{K h_{\rm cl}}{h_{\rm pre}} + \frac{h_{\rm post}}{h_{\rm pre}} \right).
\eea
The result in equation~(\ref{eq:slide_out_DeltaX_v_p}) is
plotted versus $p$ on $[0,p_{\rm max}]$ in the lower two plots of Figure~\ref{fig_SlideOut_LargeD}.
In these plots, we also show several computed solutions from the full model (open circles).  
Note that for the etafilcon A lens (lower left plot) there is no value of $\Delta X_{\rm cl}$ when $p=0.008$ for which the cumulative drug release data of Phan {\it et al.} can be matched by the full model.  For the senofilcon A lens (lower right plot) the full model values for $\Delta X_{\rm cl}$ are larger than those for the large
diffusion limit model.  We note later in Section \ref{sec:full_model_results} that  we can find solutions of the full model with $p=1$ and $\Delta X_{\rm cl}=0$ that, like these
predictions, also can match the Phan {\it et al.} \cite{phan2021development} data.

\end{itemize}

%%%%%%%%%%%%%%%%%%%%%%%%%%%%%%%%%%%%%%%%%%%%%%%%%%%%%%%%%%%%%%%%%%
%%%%%%%%%%%%%%%%%%%%%%%%%%%%%%%%%%%%%%%%%%%%%%%%%%%%%%%%%%%%%%%%%%

\subsection{Squeeze-out cases:}
Following arguments similar to the slide out case outlined above, we find that for the squeeze out post-lens flow the parameter $s$ takes the form
\bea
\label{eq:s_factor_squeezeA}
s & = & \frac{1}{ {\cal H} } \left( \frac{\Delta h_{\rm post}}{h_{\rm cl}} + f p  \frac{h_{\rm pre}}{h_{\rm cl}} \frac{K_{\rm post} }{K_{\rm pre} }\right) .
\eea
Then, given $s=s^*$ obtained from comparison to the cumulative drug release data of Phan {\it et al.} it follows that the required post-lens thickness change is
\bea
\frac{\Delta h_{\rm post}}{h_{\rm cl}} & = & s^* {\cal H} -  f p  \frac{h_{\rm pre}}{h_{\rm cl}} \frac{K_{\rm post} }{K_{\rm pre} }, \nonumber \\
 & = & s^* \left( K_{\rm post}  + \frac{h_{\rm post}}{h_{\rm cl}} \right) + (s^* - p ) f \frac{h_{\rm pre}}{h_{\rm cl}} \frac{K_{\rm post} }{K_{\rm pre} }.
\eea
Here we identify feasible values of $\Delta h_{\rm post}/h_{\rm cl}$ for known or estimated values
for the other parameters.  Several particular cases are outlined below.  Again, we assume in these cases that $K_{\rm pre} = K_{\rm post} =K$.  

\begin{figure}[h]
\centering
\vspace{-0.9in}
\includegraphics[scale=.35]{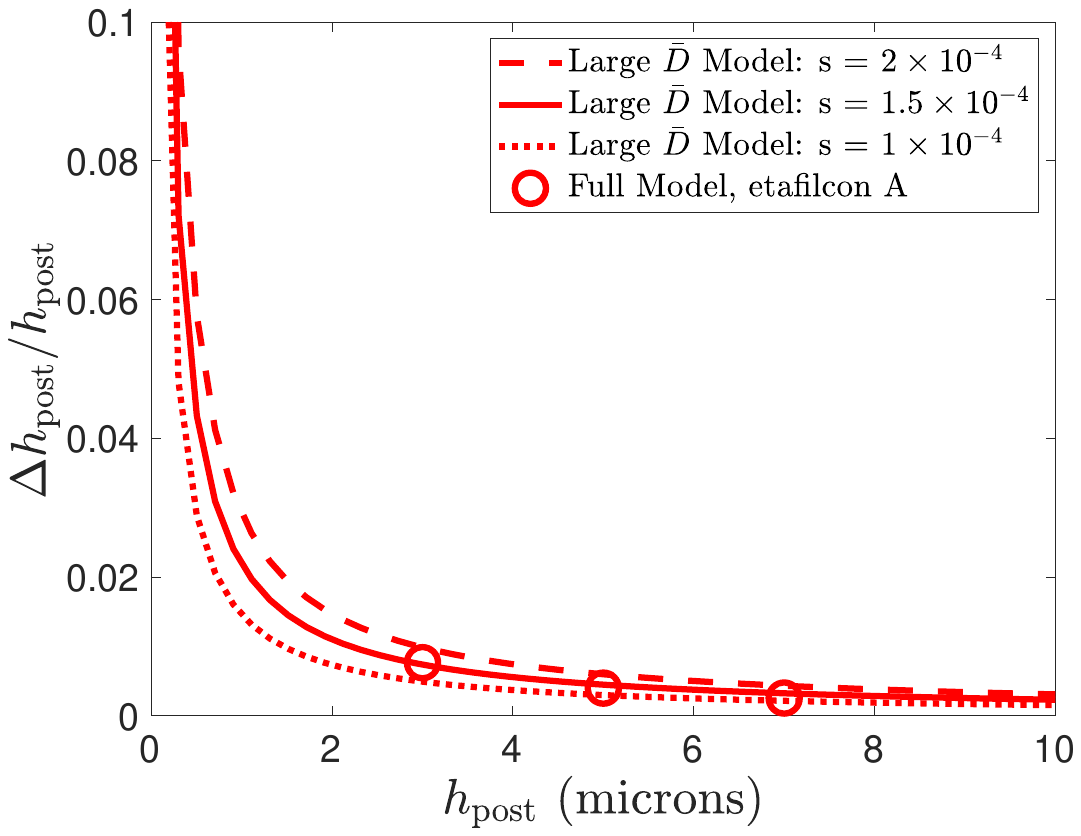}
\includegraphics[scale=.35]{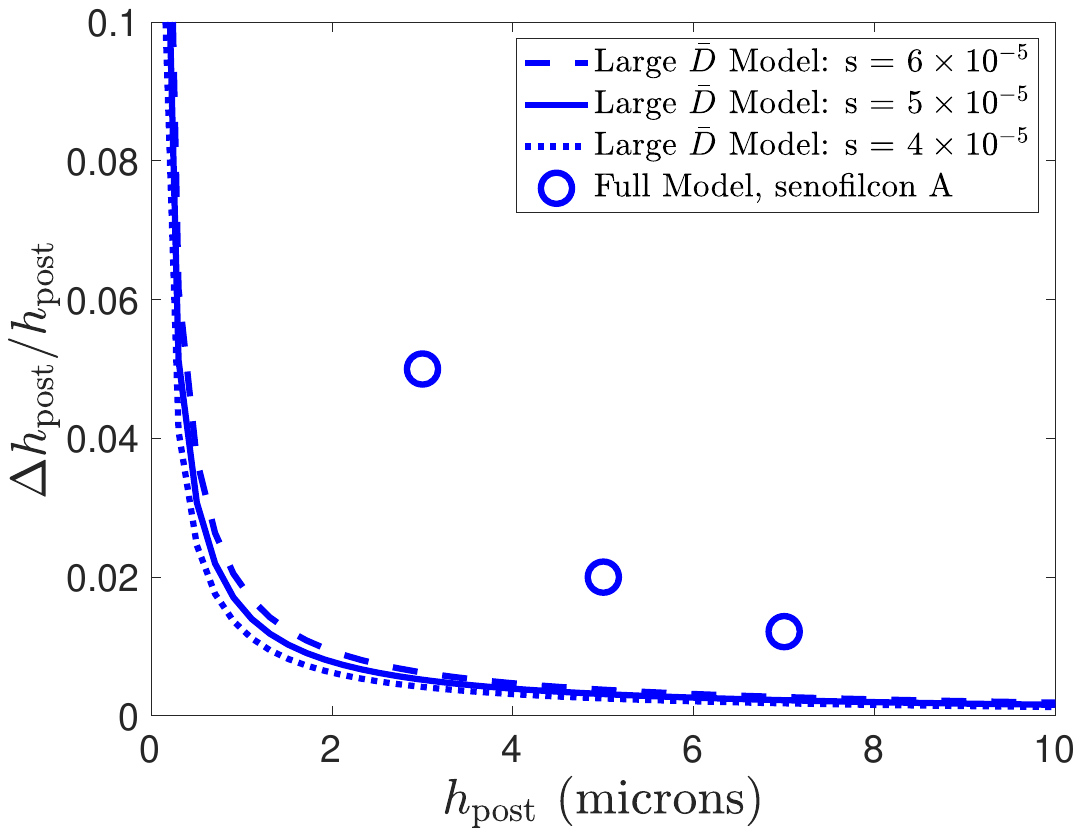} \vspace{-1.75in} \\
\includegraphics[scale=.35]{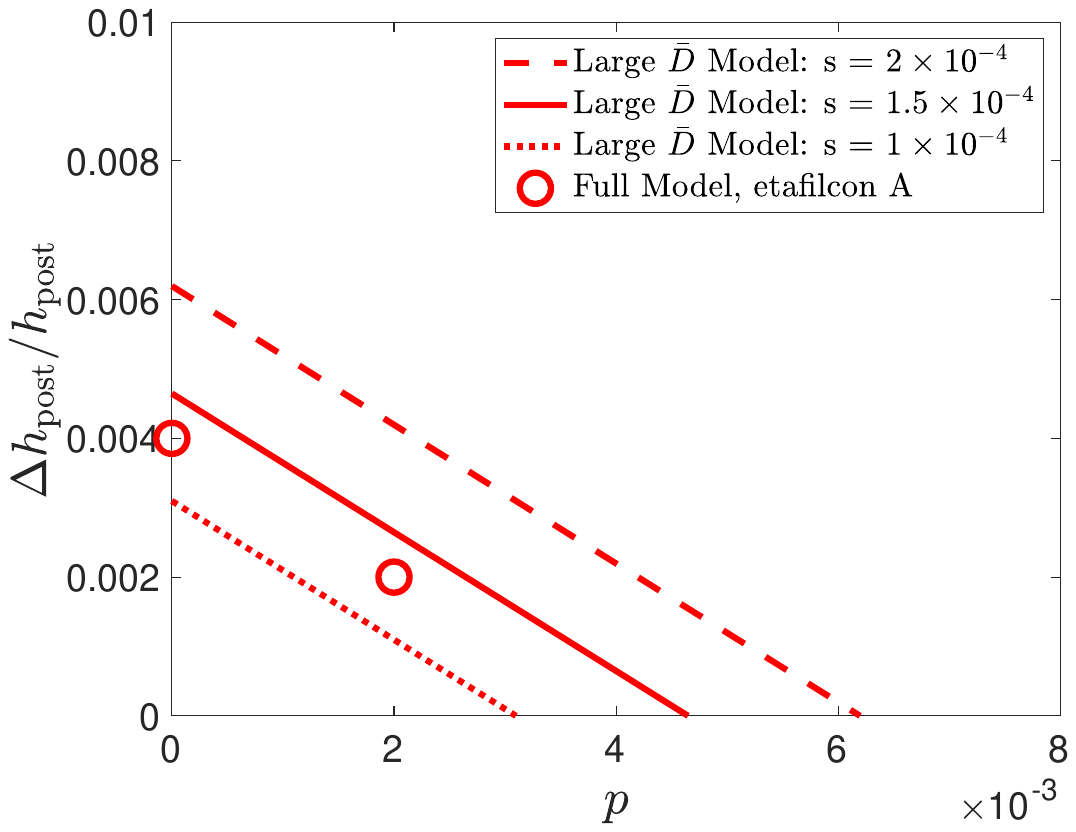}
\includegraphics[scale=.35]{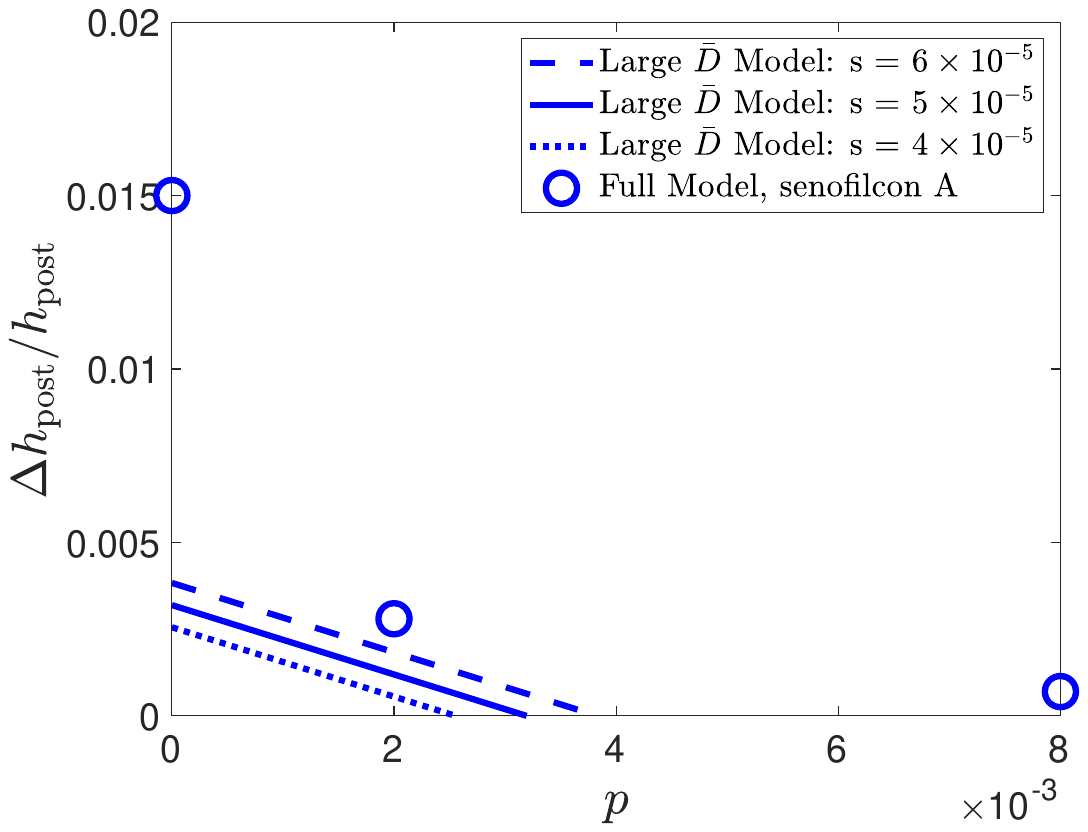}
\vspace{-0.95in}
\caption{Squeeze-Out Cases.  
 The upper plots shows the value of $\Delta h_{\rm post}/h_{\rm post}$ 
required to achieve the specified value of $s$ (and fit the Phan {\it et al.}~eye model data) for the case with no-flux into the pre-lens (upper left: etafilcon A lens; upper right: senofilcon A lens).
The lower plots show the value of $\Delta h_{\rm post}/h_{\rm post}$ required to achieve the 
specified value of $s$ with partial mass loss from the pre-lens specified by the value $p$
(lower left: etafilcon A lens; lower right: senofilcon A lens). In these cases with
$p \neq 0$ we use the partition coefficient boundary condition at the contact lens/pre-lens boundary.  
In the lower plots we have used $h_{\rm pre} = h_{\rm post} = 5$ $\mu$m.
Also shown in the plots are predictions from the full model fits (red and blue open circles).
%Further details are discussed in the text.
}
\label{fig_SqueezeOut_LargeD}
\end{figure}

\begin{itemize}
\item {\bf Post-lens squeeze-out, pre-lens no flux:} Here we examine the no-flux boundary condition on the contact-lens/pre-lens boundary, $f=0$.  We find
\bea
\frac{\Delta h_{\rm post}}{h_{\rm cl}} & = & s^* \left( K  + \frac{h_{\rm post}}{h_{\rm cl}} \right) .
\eea
Another way to write this is
\bea
\label{eq:Squeeze_Deltah_over_h}
\frac{\Delta h_{\rm post}}{h_{\rm post}} & = & s^* \left( 1 + K  \frac{h_{\rm cl}}{h_{\rm post}} \right) .
\eea
The result in equation~(\ref{eq:Squeeze_Deltah_over_h}) is shown in the upper two plots of Figure~\ref{fig_SqueezeOut_LargeD}.  In these plots, we also show several
computed solutions from the full model (open circles).  For the etafilcon A lens the full model predictions match closely with the large diffusion limit predictions while for the senofilcon A lens the full model predictions have a much larger $\Delta h_{\rm post}$ prediction.

\item {\bf Post-lens squeeze-out, partial pre-lens loss:} Here we assume $f=1$ and $p \in [0,1]$.  We find
%\bea
%\frac{\Delta h_{\rm post}}{h_{\rm cl}} & = & s^* \left( K  + \frac{h_{\rm post}}{h_{\rm cl}} \right) + (s^* - p ) \frac{h_{\rm pre}}{h_{\rm cl}} .
%\eea
%
\bea
\label{eq:squeeze_out_DeltaX_v_p}
\frac{\Delta h_{\rm post}}{h_{\rm post}} & = & s^* \left( K \frac{h_{\rm cl}}{h_{\rm post}} + 1 \right) + (s^* - p ) \frac{h_{\rm pre}}{h_{\rm post}} .
\eea
For $\Delta h_{\rm post} \ge 0$ it is required that $p \in [0,p_{\rm max}]$ where $\Delta h_{\rm post}=0$ gives $p = p_{\rm max}$ and $p_{\rm max}$ is the same as the Slide-Out result given in equation~(\ref{eq:pmax}).
%\bea
%p_{\rm max} & = & s^* \left( 1 + \frac{k H_{\rm cl}}{h_{\rm pre}} + \frac{h_{\rm post}}{h_{\rm pre}} \right),
%\eea
%which is the same $p_{\rm max}$ identified earlier.
%
The result in equation~(\ref{eq:squeeze_out_DeltaX_v_p}) is plotted versus $p$ on $[0,p_{\rm max}]$ in the lower two plots of Figure~\ref{fig_SqueezeOut_LargeD}.
In these plots, we also show several
computed solutions from the full model (open circles).  Note
that for the etafilcon A lens (lower left plot) there is no value of $\Delta h_{\rm post}$ when
$p=0.008$ for which the cumulative drug release data can be matched by the full model.

\end{itemize}

%%%%%%%%%%%%%%%%%%%%%%%%%%%%%%%%%%%%%%%%%%%%%%%%%%%%%%%%%%%%%%%%%%
%%%%%%%%%%%%%%%%%%%%%%%%%%%%%%%%%%%%%%%%%%%%%%%%%%%%%%%%%%%%%%%%%%

\subsection{Large diffusion model summary}
Within the context of this large diffusion-limit model, these results   demonstrate that different mechanisms associated with drug loss in the pre-lens, post-lens, contact lens system tied to blinking over a 24 hour period can be identified as candidates to characterize the dynamics of cumulative mass loss data of Phan {\it et al.}\cite{phan2021development}.  

\begin{figure}[h]
\centering
\includegraphics[scale=.35]{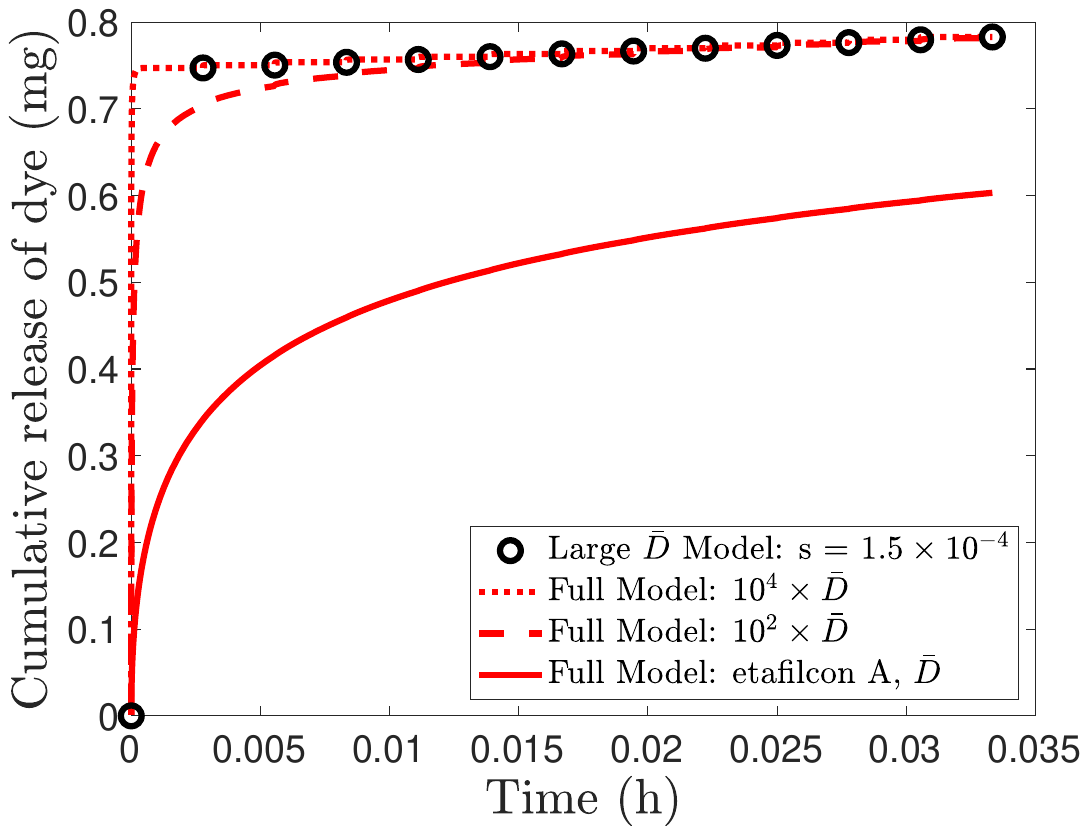}
\includegraphics[scale=.35]{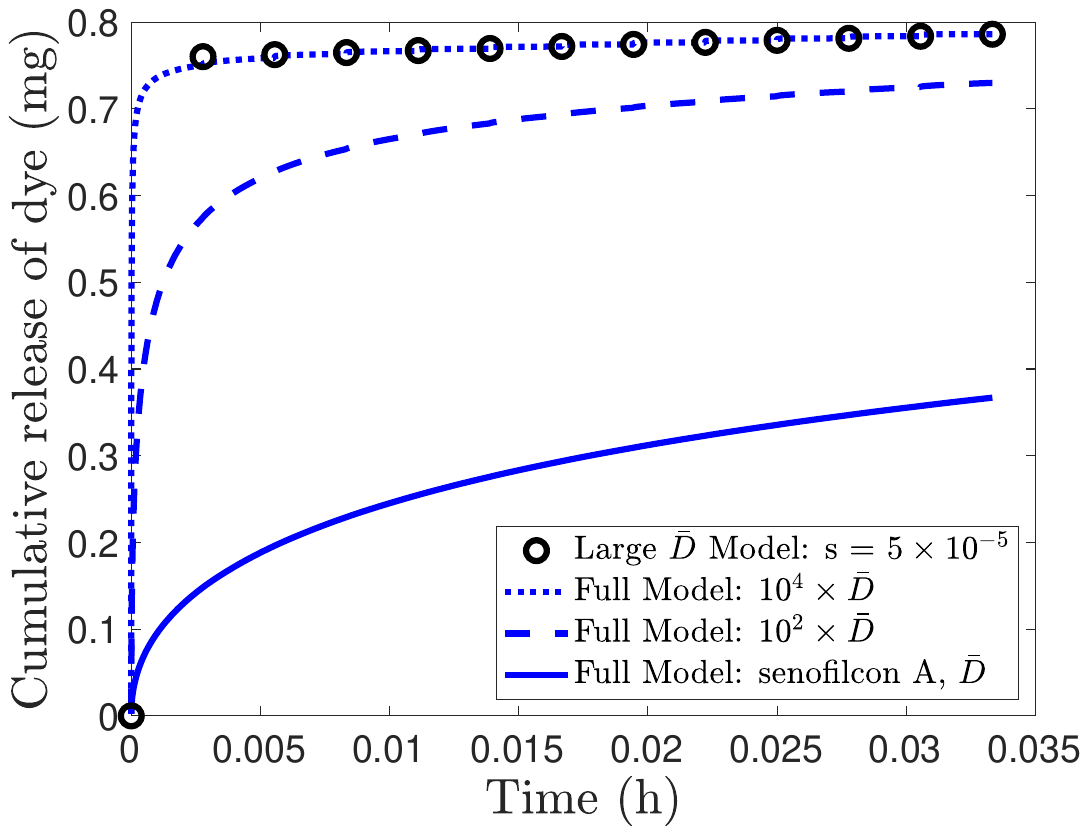} \vspace{-1.75in} \\
\includegraphics[scale=.35]{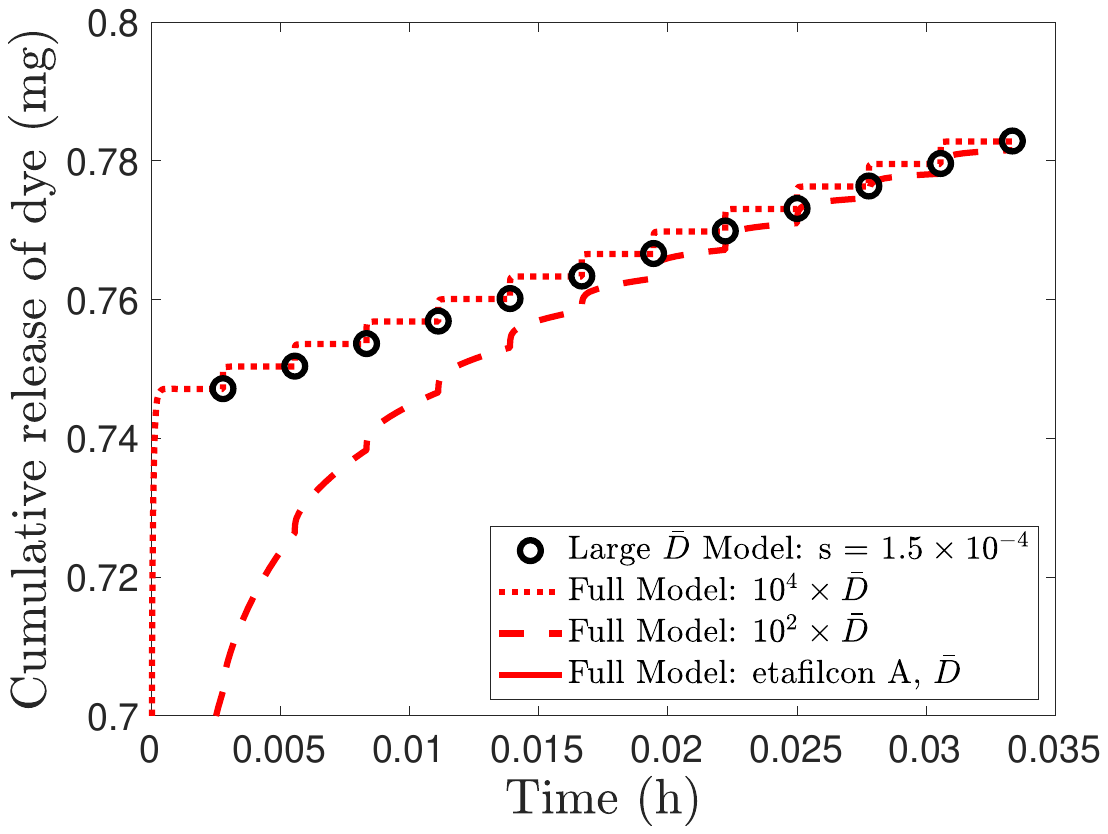}
\includegraphics[scale=.35]{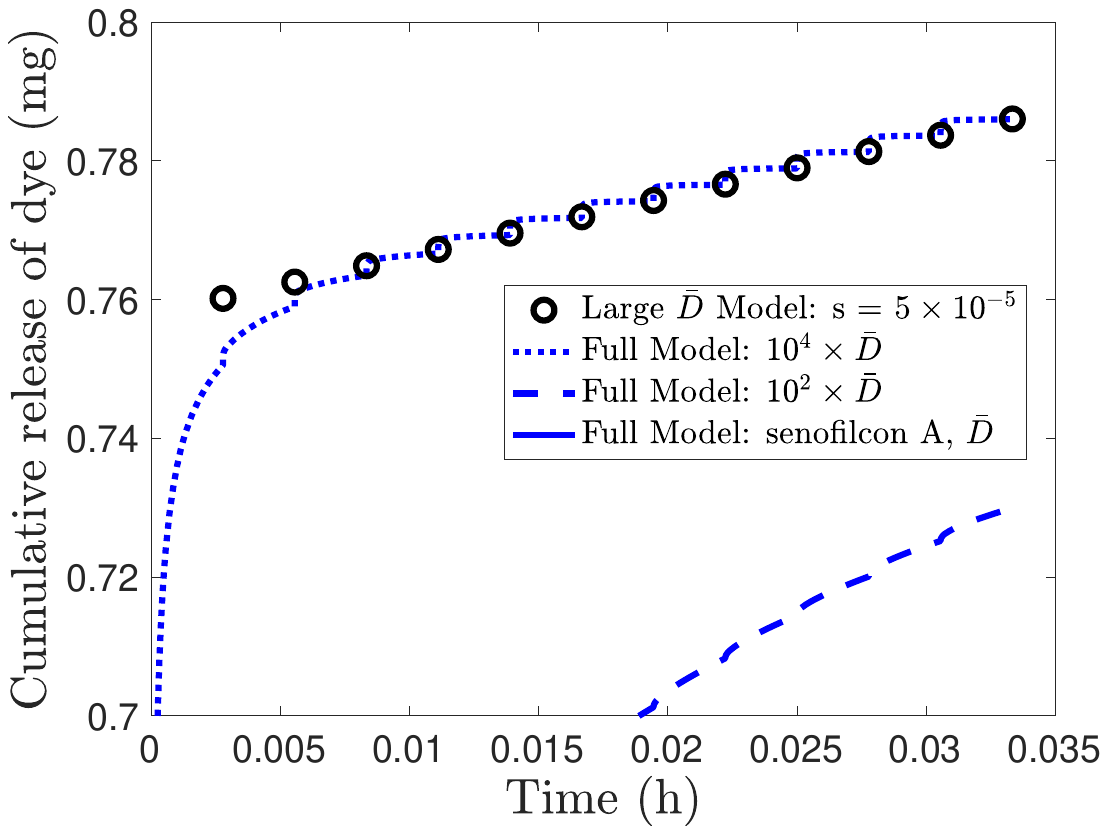} \\
\vspace{-0.75in}
\caption{Comparison of the large diffusion limit solution with full model solution for 
a case with no flux into the pre-lens tear film and slide-out in the post-lens tear film.
The left plots correspond to the etafilcon A lens and the right plots correspond to the 
senofilcon A lens.  The upper plots show cumulative drug release over a few blinks and the 
lower plots show a zoomed-in view comparing the large diffusion limit solution (black circles) 
with the full solution.  
The solid curves use $\bar{D} \approx 5 \times 10^{-4}$ and $1 \times 10^{-5}$
for the etafilcon A and senofilcon A lenses, respectively (based on Table~\ref{table-compA}).
The dashed curves use $10^2 \times \bar{D} \approx$ $5 \times 10^{-2}$ and $1 \times 10^{-3}$ 
for the two lenses.
The dotted curves use $10^4 \times \bar{D} \approx$ $5$ and $0.1$ for the two lenses.
}
\label{fig-largeD_comparison}
\end{figure}

In the next section we outline results computed from our full eye model.  As a final sneak preview to the full model results we show a comparison of the large diffusion limit solution and the full model in Figure~\ref{fig-largeD_comparison} with no-flux into the pre-lens tear film and the slide out mechanism active in the post-lens tear film.  These plots show the early time predictions for cumulative drug release
for the two contact lenses.  Three values of the diffusion coefficient
are used in the full model -- $\bar{D}$, $10^2 \bar{D}$, and $10^4 \bar{D}$ -- along with the large diffusion limit solution for the first twelve blinks.  It is clear in the upper plots as well as the lower, zoomed-in plots, that the full model solutions with sufficiently large diffusion coefficient agree extremely well with the discrete (individual blink) time predictions of the large diffusion model.  It appears that the large diffusion limit analysis gives at least qualitatively good information about potential drug release and transport mechanisms in the eye model configuration.

%%%%%%%%%%%%%%%%%%%%%%%%%%%%%%%%%%%%%%%%%%%%%%%%%%%%%%%%%%%%%%%%%%
%%%%%%%%%%%%%%%%%%%%%%%%%%%%%%%%%%%%%%%%%%%%%%%%%%%%%%%%%%%%%%%%%%

%%%%%%%%%% 
% Dan (12-29-2023) removed the matched asymptotics summary for the contact lens blink reset since it sounds like our numerics are not having an issue at this boundary.
%%%%%%%%%%

%%%%%%%%%%%%%%%%%%%%%%%%%%%%%%%%%%%%%%%%%

 \section{Eye model results}
\label{sec:full_model_results}

 Here we present results from numerical solutions of the 
 eye model summarized in Section~\ref{sec:nondim_eye_model}.  Our numerical 
 approach involves the spatial discretization of the 
 diffusion equation in the contact lens which leads to a coupled system
 of ordinary differential equations for
 the pre-lens, post-lens, and eyelid compartments and solved using {\tt Matlab's}
 {\tt ode15s} solver.  Both equally-spaced finite differences as well as Chebyshev spectral methods have been implemented for the spatial discretization.  
 These results have been validated using the large
 diffusion limit solution identified in the previous section.

 In our model, the drug mass is initially located only 
 in the contact lens.  In general, at any later time, nonzero drug mass may reside
 in one of five compartments -- the contact lens, the pre-lens, the eyelid, 
 the post-lens, and the cornea -- or have been lost from the system altogether 
 by one or more of several mechanisms.  
  Our primary comparison will be with the experimental eye model data from Phan \textit{et al.}, which
 reports cumulative drug loss (red dye) from two different contact lenses in 24 hours
 of simulated wear.   In our model there are two pathways for the drug to leave
 the contact lens; through the pre-lens tear film and/or through the post-lens tear film.   Drug that reaches the pre-lens tear film can be subsequently absorbed
 by the eyelid and/or lost directly out of the pre-lens due to blinking.
 Drug that reaches the post-lens tear film can be lost out of the post-lens
 due to contact lens motion-driven transport (i.e.~slide out or squeeze out mechanisms).
 Although in clinical settings the cornea can be a desired target for drug delivery, we note that the experimental eye in the Phan {\it et al.} configuration is not permeable to the dye and in the results we present in this paper there is no transport of 
 drug from the post-lens into the cornea.

 There are several candidate mechanisms that may explain the cumulative drug
 release observations of Phan {\it et al.}  We organize these in terms of
 pathways through the pre-lens (and subsequently into the eyelid) and pathways
 through the post-lens.  In our model there is no pathway
 for the drug to reach the eyelid through the post-lens tear film.  
 There are a host of parameters that must be specified in order to compute
 solutions of our model.  With some exceptions noted below, we hand-tune model parameters to compare with 
 experimental data from Phan \textit{et al.} 
 \cite{phan2021development}. 
 We outline our choices and strategy below and attempt to 
 focus on parameters that most directly characterize various
 drug loss pathways:

 \begin{itemize}
     \item {\it Fixed parameters from vial configuration:}
     We use the equilibrium balance-calculated partition coefficients, $K$, and the diffusion coefficients, $D$, found via the single-variable optimization of the vial model (Section \ref{sec:vial_fitting}).
      \item {\it Pre-lens drug loss parameters/settings:} We consider either (a) a no-flux boundary condition at the contact lens/pre-lens
     boundary, in which case no drug escapes the contact lens via the pre-lens,
     or (b) a partition coefficient boundary condition that allows drug transport
     into the pre-lens tear film.  The former case was identified by setting $f=0$ in 
     the large diffusion limit model (Section \ref{sec:large_diff}).  
     In the latter case, we use parameter $p\in[0,1]$ to
     measure the proportion of drug lost from the pre-lens due to a blink. Recall that
     $p=0$ means no drug is swept out  and $p=1$ means all drug is swept out
     as the result of a blink.  The pre-lens tear
     film thickness $h_{\rm pre}^{\rm init}$ will be fixed at 5 $\mu$m.  The evaporation rate $J_E$ is set to zero; explorations in which it is allowed to vary indicate little affect on the resulting solutions.
     We comment that the use of a no-flux boundary condition
     on the contact lens/pre-lens boundary can be viewed
     as a proxy for the case of rapid evaporation where the pre-lens disappears too quickly for appreciable drug diffusion from the contact lens.
     \item {\it Eyelid absorption parameter:} For pre-lens cases with $f \neq 0$ there
     is a drug pathway from the contact lens to the eyelid that is controlled by the lid permeability.
     Mathematically, we allow this lid permeability constant $k_{\rm lid}$ to vary between the two 
     lenses in order to achieve reasonable fits.  However, the need for a lens-dependent $k_{\rm lid}$ indicates that the combination of mechanisms may not feasibly explain the dynamics since there
     is no reason to believe the contact lens should influence eyelid permeability.
     \item {\it Post-lens drug loss parameters:} Drug loss out of the post-lens tear film is characterized by either the slide out or squeeze out mechanism.
     The primary control parameters are $\Delta X_{\rm cl}$  in the former case and
      $\Delta h_{\rm post}$ in the latter.
     In general, these two parameters depend on post-lens tear film thickness
     although, except where noted, we assume $h_{\rm post}^{\rm init} = 5 \mu$m.
     Either $\Delta X_{\rm cl}$ and/or $\Delta h_{\rm post}$ could in general
     vary from one lens to the next due to the fit of the lens on the eye,
     its composition and material/elastic properties, as well as the presence
     of coatings that may influence friction.
 \end{itemize}

% \subsection{In vitro}
% \label{sec:results_in_vitro}

% \subsubsection{Comparison to experimental measurements}

% We hand-tune model parameters to compare with experimental data from Phan \textit{et al.} 
 %\cite{phan2021development}. We use the equilibrium balance-calculated partition coefficients and the diffusion coefficients found via the single-variable optimization (Section \ref{sec:vial_fitting}). %Unless otherwise mentioned, in these comparisons, the evaporation rate is set to zero, $h_{\rm pre}^{\rm init} = h_{\rm post}^{\rm init} = 5 \  \mu$m, $\Delta h = 0.033 \ \mu$m, $\Delta x = 1$ mm, and $k_{\rm lid} = 1 \ \mu$m/s. 
 %The initial pre- and post-lens thicknesses $h_{\rm pre}^{\rm init}$ (5 $\mu$m) and $h_{\rm post}^{\rm init}$ (5 $\mu$m) are held constant between the two lens types. The evaporation rate $J_E$ is set to zero; explorations in which it is allowed to vary indicate little affect on the resulting solutions. The squeeze out or slide out amounts $\Delta h$ and $\Delta x$ may differ between the lenses, as the lens composition may affect these quantities. As an example, the effect of friction is accounted for in the term $\Delta x$ and may vary between lens types due to their coatings. We also allow   the lid permeability constant $k_{\rm lid}$ to vary between the two lenses in order to achieve reasonable fits, but needing to do so indicates that the combination of mechanisms may not feasibly explain the dynamics since this should be an invariant quality of the eyelid.

\begin{figure}[H]
 \centering
\subfloat[][Contact lens concentration]{\includegraphics[scale=0.28]{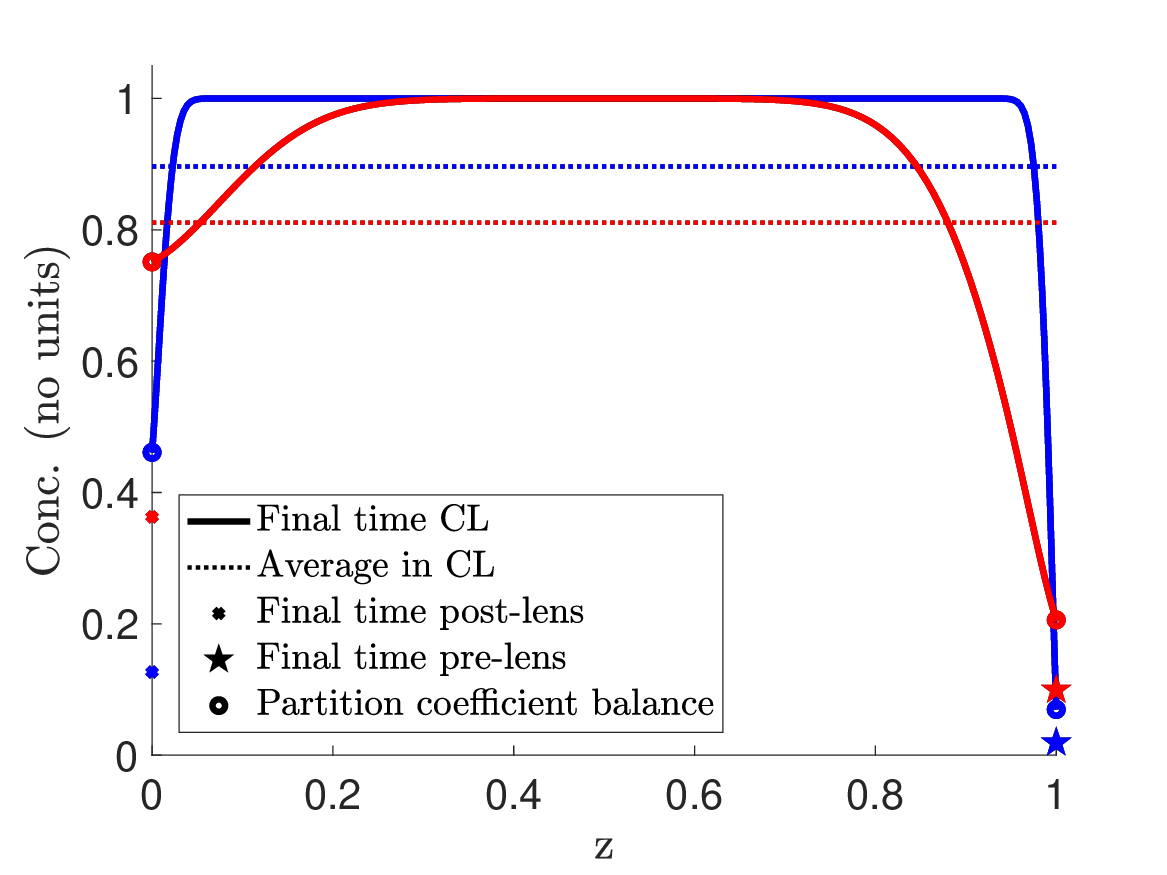}}
\subfloat[][Pre- and post-lens concentration]{\includegraphics[scale=0.28]{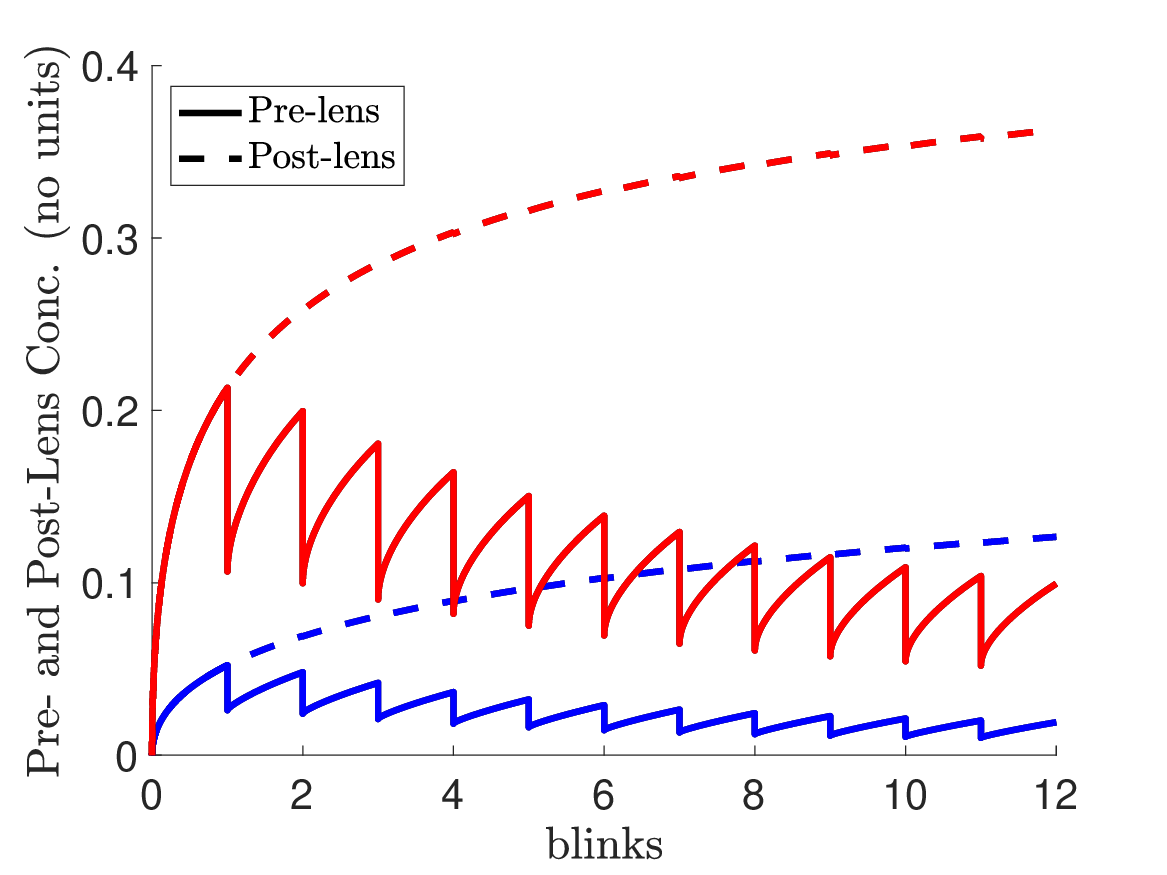}}
\subfloat[][Drug release]{\includegraphics[scale=0.28]{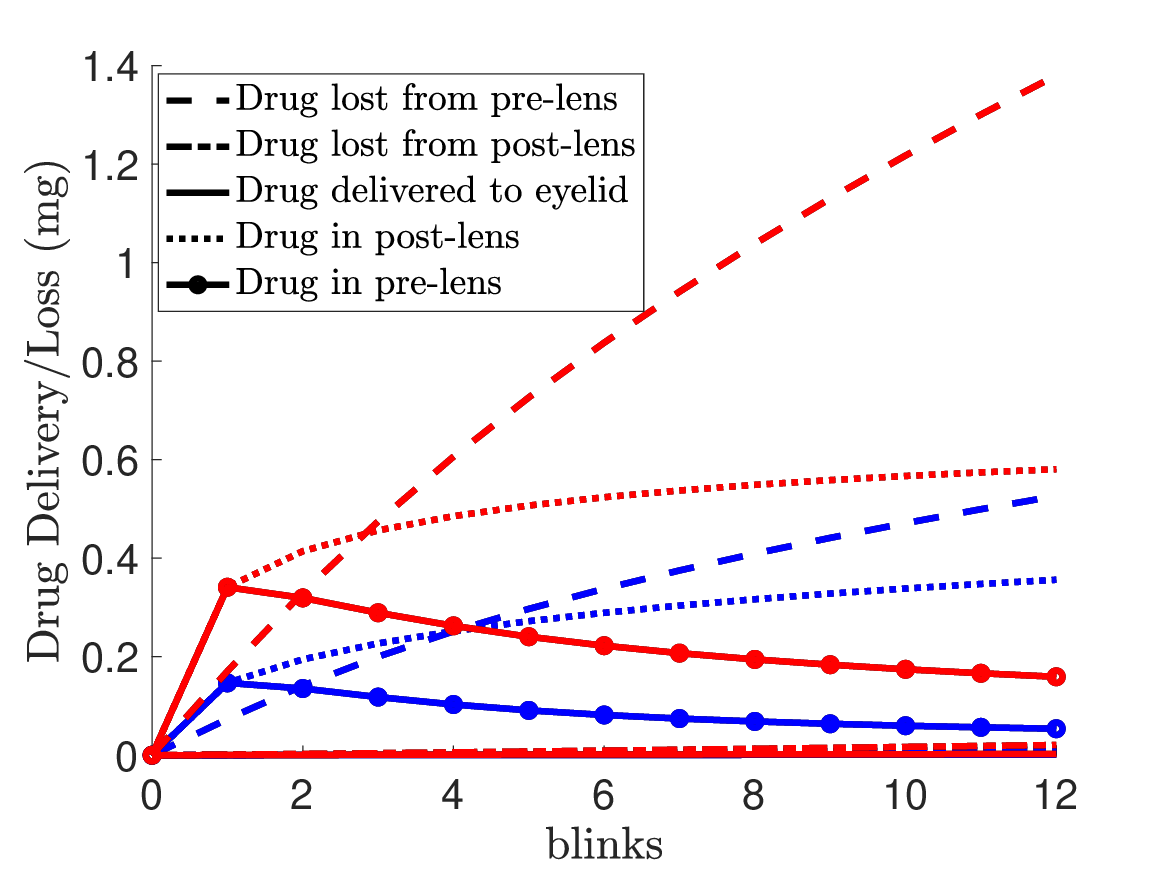}}
%\subfloat[][Cumulative drug release]{\includegraphics[scale=0.35]{Figs/1_29_24_CDR_model_sols_slide.eps}}
\caption{Representative model solutions for the first twelve blinks with the slide out contact lens motion option for the etafilcon A lens (red curves) and senofilcon A lens (blue curves).}
\label{fig:slide_model_sols}
 \end{figure}

 In Figure \ref{fig:slide_model_sols}, we show representative model solutions with the slide out contact lens motion option for the final time contact lens concentration, pre- and post-lens concentrations, and drug delivery/loss to/from various model compartments. Red and blue curves correspond to the etafilcon A and senofilcon A lenses, respectively. The model is simulated for the first 12 blinks to capture the initial release of drug. We use partition coefficient boundary conditions on the pre- and post-lens sides, $p = 0.5$ so that half of the drug in the pre-lens is swept out by a blink, $h_{\rm pre}^{\rm init} = h_{\rm post}^{\rm init} = 5 \ \mu$m, $k_{\rm lid} = 3 \times 10^{-9}$ m/s, and $\Delta X_{\rm cl}= 0.05$ mm. Squeeze out solutions with $\Delta h_{\rm post} = 0.015 \ \mu$m (not shown) are nearly identical. 
 
 Figure \ref{fig:slide_model_sols}(a) shows the contact lens concentration after 12 blinks. Unlike the vial setting (see Figure \ref{fig:vial_conc}), the profiles are not symmetric across the lens, but similar to the vial case, the etafilcon A lens loses more drug than  senofilcon A  due to its larger diffusion coefficient. The pre- and post-lens concentrations are shown by X and star markers at $x = 0$ and $x = 1$, and the partition coefficient balances at those boundaries are shown by large circular markers. Figure \ref{fig:slide_model_sols}(b) shows the pre- and post-lens concentrations over the 12 blinks. Due to the reset conditions at the start of each interblink, the concentrations drop rapidly and then increase between blinks. Figure \ref{fig:slide_model_sols}(c) breaks down the drug released from the contact lens into the avenues of drug delivery/loss.  The drug in the post-lens and the drug delivered to the eyelid for both lenses are very close to zero.

 There are four  categories that we organize our results into based on pre-lens drug loss or lack thereof:

 \begin{itemize}
\item Full pre-lens drug loss ($p=1$).
\item No flux into the pre-lens.
\item No pre-lens drug loss ($p = 0$).
\item Partial pre-lens drug loss ($0 < p < 1$).
 \end{itemize}
In all but the second category, we assume a partition coefficient boundary condition at the lens/tear film interface for both the pre- and post-lens. 
 There is a natural ordering of these four options and they are discussed in detail in the next subsections. The aggregate results of the hand-tuned parameters for each model option simulation, including subcases where appropriate, are summarized in Table \ref{table:results}.

 \begin{table}[h]
\centering
\begin{tabular}{|c|c|c|c|r|r|r|r|}
\hline
\textbf{\begin{tabular}[c]{@{}c@{}}Allow flux\\ on pre- \\lens side?\end{tabular}} & \textbf{\begin{tabular}[c]{@{}c@{}}\% drug swept\\ out of pre-lens \\ by blink\end{tabular}} & \textbf{\begin{tabular}[c]{@{}c@{}}Lens motion\\ option\end{tabular}} & \textbf{\begin{tabular}[c]{@{}c@{}}Lens\\ type\end{tabular}} & \multicolumn{1}{c|}{\textbf{\begin{tabular}[c]{@{}c@{}}$\bm{k_{\rm lid}}$\\ (m/s)\end{tabular}}} & \multicolumn{1}{c|}{\textbf{\begin{tabular}[c]{@{}c@{}}$\bm{\Delta h_{\rm post}}$\\ ($\bm{\mu}$m)\end{tabular}}} & \multicolumn{1}{c|}{\textbf{\begin{tabular}[c]{@{}c@{}}$\bm{\Delta X_{\rm cl}}$\\ (mm)\end{tabular}}}   &  \textbf{Figure}  \\ \hline
Yes & 100 & None & Eta & $700 \times 10^{-9}$ &  & & \multirow{2}{*}{\ref{fig:noCLmot_fit_p1}} \\ \cline{1-7} 
Yes & 100 & None & Seno & $700 \times 10^{-9}$ &  & & \\ \hline
No & N/A & Slide out & Eta & N/A &  & 0.06 & \multirow{4}{*}{\ref{fig:slide_squeeze_noflux_fit}} \\ \cline{1-7} 
No & N/A & Slide out & Seno & N/A &  & 0.35 & \\ \cline{1-7}
No & N/A & Squeeze out & Eta & N/A & 0.02 & &  \\ \cline{1-7}
No & N/A & Squeeze out & Seno & N/A & 0.1 & & \\ \hline
Yes & 0 & Slide out & Eta & $3 \times 10^{-9}$ &  & 0.04 & \multirow{2}{*}{\ref{fig:slide_fit}} \\ \cline{1-7} 
Yes & 0 & Slide out & Seno & $3 \times 10^{-9}$ &  & 0.065 & \\ \hline
Yes & 0 & Squeeze out & Eta & $3 \times 10^{-9}$ & 0.015 &  & \multirow{2}{*}{\ref{fig:squeeze_fit}} \\ \cline{1-7} 
Yes & 0 & Squeeze out & Seno & $3 \times 10^{-9}$ & 0.025 & & \\ \hline
Yes & 0 & None & Eta & $11.5 \times 10^{-9}$ &  & & \multirow{2}{*}{\ref{fig:noCLmot_fit}} \\ \cline{1-7} 
Yes & 0 & None & Seno & $35 \times 10^{-9}$ &  &  &\\ \hline
Yes & 0 & Slide out & Eta & 0 &  & 0.06 & \multirow{4}{*}{\ref{fig:slide_squeeze_fit_klid0}} \\ \cline{1-7} 
Yes & 0 & Slide out & Seno & 0 &  & 0.2 & \\ \cline{1-7}
Yes & 0 & Squeeze out & Eta & 0 & 0.018 & & \\ \cline{1-7}
Yes & 0 & Squeeze out & Seno & 0 & 0.075 & & \\ \hline
Yes & 0.2 & Slide out & Eta & 0 &  & 0.03 & \multirow{4}{*}{\ref{fig:slide_squeeze_fit_p01}} \\ \cline{1-7}
Yes & 0.2 & Slide out & Seno & 0 &   & 0.04 & \\ \cline{1-7}
Yes & 0.2 & Squeeze out & Eta & 0 & 0.01 & & \\ \cline{1-7}
Yes & 0.2 & Squeeze out & Seno & 0 & 0.014 & & \\ \hline
\end{tabular}
\caption{Results from hand-tuning parameters. The partition coefficients and diffusion coefficients used are those from the vial setting. The initial pre- and post-lens thicknesses $h_{\rm pre}^{\rm init} = h_{\rm post}^{\rm init} =  5 \ \mu$m are held constant and the evaporation rate is set to zero.  An N/A in the table means the related mechanism is not active for that version of the model. A zero in the table means that parameter has been set to zero.  We use the shorthand Eta and Seno to stand for etafilcon A and senofilcon A lenses.}
\label{table:results}
\end{table}

%{\color{red}{DMA - should we advertise/list the aggregate table up front to indicate parameter values for the various plots ?}}

 %\subsection{Pre-lens and Post-lens partition coefficient balance}
 \subsection{Full pre-lens drug loss}

\begin{figure}[H]
\centering
\subfloat[][Etafilcon A lens]{\includegraphics[scale=0.36]{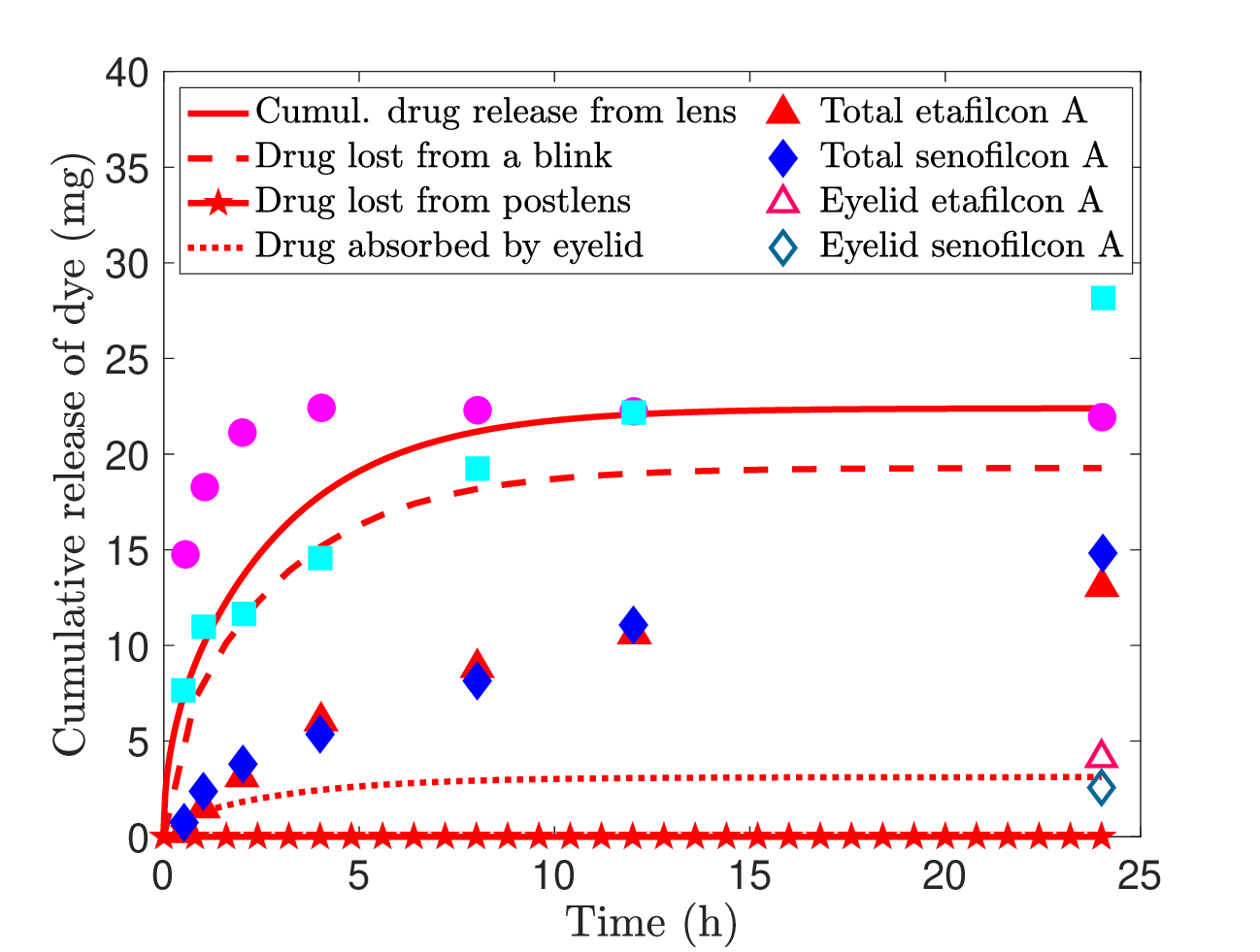}}
\subfloat[][Senofilcon A lens]{\includegraphics[scale=0.36]{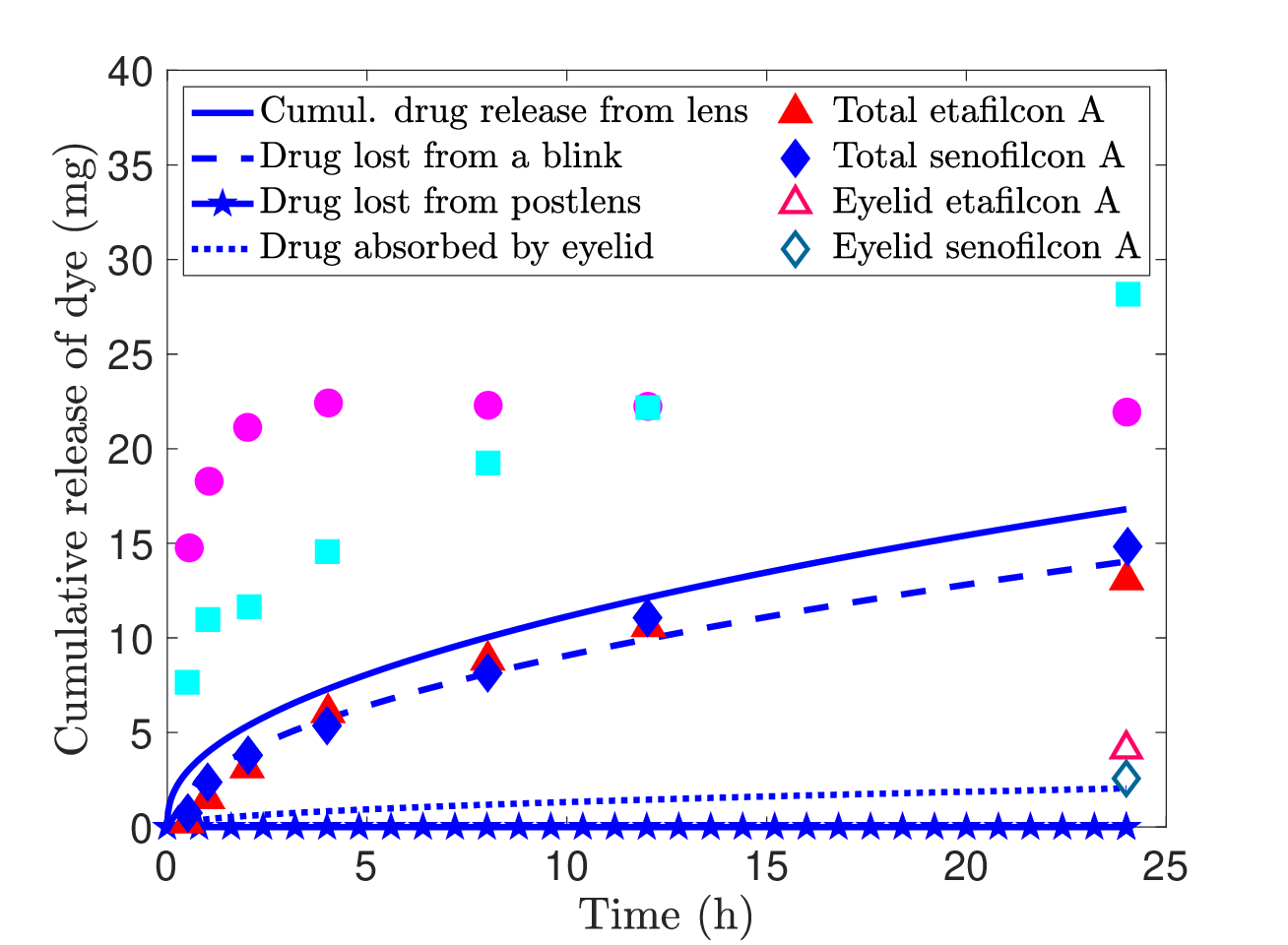}} 
\caption{Comparison of model output with Phan \textit{et al.} \cite{phan2021development} data.  The no contact lens motion option is selected with $p = 1$ so that all of the drug is swept away by the blink through the pre-lens tear film.  %The hand-tuned parameters used are  $k_{\rm lid} = 3\times 10^{-9}$ m/s,  $\Delta h_{\rm eta} = 0.015 \ \mu$m and $\Delta h_{\rm seno} = 0.025 \ \mu$m. 
}
\label{fig:noCLmot_fit_p1}
 \end{figure}

We use pre-lens and post-lens partition coefficient balance boundary conditions which means that the drug can diffuse into both the pre-lens and the post-lens tear films.
 %\subsubsection{Pre-lens Drug Loss; Eyelid Absorption; No Post-lens Motion}
%{\color{red}{I think this might be a good first case to start with in the results section ... it forces the consideration of either $p=0$ or no-flux on the pre-lens side and consideration of post-lens motion.}}
% \paragraph{Partition coefficient balance on both pre- and post-lens sides, $\bm{p = 1}$}
 %\underline{Slide out}:
 %\underline{Squeeze out}:
% \underline{No lens motion}:
We assess the possibility that each blink sweeps away 
all drug in the pre-lens and there is no motion of the contact
lens (so that there is no drug loss out of the post-lens tear film). Figure \ref{fig:noCLmot_fit_p1} shows this case
in comparison with the Phan \textit{et al.} data for the two lenses. The cumulative drug release corresponding to the model solutions gives a reasonable fit to the experimental data for the senofilcon A lens (right plot), which might at first suggest a drug loss pathway through the pre-lens as a dominant mechanism behind
the observed drug release.  
However, the corresponding predicted drug release for the etafilcon A lens reaches that of the vial setting by the final time and overall the drug release rate for this lens is much too large (left plot). 
In the absence of a reason why a blink mechanism would clear out all pre-lens drug mass for one
lens and not another, it seems that another drug release pathway should be considered.
This leads us to examine either no-flux on the pre-lens side or $p=0$ (no pre-lens drug loss due to blinking) and options for post-lens motion in order to lessen the amount of pre-lens drug lost from a blink.

 \subsection{Post-lens partition coefficient balance, pre-lens no flux}

Since in the previous subsection the drug loss through the pre-lens was too large to explain the etafilcon A lens model eye data,  
we next investigate a version of the model in which no drug is lost from a blink because no drug is allowed to diffuse into the pre-lens. %This version of the model may help match the experimental data from Phan {\it et al.} \cite{phan2021development}, which shows much slower cumulative release of drug in the model eye vs. the vial setting.
Here the drug can only leave the contact lens via the post-lens tear film.
In this instance our model does not allow any eyelid drug absorption; we acknowledge that we cannot fit the final time eyelid measurement from Phan \textit{et al.} \cite{phan2021development}.   The no flux condition formulation of the model is described in Li and Chahaun (2006) \cite{li2006modeling}. One can assume that the tear film evaporates instantly, so that there is no pre-lens, and therefore there is nowhere for the drug to go in the upwards direction. In this way, this version of the model acts as a proxy for very high evaporation, and sets the evaporation rate aside as a parameter that we do not consider in this paper.  

\begin{figure}[H]
\centering
  \subfloat[][Etafilcon A lens, slide out]{\includegraphics[scale=0.36]{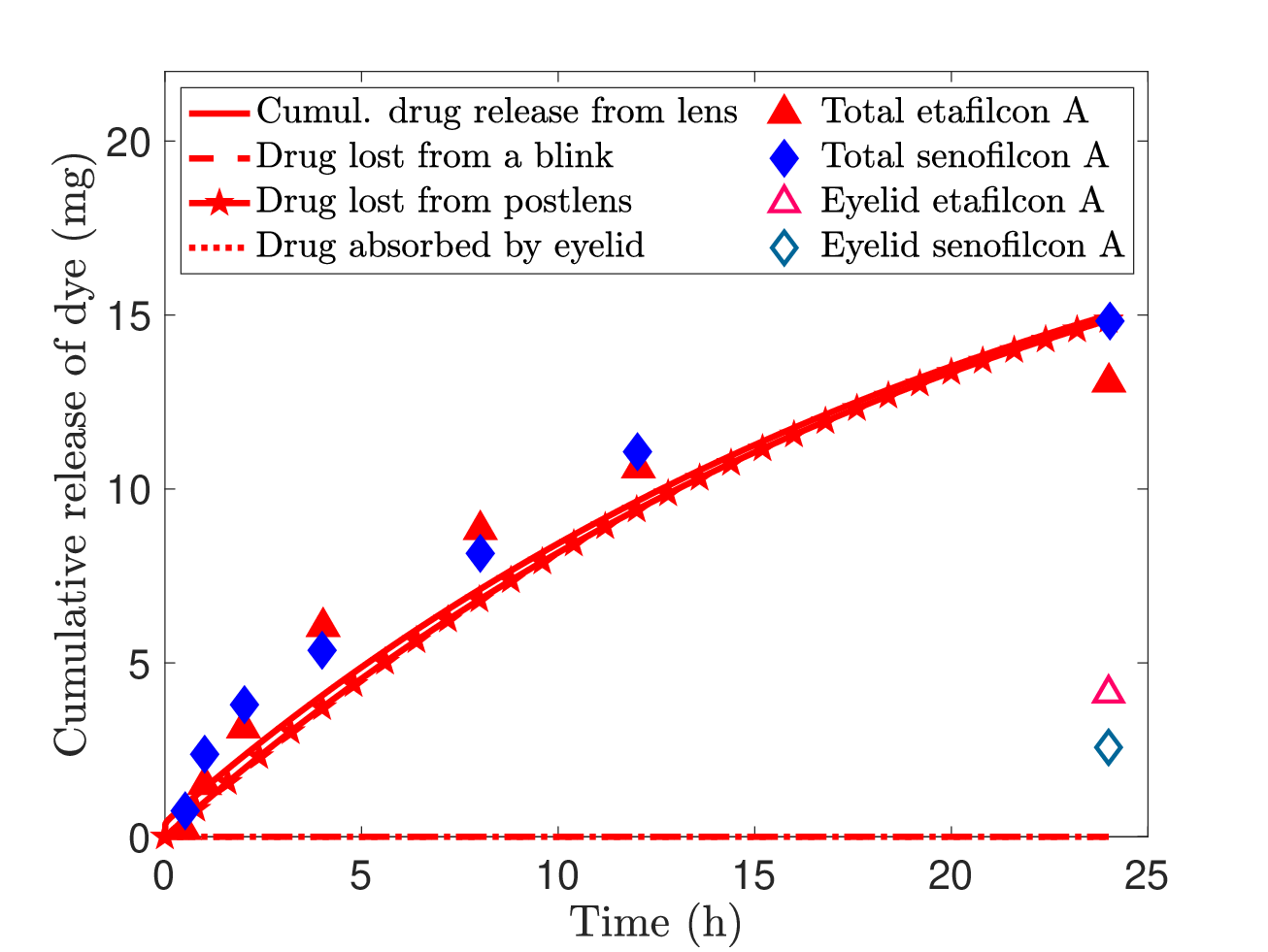}}
\subfloat[][Senofilcon A lens, slide out]{\includegraphics[scale=0.36]{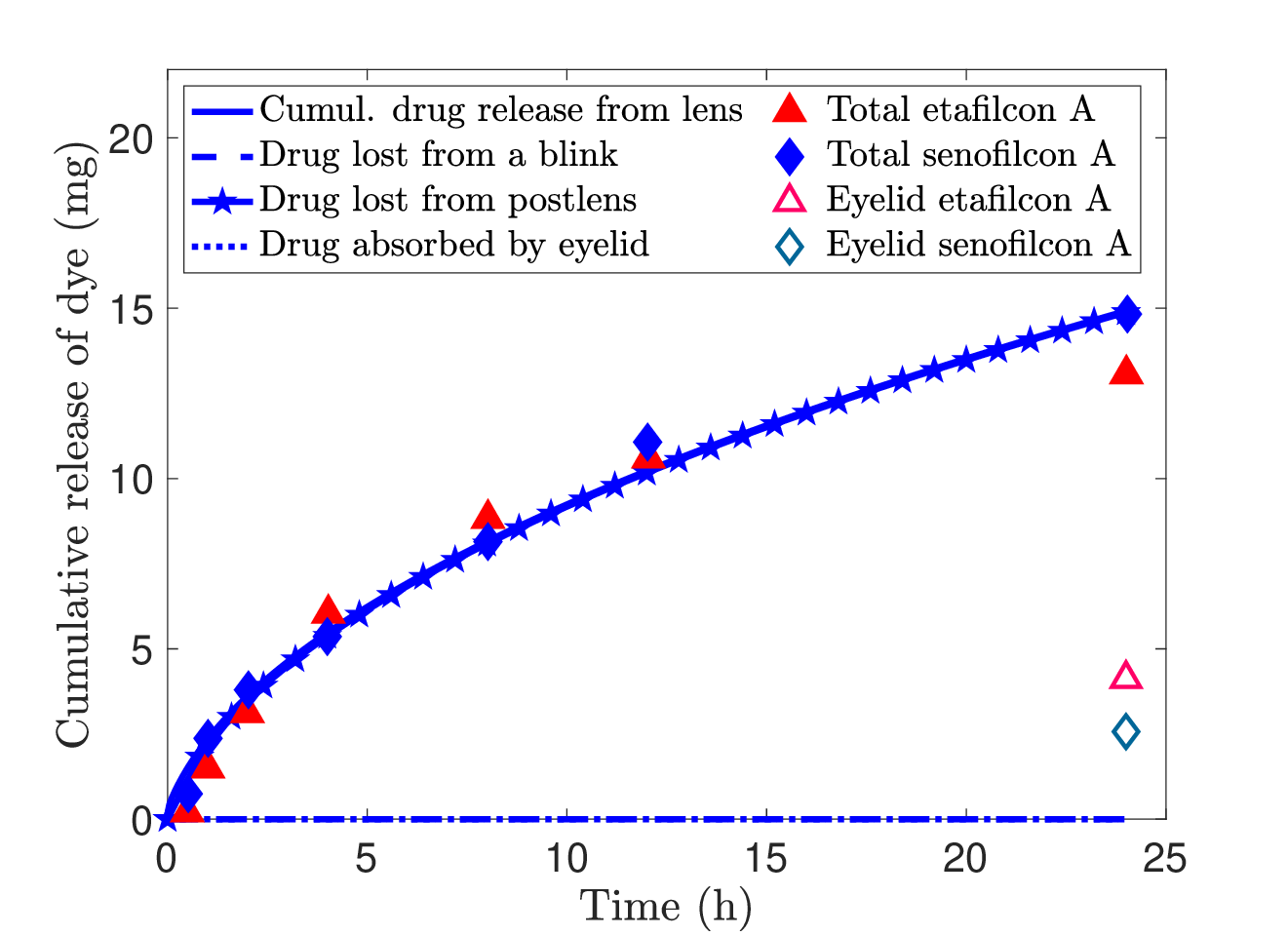}} \\
 \subfloat[][Etafilcon A lens, squeeze out]{\includegraphics[scale=0.36]{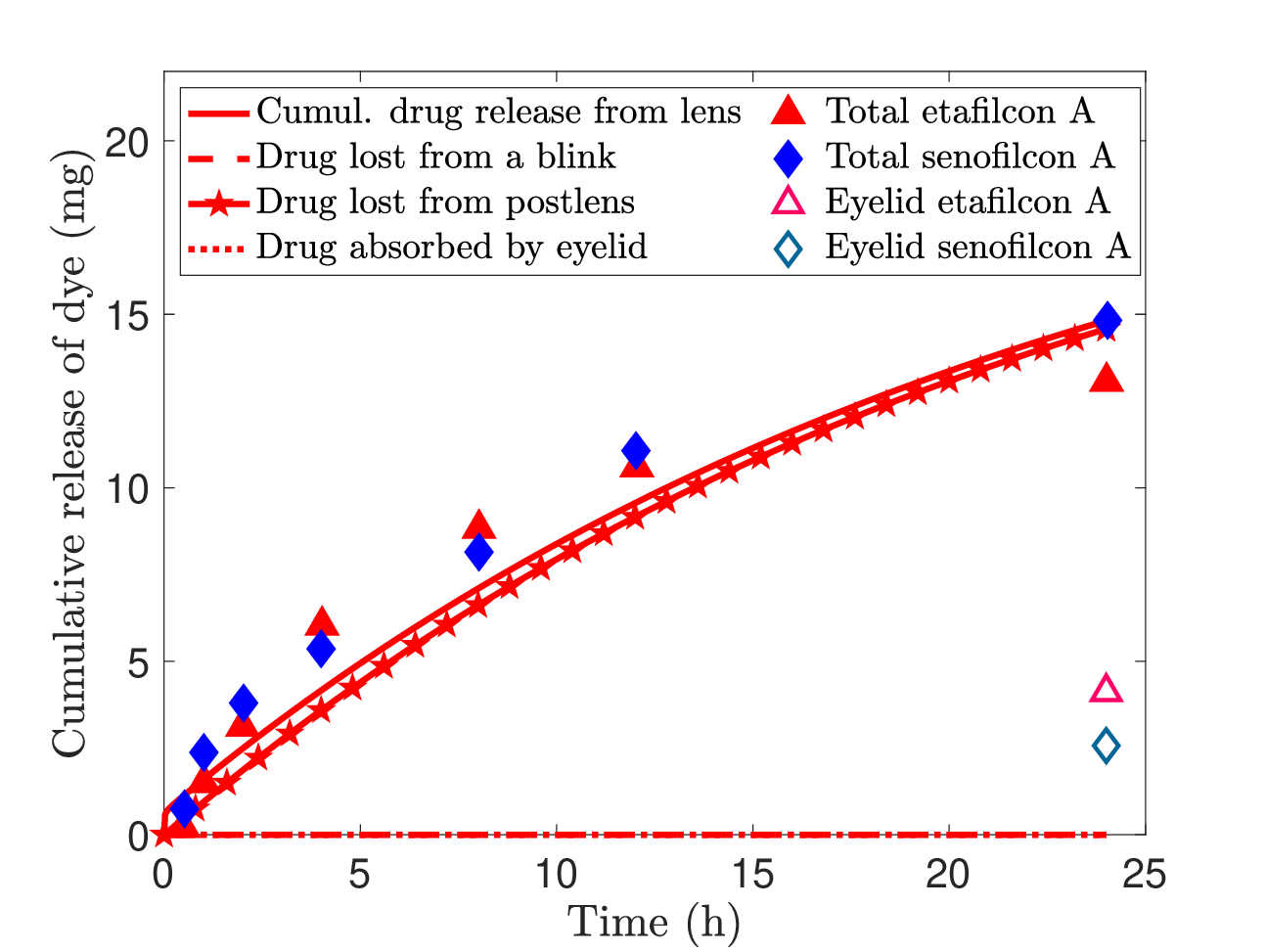}}
\subfloat[][Senofilcon A lens, squeeze out]{\includegraphics[scale=0.36]{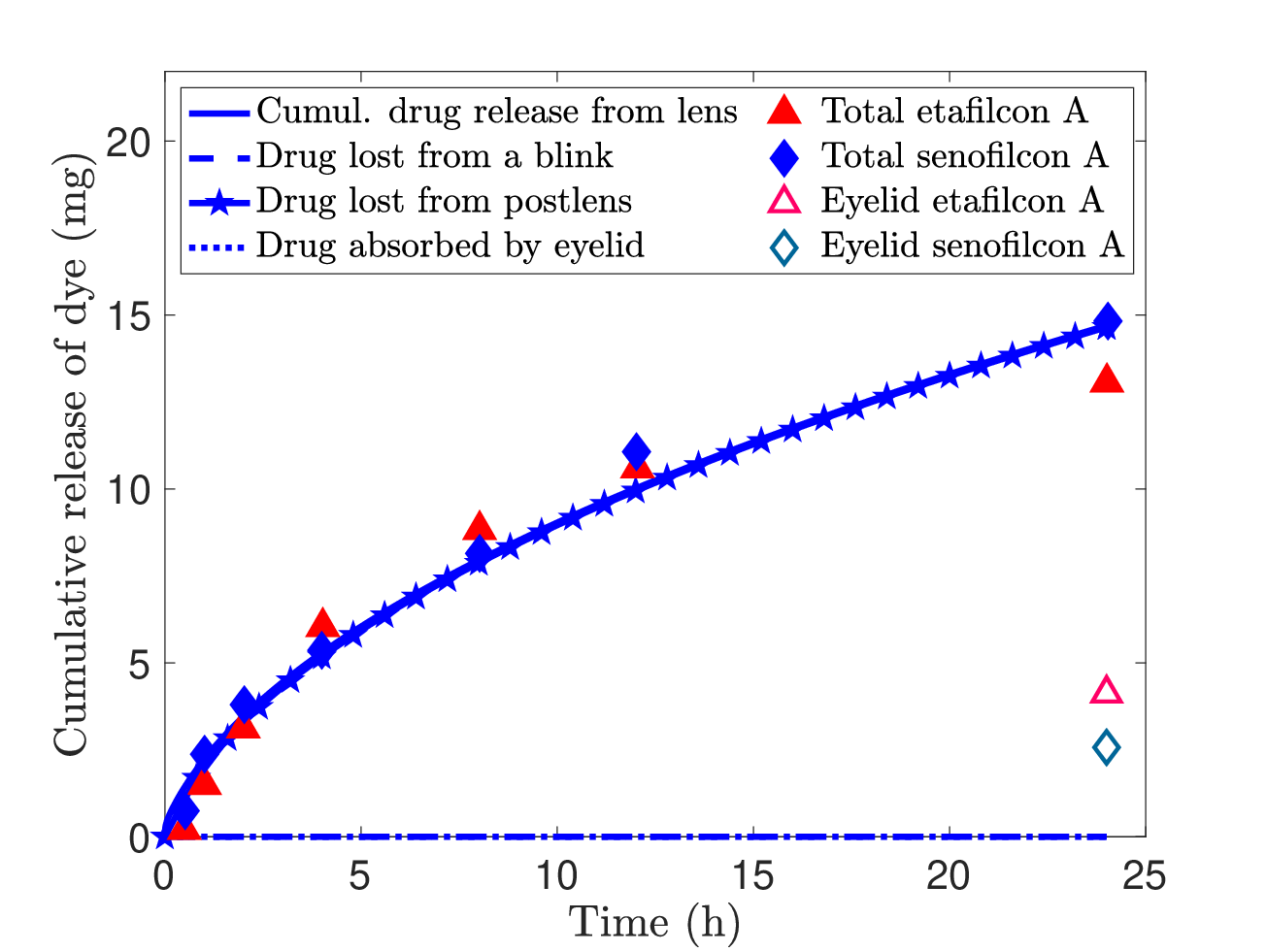}}
\caption{Comparison of model solutions with no flux on the pre-lens side with Phan \textit{et al.} \cite{phan2021development} data. %The hand-tuned parameters used are  $k_{\rm lid} = 3\times 10^{-9}$ m/s, $\Delta x^{\rm eta} = 0.045$ mm and $\Delta x^{\rm seno} = 0.065$ mm.
}
\label{fig:slide_squeeze_noflux_fit}
 \end{figure}

Full model solutions for the pre-lens no-flux and post-lens slide out and squeeze out cases are shown in Figure~\ref{fig:slide_squeeze_noflux_fit} and
reveal excellent agreement with the cumulative drug release data of Phan 
{\it et al.} In the squeeze out case, the post-lens film depression amounts correspond to blink pressures of  59.6 Pa and 306 Pa for the etafilcon A and senofilcon A lenses. We also run simulations of this version of the model with the no contact lens motion option (not shown), but the model predicts a very small cumulative release from the lens that is essentially constant after an initial increase; this is far from the experimental data.

Example solutions of this case were shown also in the context of the large diffusion
limit.  In particular, for the pre-lens no-flux and post-lens slide out case, three different values of $\Delta X_{\rm cl}$ corresponding to different values of $h_{\rm post}$ were shown in the upper plots of Figure~\ref{fig_SlideOut_LargeD} that achieve a good match to the cumulative drug release data of
Phan {\it et al.}  Notably for the etafilcon A lens (upper left plot
of Figure~\ref{fig_SlideOut_LargeD}) the full model predictions for $\Delta X_{\rm cl}$ align very well with the large diffusion predictions, despite the fact that the full model seems to have $\bar{D} \ll 1$ rather than $\bar{D} \gg 1$.
The comparison between the full model predictions and the large diffusion
limit is not as close for the senofilcon A lens (upper right plot
of Figure~\ref{fig_SlideOut_LargeD}) perhaps due to $\bar{D}$ being even smaller for this lens.  The larger $\Delta X_{\rm cl}$ values predicted by the full model appear to be consistent with the fact that to achieve the same drug release, larger post-lens drug loss due to lens motion is
required for reduced diffusive flux out of the contact lens. 
Of additional note is that the observed
trend of decreasing $\Delta X_{\rm cl}$ for increasing $h_{\rm post}$ seems to be robust for both the large diffusion model and the full model.

For the pre-lens no-flux and post-lens squeeze out case, three different values of $\Delta h_{\rm post}$
corresponding to different values of $h_{\rm post}$ were shown in the upper plots of
Figure~\ref{fig_SqueezeOut_LargeD} that achieve a good match to the cumulative drug release data of Phan {\it et al.}  Similar observations
here can be made regarding the comparison between the large diffusion model predictions and the full model predictions.

\subsection{No pre-lens drug loss due to blinking}

We now consider a setting in which drug is lost into both the pre- and post-lens tear films. Here, a partition coefficient balance controls diffusion of drug into both the pre-lens and post-lens.  However, in contrast to the first option explored, in which $p = 1$ so that all drug is swept out by a blink, we investigate the setting where $p = 0$ so that no drug is lost via blinking. 
In terms of the blink effect on the pre-lens tear film, eyelid absorption, 
 and post-lens drug loss, we study the following cases:
\begin{itemize} 
 \item %No pre-lens drug loss ($p=0$). 
 Eyelid absorption.  Post-lens slide out.
\item %No pre-lens drug loss ($p=0$). 
Eyelid absorption.  Post-lens squeeze out.
 \item %No pre-lens drug loss ($p=0$). 
 Eyelid absorption.  No post-lens drug loss (no contact lens motion).
 \item %No pre-lens drug loss ($p=0$). 
 No eyelid absorption.  Post-lens slide out or squeeze out.
 \end{itemize}

 \subsubsection{No pre-lens drug loss; eyelid absorption; post-lens: slide out}
 
\begin{figure}[H]
\centering
  \subfloat[][Etafilcon A lens]{\includegraphics[scale=0.36]{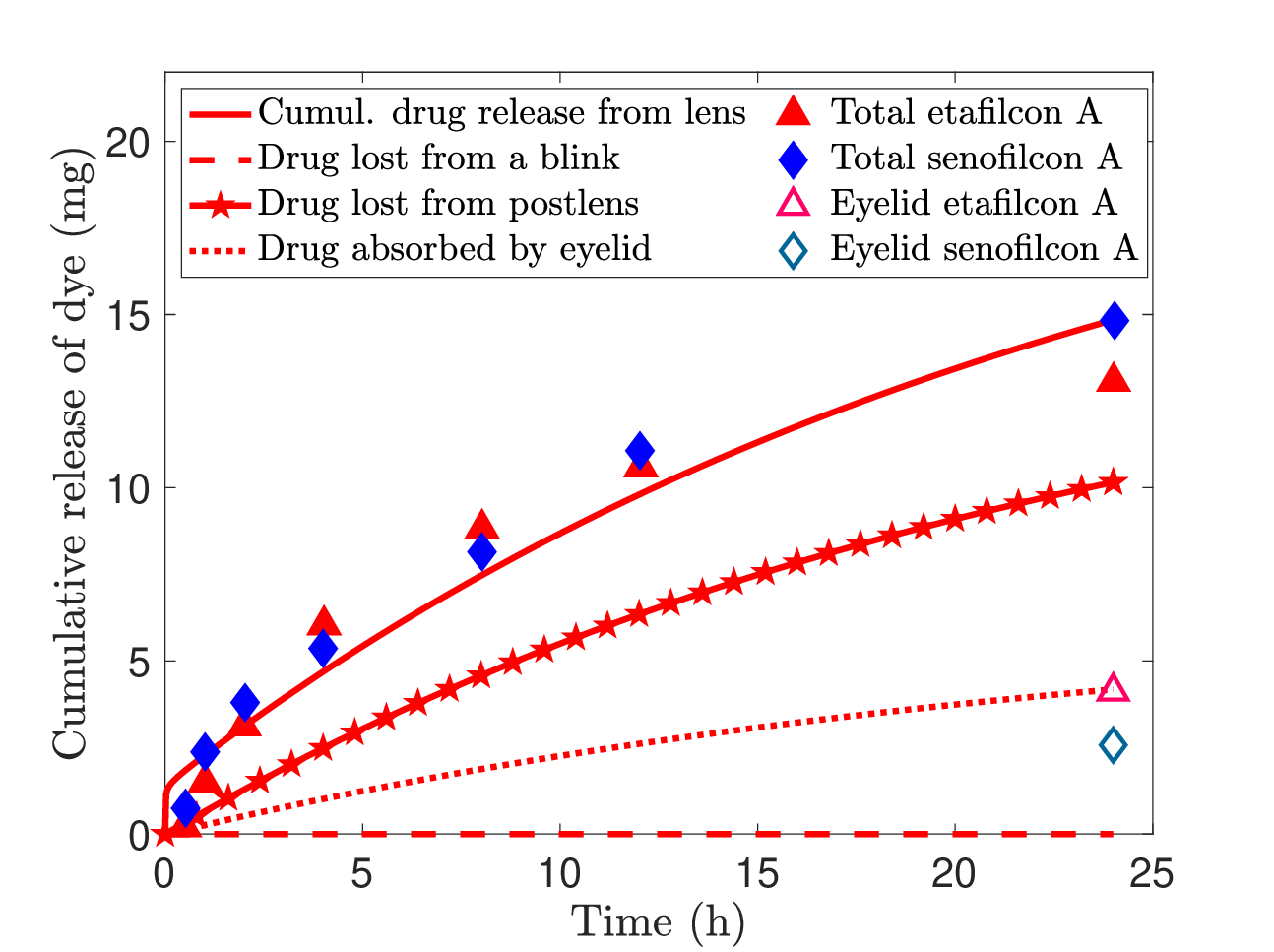}}
\subfloat[][Senofilcon A lens]{\includegraphics[scale=0.36]{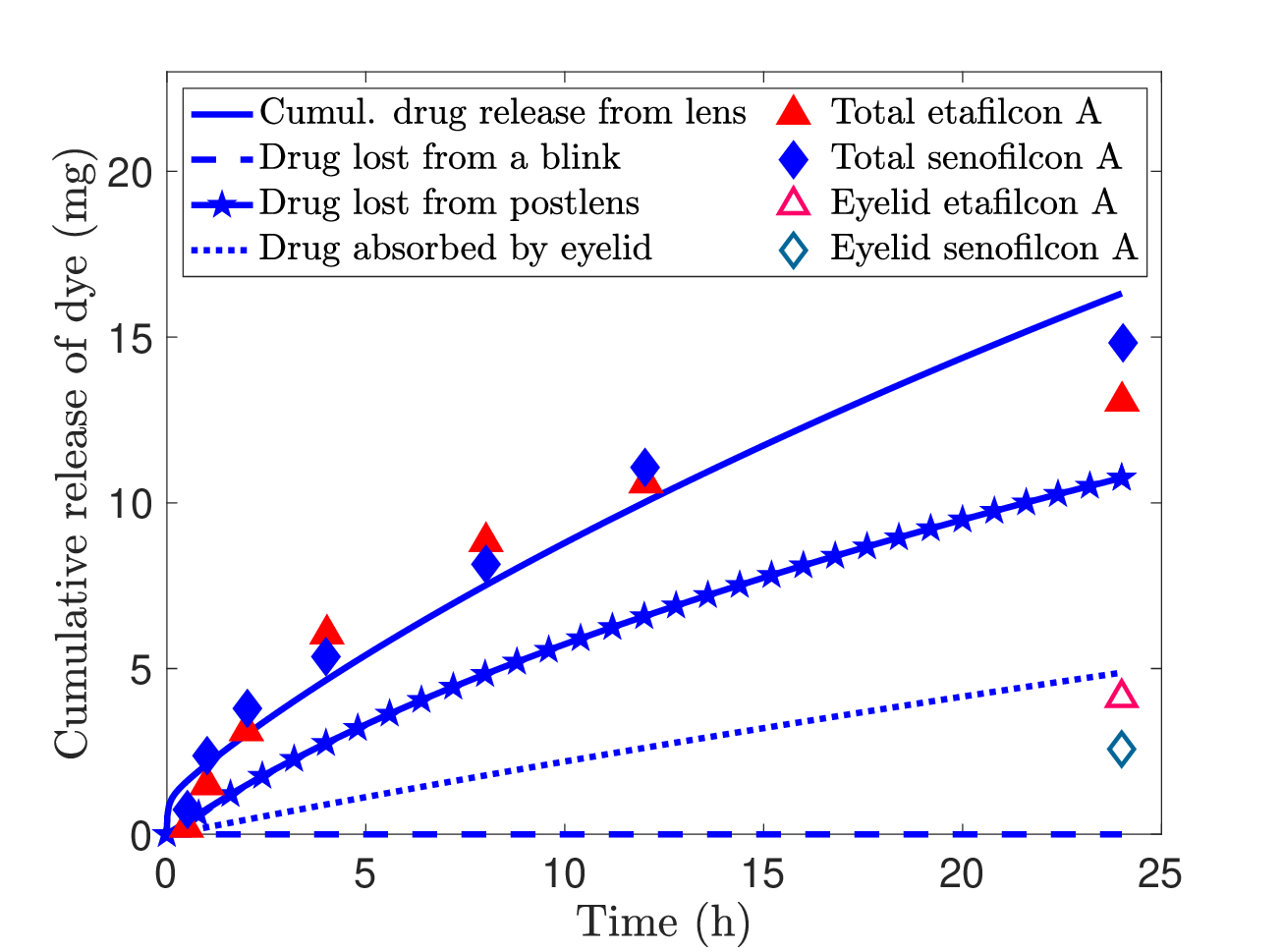}} 
\caption{Comparison of model output with Phan \textit{et al.} \cite{phan2021development} data. The model uses the slide out option with $p = 0$ so that no drug is swept away by the blink. %The hand-tuned parameters used are  $k_{\rm lid} = 3\times 10^{-9}$ m/s, $\Delta x^{\rm eta} = 0.045$ mm and $\Delta x^{\rm seno} = 0.065$ mm.
}
\label{fig:slide_fit}
 \end{figure}

%\underline{Slide out}: 
Motivated by the findings of Section \ref{sec:large_diff}, namely, that the proportion of pre-lens mass lost as a result of a blink $p$ may need to be much closer to 0 than to 1 in order to match the 
Phan \textit{et al.} \cite{phan2021development} data, we examine full model simulations with $p=0$. We observe that both the calculated total cumulative drug release from the lens and the eyelid absorption profile from the model match reasonably well with the experimental data (see
Figure~\ref{fig:slide_fit}). Most of the total drug release is lost from the system through the post-lens due to the slide out action. No drug is lost from a blink due to the assumption that $p = 0$.  The cumulative drug release profile for the etafilcon A lens is more curved than that of the senofilcon A lens. This is in part due to the larger diffusion coefficient of the etafilcon A lens, which causes more drug to be released in the first few hours, and then drug release slows as concentration gradients lessen and less drug remains to be lost.
 
 In this regime of lid permeability constant and slide out amount $\Delta X_{\rm cl}$, we find that the final time contact lens concentration profile is roughly uniform for the etafilcon A lens %(Figure \ref{fig:slide_model_sols_eta}a) 
 but not so for the senofilcon A lens (not shown).
 %(Figure \ref{fig:slide_model_sols_seno}a)
 At least for the etafilcon A lens, this provides support for the assumption that $p = 0$ following the large diffusion limit in Section \ref{sec:large_diff}. The pre- and post-lens concentration profiles are very similar for the etafilcon A lens, but the post-lens values are much smaller than the pre-lens for the senofilcon A lens. 
 The slide out amount $\Delta X_{\rm cl}^{\rm eta}$ is smaller than $\Delta X_{\rm cl}^{\rm seno}$, but they are both smaller than the range of 0.1-4 mm from studies cited earlier.

% \begin{figure}[H]
%     \centering
%\subfloat[][Contact lens concentration]{\includegraphics[scale=0.27]{Figs/1_29_24_C_eta_2.eps}}
%\subfloat[][Pre- and post-lens concentration]{\includegraphics[scale=0.27]{Figs/1_29_24_prepost_eta_2.eps}}
%\subfloat[][Drug release]{\includegraphics[scale=0.27]{Figs/1_29_24_drug_eta_2.eps}}
%\caption{Model solutions for the etafilcon A lens for the slide out option with $p = 0$ so that no drug is swept away by the blink. In (c), the drug in the post-lens (not shown) nearly overlaps that of the drug in the pre-lens. %The hand-tuned parameters used are $k_{\rm lid} = 3\times 10^{-9}$ m/s, and $\Delta x = 0.045$ mm.
%}
%\label{fig:slide_model_sols_eta}
% \end{figure}

 % \begin{figure}[H]
 %    \centering
%\subfloat[][Contact lens concentration]{\includegraphics[scale=0.27]{Figs/1_16_24_C_seno_2.eps}}
%\subfloat[][Pre- and post-lens concentration]{\includegraphics[scale=0.27]{Figs/1_16_24_prepost_seno_2.eps}}
%\subfloat[][Drug release]{\includegraphics[scale=0.27]{Figs/1_16_24_drug_seno_2.eps}}
%\caption{Model solutions for the senofilcon A lens for the slide out option with $p = 0$ so that no drug is swept away by the blink. %The hand-tuned parameters used are  $k_{\rm lid} = 3\times 10^{-9}$ m/s, and $\Delta x = 0.065$ mm. 
%}
%\label{fig:slide_model_sols_seno}
% \end{figure}

\subsubsection{No pre-lens drug loss; eyelid absorption; post-lens: squeeze out}
\begin{figure}[H]
\centering
\subfloat[][Etafilcon A lens]{\includegraphics[scale=0.36]{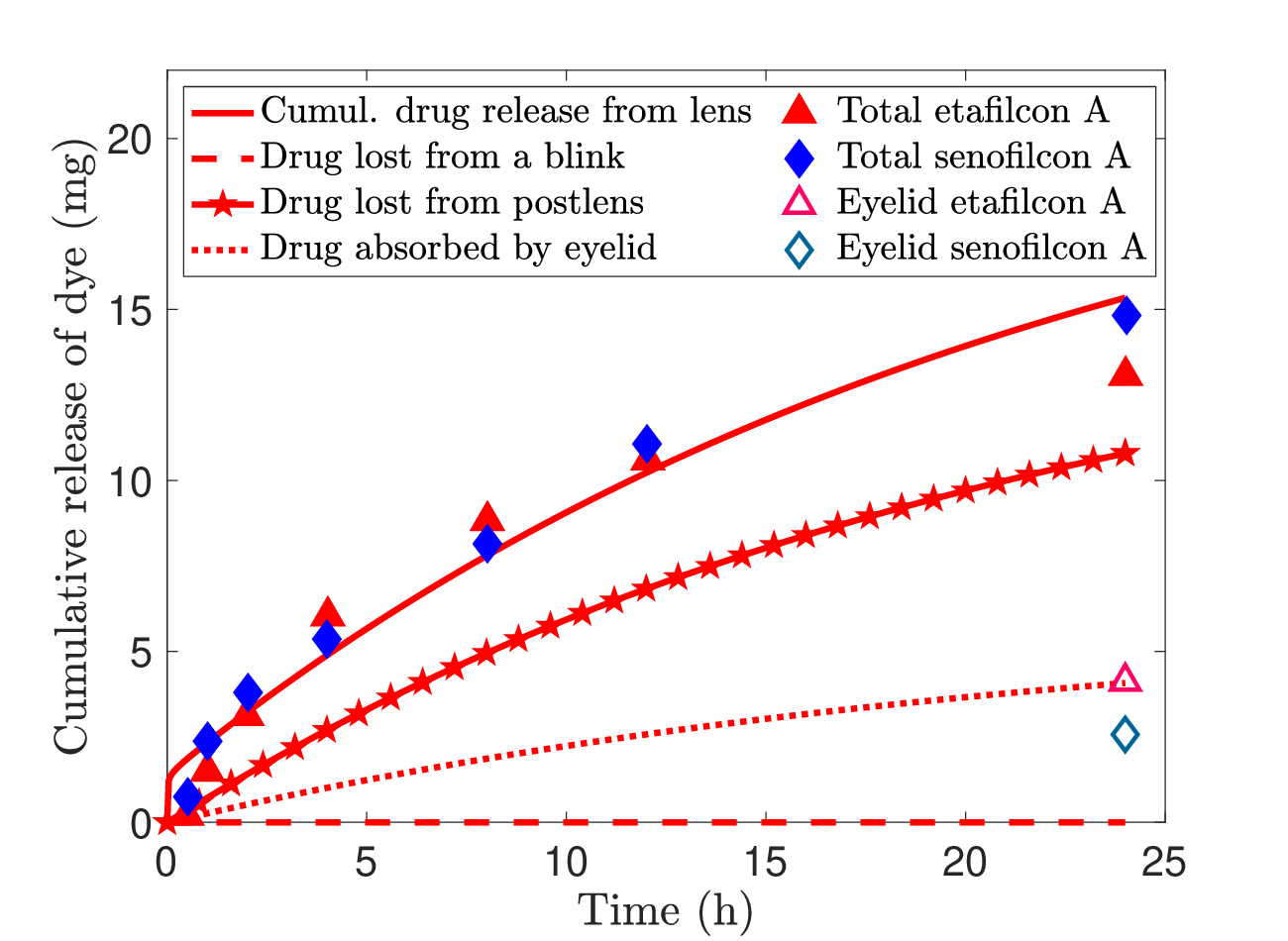}}
\subfloat[][Senofilcon A lens]{\includegraphics[scale=0.36]{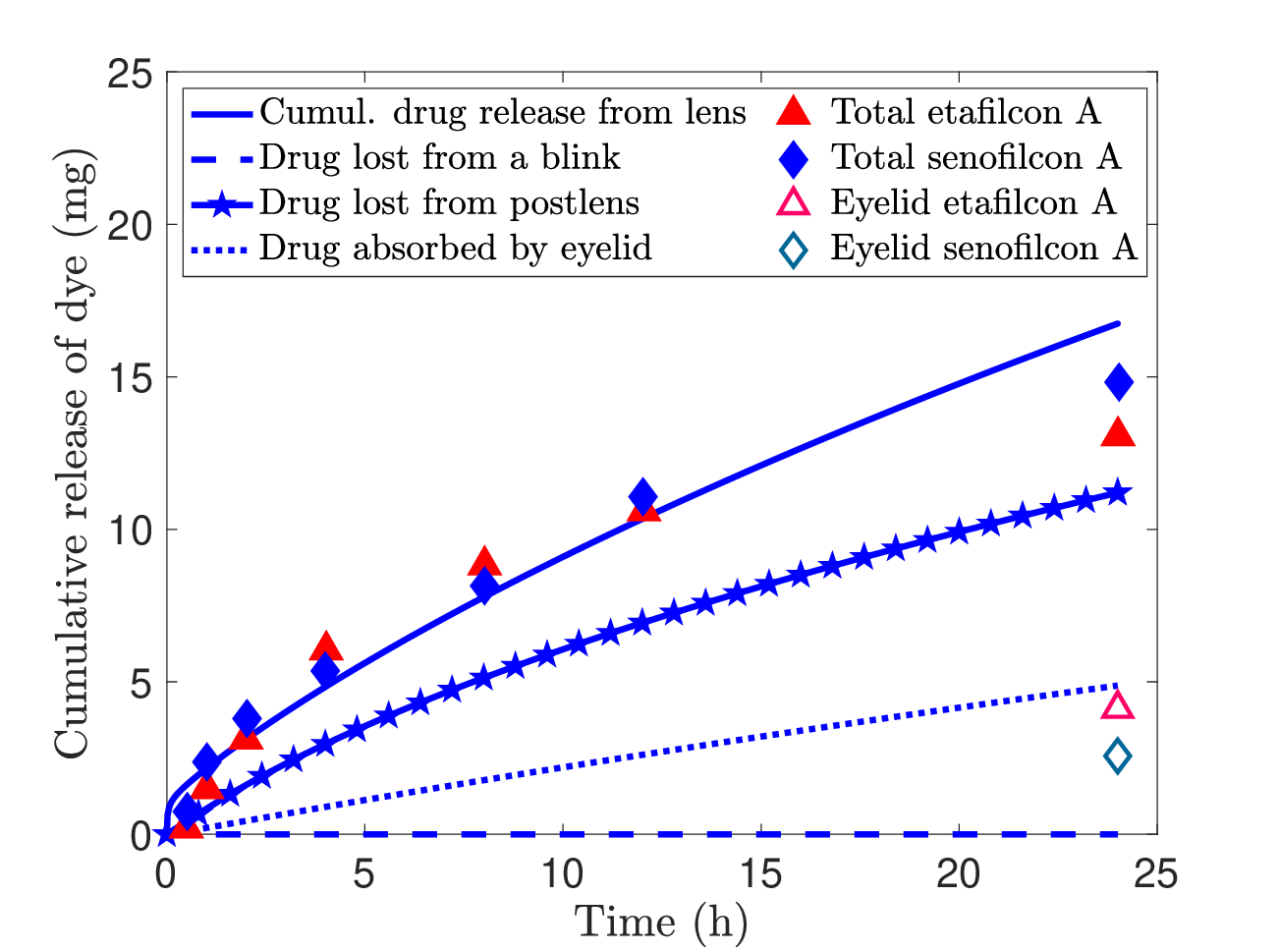}} 
\caption{Comparison of model output with Phan \textit{et al.} \cite{phan2021development} data. The model uses the squeeze out option with $p = 0$ so that no drug is swept away by the blink.  %The hand-tuned parameters used are  $k_{\rm lid} = 3\times 10^{-9}$ m/s,  $\Delta h_{\rm eta} = 0.015 \ \mu$m and $\Delta h_{\rm seno} = 0.025 \ \mu$m. 
}
\label{fig:squeeze_fit}
 \end{figure}

%\underline{Squeeze out}: 
 The model simulations look similar to that of the slide out scenario and match the experimental data from Phan \textit{et al.} \cite{phan2021development} fairly well (see
 Figure~\ref{fig:squeeze_fit}). Most of the drug is lost due to the squeeze out mechanism. The etafilcon A final time contact lens concentration is more uniform than that of the senofilcon A lens, and the pre- and post-lens concentration profiles are more similar for the etafilcon A lens.  The post-lens thickness depression amounts correspond to blink pressures of 44.7 Pa and 74.7 Pa for the etafilcon A and senofilcon A lenses.

 %\begin{figure}[H]
 %    \centering
%\subfloat[][Contact lens concentration]{\includegraphics[scale=0.27]{Figs/1_29_24_C_eta_1.eps}}
%\subfloat[][Pre- and post-lens concentration]{\includegraphics[scale=0.27]{Figs/1_29_24_prepost_eta_1.eps}}
%\subfloat[][Drug release]{\includegraphics[scale=0.27]{Figs/1_29_24_drug_eta_1.eps}}
%\caption{Model solutions for the etafilcon A lens for the squeeze out option with $p = 0$ so that no drug is swept away by the blink. In (c), the drug in the post-lens (not shown) nearly overlaps that of the drug in the pre-lens. %The hand-tuned parameters used are  $k_{\rm lid} = 3\times 10^{-9}$ m/s, and $\Delta h = 0.015 \ \mu$m.
%}
%\label{fig:squeeze_model_sols_eta}
 %\end{figure}

%  \begin{figure}[H]
%     \centering
%\subfloat[][Contact lens concentration]{\includegraphics[scale=0.27]{Figs/1_16_24_C_seno_1.eps}}
%\subfloat[][Pre- and post-lens concentration]{\includegraphics[scale=0.27]{Figs/1_16_24_prepost_seno_1.eps}}
%\subfloat[][Drug release]{\includegraphics[scale=0.27]{Figs/1_16_24_drug_seno_1.eps}}
%\caption{Model solutions for the senofilcon A lens for the squeeze out option with $p = 0$ so that no drug is swept away by the blink. %The hand-tuned parameters used are $k_{\rm lid} = 3\times 10^{-9}$ m/s, and $\Delta h = 0.025 \ \mu$m. 
%}
%\label{fig:squeeze_model_sols_seno}
% \end{figure}

\subsubsection{No pre-lens drug loss; eyelid absorption; no post-lens motion}
\begin{figure}[H]
\centering
\subfloat[][Etafilcon A lens]{\includegraphics[scale=0.36]{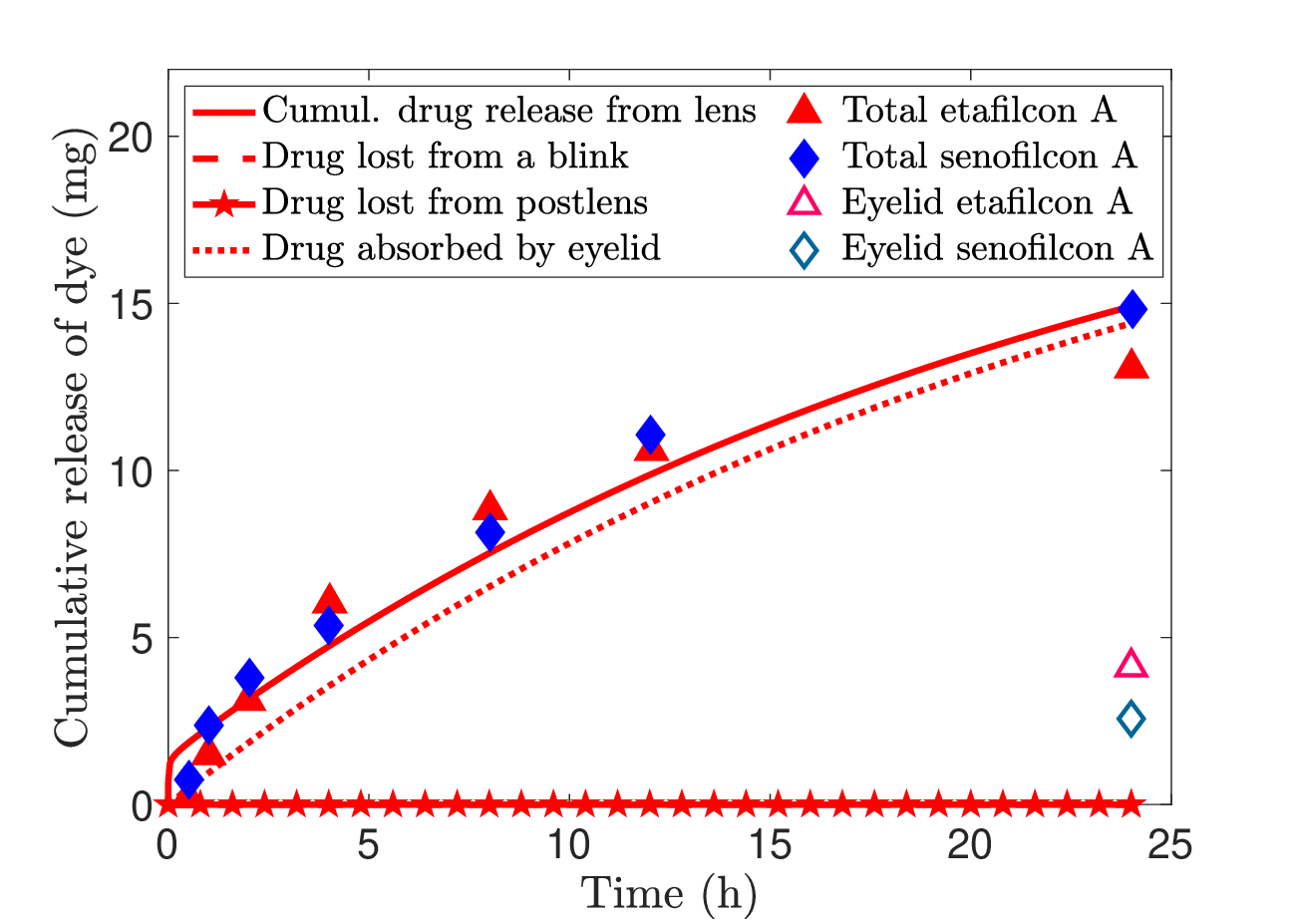}}
\subfloat[][Senofilcon A lens]{\includegraphics[scale=0.36]{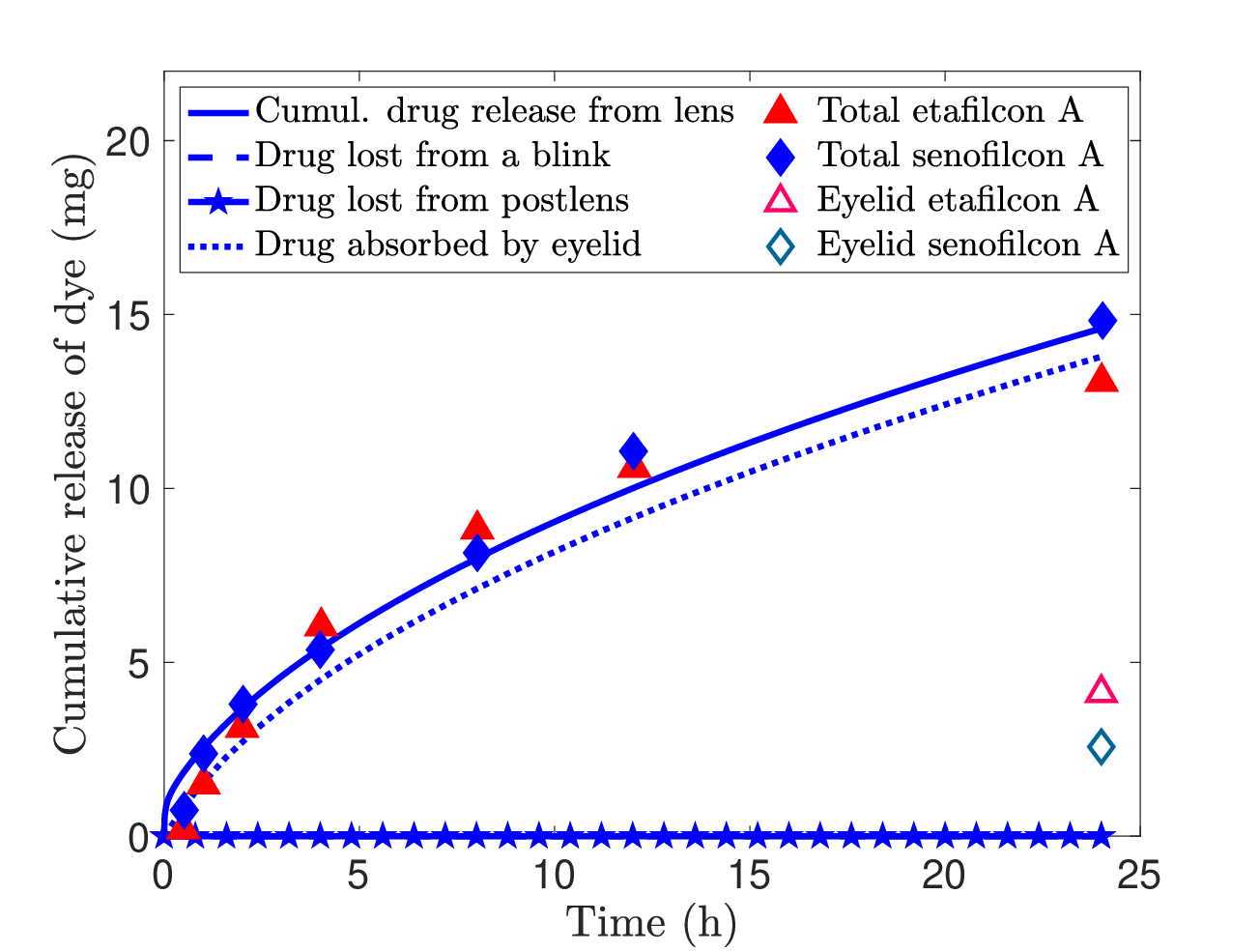}} 
\caption{Comparison of model output with Phan \textit{et al.} \cite{phan2021development} data. The model uses the no lens motion option with $p = 0$ so that no drug is swept away by the blink. %The hand-tuned parameters used are   $k_{\rm lid}^{\rm eta} = 1.25\times 10^{-8}$ m/s, and $k_{\rm lid}^{\rm seno} = 3.5\times 10^{-8}$ m/s.
}
\label{fig:noCLmot_fit}
 \end{figure}

%\underline{No lens motion}: 
The etafilcon A and senofilcon A cumulative drug release profiles look very similar to each other in this case. Neither can simultaneously match the experimental measurements of cumulative drug release and eyelid absorption from Phan \textit{et al.} \cite{phan2021development}. We also note that the model fits in Figure \ref{fig:noCLmot_fit} required two different lid permeability constants, which is infeasible in reality. Thus, the fit suggests that this combination of mechanisms does not describe the dynamics of the drug release. In this setting, which allows flux on the pre-lens side and sets $p = 0$, we find that contact lens motion is required in order to separate cumulative drug release and eyelid absorption. Unlike the slide out and squeeze out scenarios, the pre-lens drug concentration is greater than that of the post-lens. %Note that the drug in the post-lens is not shown in Figure \ref{fig:noCLmot_model_sols_eta} because it matches almost perfectly with the drug in the pre-lens.

% \begin{figure}[H]
%     \centering
%\subfloat[][Contact lens concentration]{\includegraphics[scale=0.27]{Figs/1_29_24_C_eta_3.eps}}
%\subfloat[][Pre- and post-lens concentration]{\includegraphics[scale=0.27]{Figs/1_29_24_prepost_eta_3.eps}}
%\subfloat[][Drug release]{\includegraphics[scale=0.27]{Figs/1_29_24_drug_eta_3.eps}}
%\caption{Corresponding model solutions for the etafilcon A lens for the no lens motion option with $p = 0$ so that no drug is swept away by the blink. In (c), the drug in the post-lens (not shown) nearly overlaps that of the drug in the pre-lens.
%The hand-tuned parameters used are  $k_{\rm lid}^{\rm eta} = 1.25\times 10^{-8}$ m/s and $k_{\rm lid}^{\rm seno} = 3.5\times 10^{-8}$ m/s. 
%}
%\label{fig:noCLmot_model_sols_eta}
 %\end{figure}

%  \begin{figure}[H]
%     \centering
%\subfloat[][Contact lens concentration]{\includegraphics[scale=0.27]{Figs/1_17_24_C_seno_3.eps}}
%\subfloat[][Pre- and post-lens concentration]{\includegraphics[scale=0.27]{Figs/1_17_24_prepost_seno_3.eps}}
%\subfloat[][Drug release]{\includegraphics[scale=0.27]{Figs/1_17_24_drug_seno_3.eps}}
%\caption{Corresponding model solutions for the senofilcon A lens for the squeeze out option with $p = 0$ so that no drug is swept away by the blink. 
%The hand-tuned parameters used are  $k_{\rm lid}^{\rm eta} = 1.25\times 10^{-8}$ m/s, and $k_{\rm lid}^{\rm seno} = 3.5\times 10^{-8}$ m/s. 
%}
%\label{fig:noCLmot_model_sols_seno}
% \end{figure}

\subsubsection{No pre-lens drug loss; no eyelid absorption; post-lens: slide out or squeeze out}
% \paragraph{Partition coefficient balance on both pre- and post-lens sides, $\bm{p = 0}$, $\bm{k_{\rm lid} = 0}$}

\begin{figure}[H]
\centering
\subfloat[][Etafilcon A lens, slide out]{\includegraphics[scale=0.36]{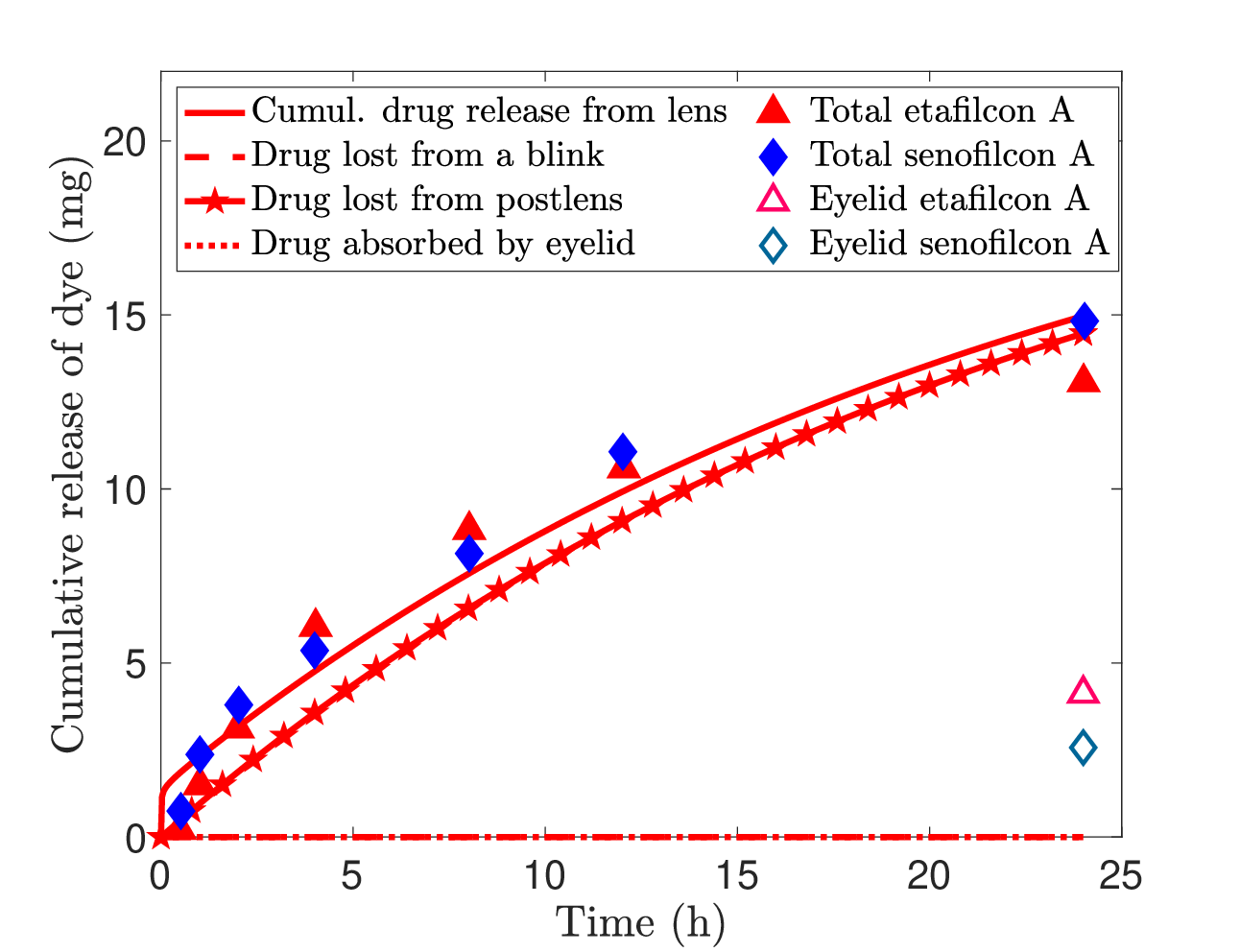}}
\subfloat[][Senofilcon A lens, slide out]{\includegraphics[scale=0.36]{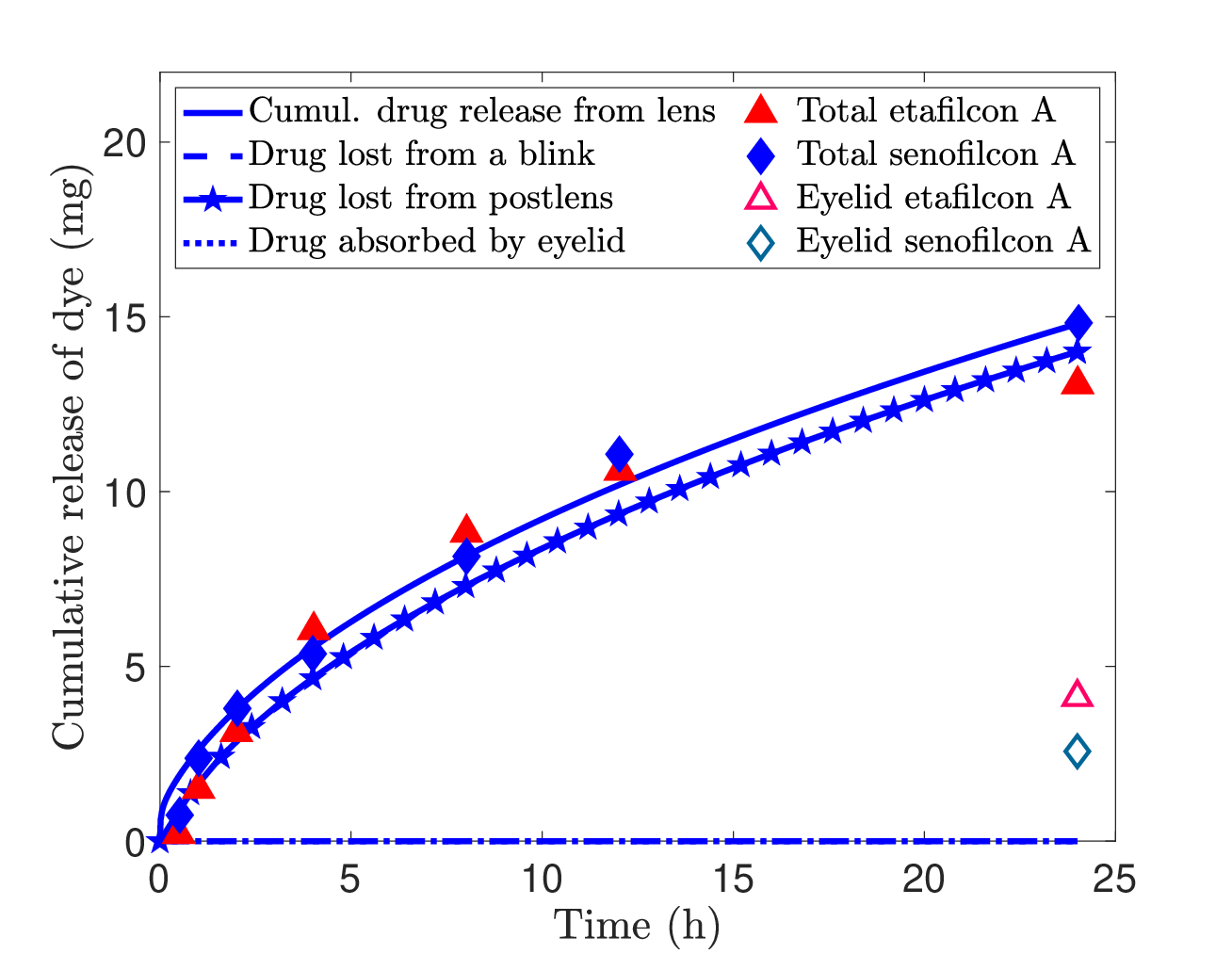}} \\
\subfloat[][Etafilcon A lens, squeeze out]{\includegraphics[scale=0.36]{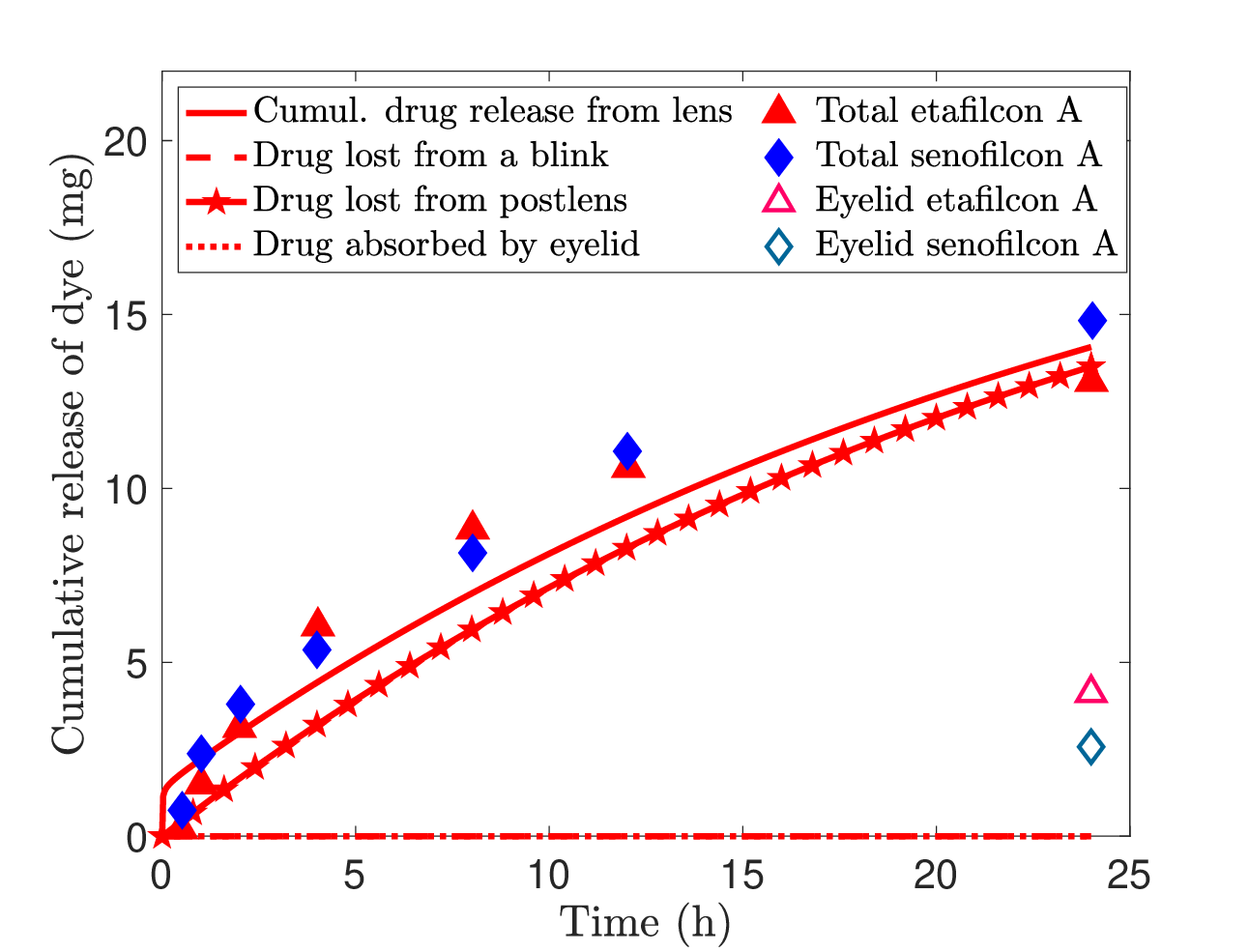}}
\subfloat[][Senofilcon A lens, squeeze out]{\includegraphics[scale=0.36]{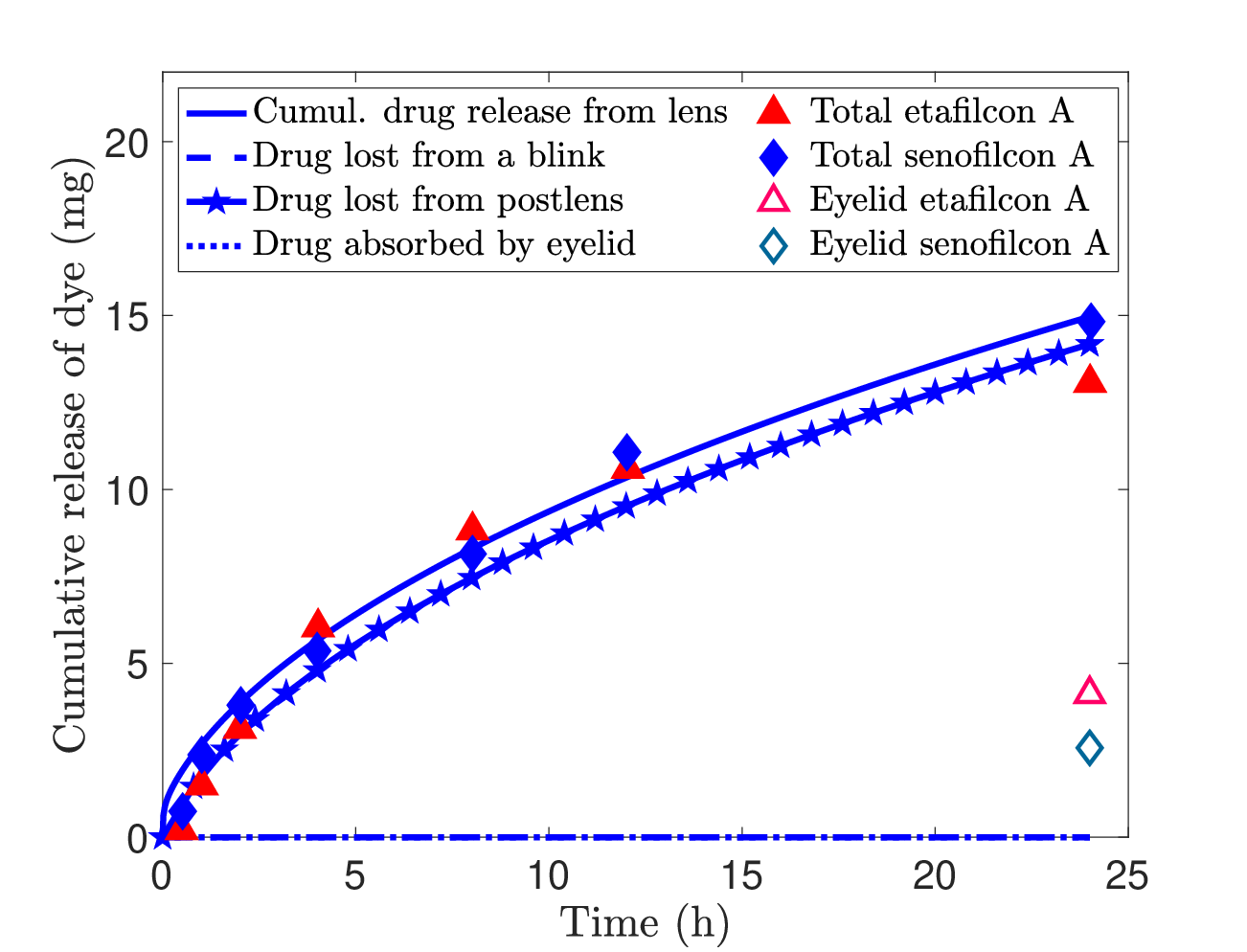}} 
\caption{Comparison of model output with Phan \textit{et al.} \cite{phan2021development} data.  We use $p = 0$ so that no drug is swept away by the blink. The model uses $k_{\rm lid} = 0$ to ignore eyelid absorption. %The hand-tuned parameters used are  $k_{\rm lid} = 3\times 10^{-9}$ m/s,  $\Delta h_{\rm eta} = 0.015 \ \mu$m and $\Delta h_{\rm seno} = 0.025 \ \mu$m. 
}
\label{fig:slide_squeeze_fit_klid0}
 \end{figure}

 Here $p=0$ and $k_{\rm lid}=0$.  
 In order to compare to the results of Section \ref{sec:large_diff}, where eyelid absorption was ignored, we set $k_{\rm lid} = 0$ and repeat the hand-tuned fitting for the slide out and squeeze out cases with $p = 0$. 
 Details are shown in Figure~\ref{fig:slide_squeeze_fit_klid0}.
 We also find ranges of parameter values that give reasonable fits with which to compare to the large diffusion limit scenario. The majority of drug loss will be due to either the slide out or squeeze out mechanism, and a small amount of drug will remain in the pre- and post-lens tear films.
 In the slide out case, the slide out amount $\Delta X_{\rm cl}$ is larger for both lenses as compared to the version of the model with $k_{\rm lid} \neq 0$, as expected. %In order to compare to the results in Section \ref{sec:large_diff}, we find ranges of $\Delta x$
 We repeat for the squeeze out case. The depression amount $\Delta h_{\rm post}$ is larger than the model with $k_{\rm lid} \neq 0$, as in the slide out case. The blink pressures corresponding to $\Delta h_{\rm post}$ are 53.6 Pa and 227 Pa for the etafilcon A and senofilcon A lenses.
%We find ranges of $\Delta h$ that provide fits that lie above and below the cumulative release data for both lenses. 

% \begin{figure}[H]
%\centering
%\subfloat[][Etafilcon A lens]{\includegraphics[scale=0.36]{Figs/1_22_24_phan_eta_5.eps}}
%\subfloat[][Senofilcon A lens]{\includegraphics[scale=0.36]{Figs/1_22_24_phan_seno_5.eps}} 
%\caption{Comparison of model output with Phan et al. \cite{phan2021development} data.  The slide out option is selected with $p = 0$ so that no drug is swept away by the blink. The model uses $k_{\rm lid} = 0$ to ignore eyelid absorption. %The hand-tuned parameters used are  $k_{\rm lid} = 3\times 10^{-9}$ m/s,  $\Delta h_{\rm eta} = 0.015 \ \mu$m and $\Delta h_{\rm seno} = 0.025 \ \mu$m. 
%}
%\label{fig:slide_fit_klid0}
 %\end{figure}

 \subsection{Partial pre-lens drug loss; no eyelid absorption; post-lens motion}

\begin{figure}[H]
\centering
\subfloat[][Etafilcon A lens, slide out]{\includegraphics[scale=0.36]{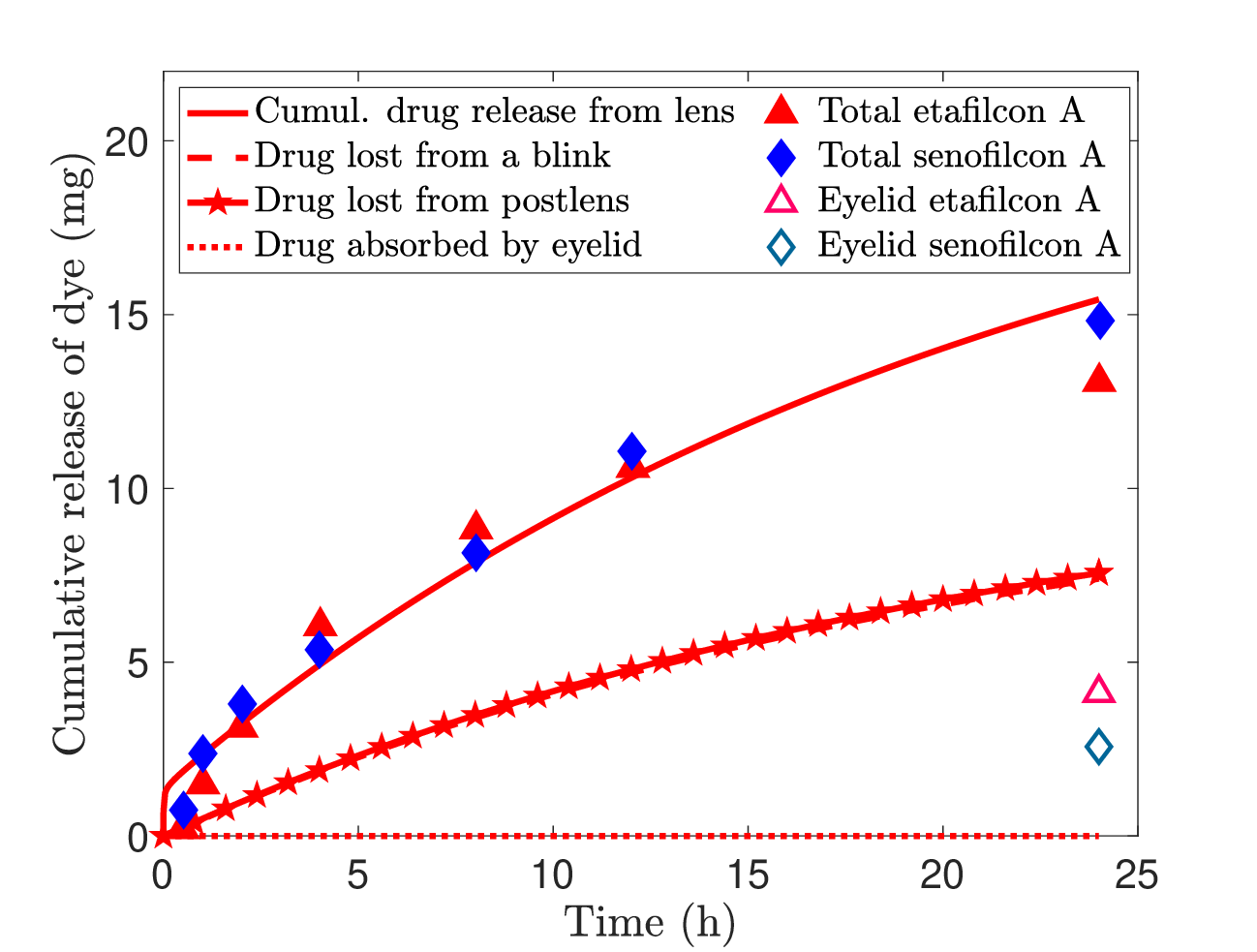}}
\subfloat[][Senofilcon A lens, slide out]{\includegraphics[scale=0.36]{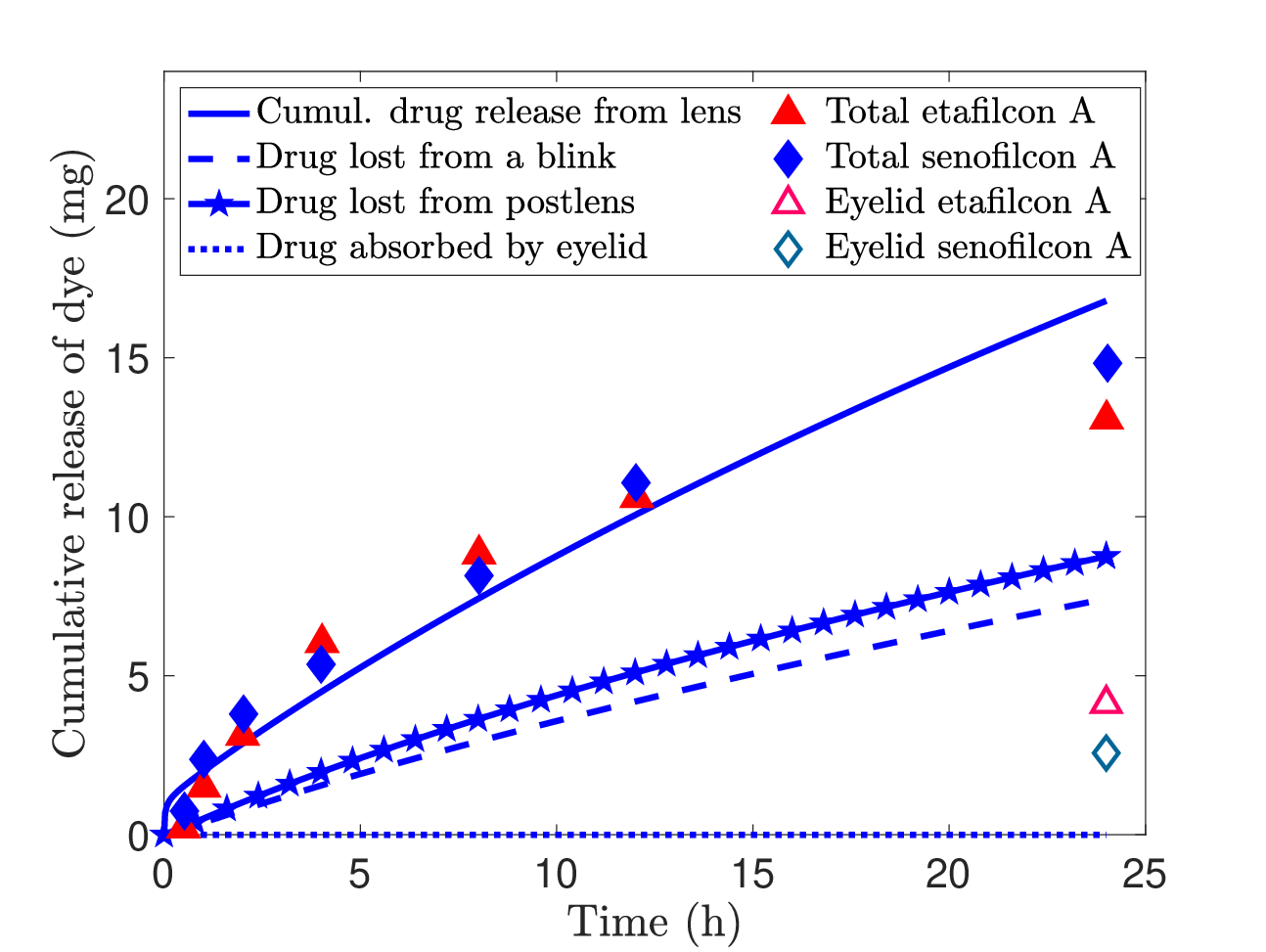}} \\
\subfloat[][Etafilcon A lens, squeeze out]{\includegraphics[scale=0.36]{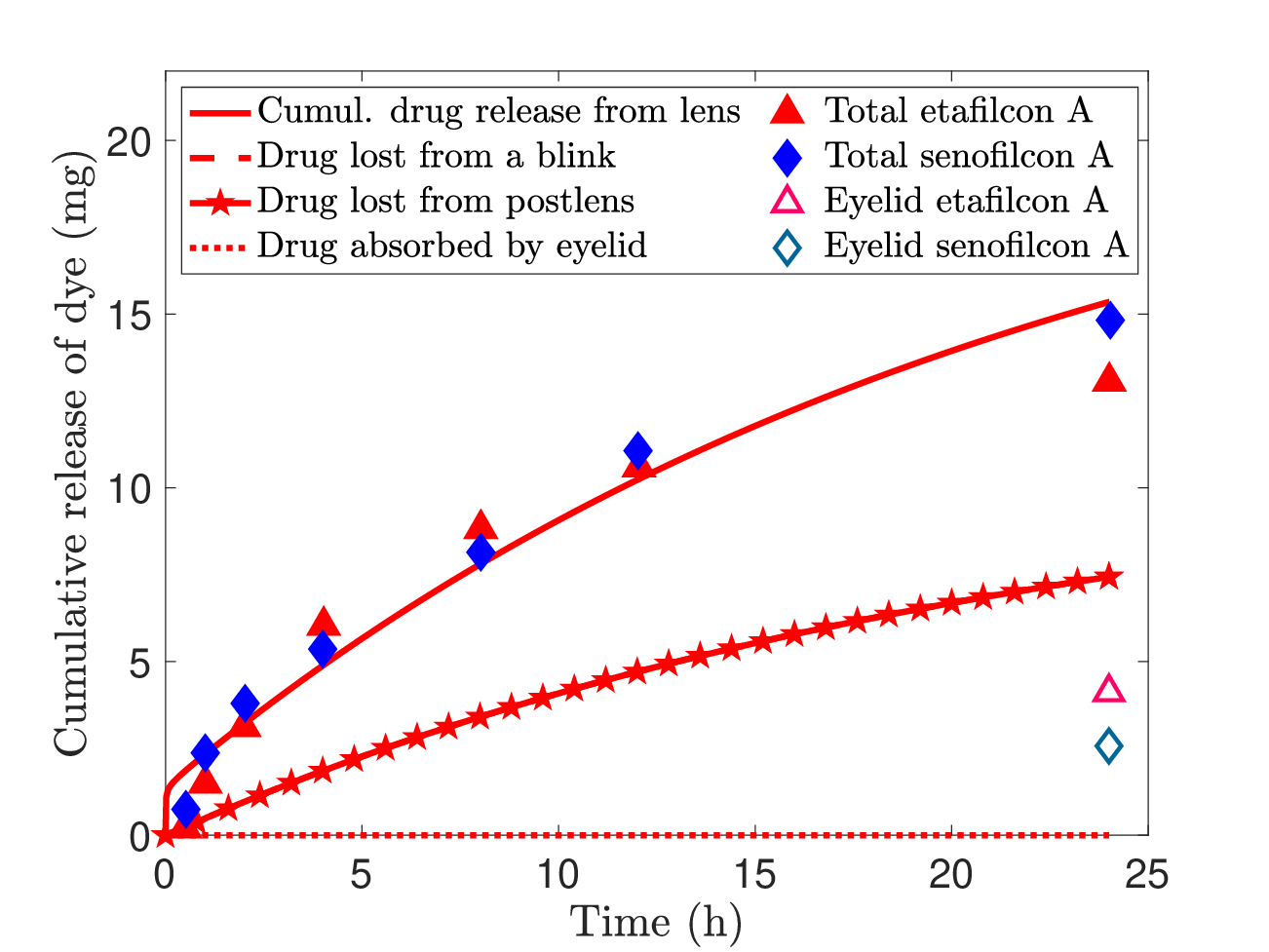}}
\subfloat[][Senofilcon A lens, squeeze out]{\includegraphics[scale=0.36]{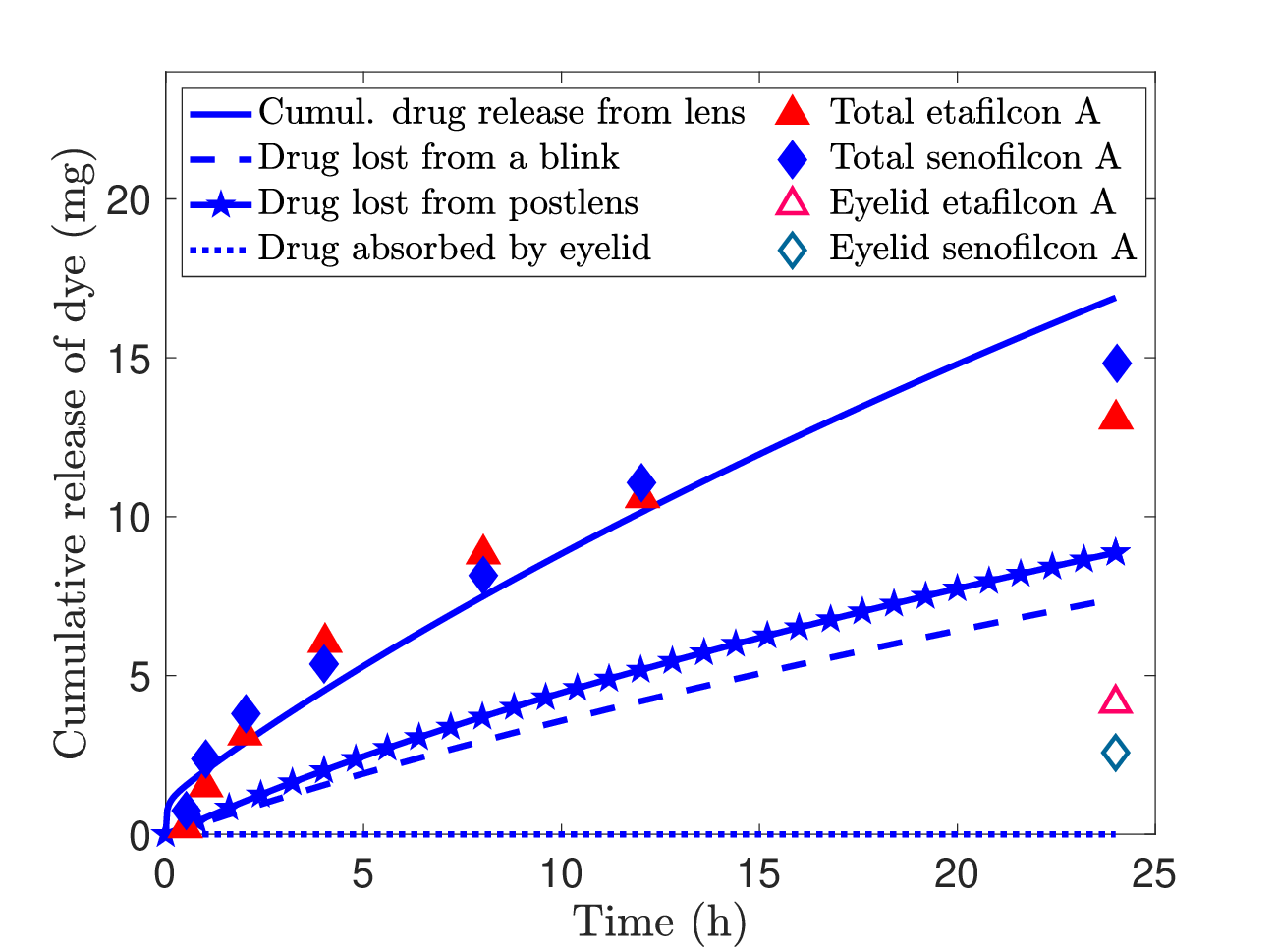}} 
\caption{Comparison of model output with Phan \textit{et al.} \cite{phan2021development} data. The model uses $p = 0.002$ so that a partial amount of drug is swept away by the blink. In (a) and (c) the curves for the drug lost from a blink and the drug lost from the post-lens are nearly identical.%The hand-tuned parameters used are   $k_{\rm lid}^{\rm eta} = 1.25\times 10^{-8}$ m/s, and $k_{\rm lid}^{\rm seno} = 3.5\times 10^{-8}$ m/s.
}
\label{fig:slide_squeeze_fit_p01}
 \end{figure}

As a final example, we now study partial pre-lens drug loss ($ 0 < p < 1$) so that a fraction of the drug in the pre-lens is swept away with each blink. We investigate the cases of post-lens slide out or squeeze out. Our comparison with the Phan \textit{et al.} experimental data is shown in Figure \ref{fig:slide_squeeze_fit_p01}. We study both the squeeze out and slide out contact lens motion scenarios and use $p = 0.002$. We set $k_{\rm lid} = 0$ in order to compare with the large diffusion limit setting (see open circles plotted in the 
lower two plots of Figure~\ref{fig_SlideOut_LargeD} and Figure~\ref{fig_SqueezeOut_LargeD}). The blink pressures corresponding to the values of $\Delta h_{\rm post}$ used are 29.7 Pa and 41.7 Pa for the etafilcon A and senofilcon A lenses.

%{\color{red}{Could we summarize here all the possible situations that could describe the Phan {\it et al.} data and all the situations that cannon ... or maybe that is the framework for the discussion section??}}

  \section{Discussion}

  \subsection{Vial model, eye model, and pre-lens/post-lens drug pathways}

  The results that we have identified for the drug release in the vial setting follow approaches commonly used in various related studies.  The experimental vial setting, typically by design, is one well-described by the so-called perfect sink conditions, which we also think of as a max flux situation.  With known geometrical properties of the lens (e.g. surface area and lens thickness) along with known initial drug content in the contact lens, if one assumes one dimensional Fickian diffusion in the contact lens, then only the diffusion coefficient remains to be determined
   to characterize drug release dynamics.  This perfect sink setting has contact 
  lens drug concentration $C=0$ at the lens boundaries (corresponding to $K C_{\rm vial} \approx 0$).  

  The drug release prediction for the vial setting under perfect sink conditions actually gives a first prediction one might contemplate for drug release during contact lens wear.  Specifically, if one assumes in the lens that $C=0$ at the boundaries with the pre-lens and post-lens tear films (obtained, for example, by assuming that $K=0$), then the drug release prediction for this case is exactly that of the vial release. That is, if one assumes perfect sink conditions
  apply in the eye model, then the contact lens releases the drug at a rate independent of any processes external to it.  
  In this extreme limit of the {\it in vitro} eye model there is no chance for the drug that reaches the pre-lens or post-lens regions to return to the lens, as the flux term $D \partial C/\partial z$ will never be directed into the contact lens.

\begin{figure}[h]
\centering
\subfloat[][Etafilcon A lens]{\includegraphics[scale=0.36]{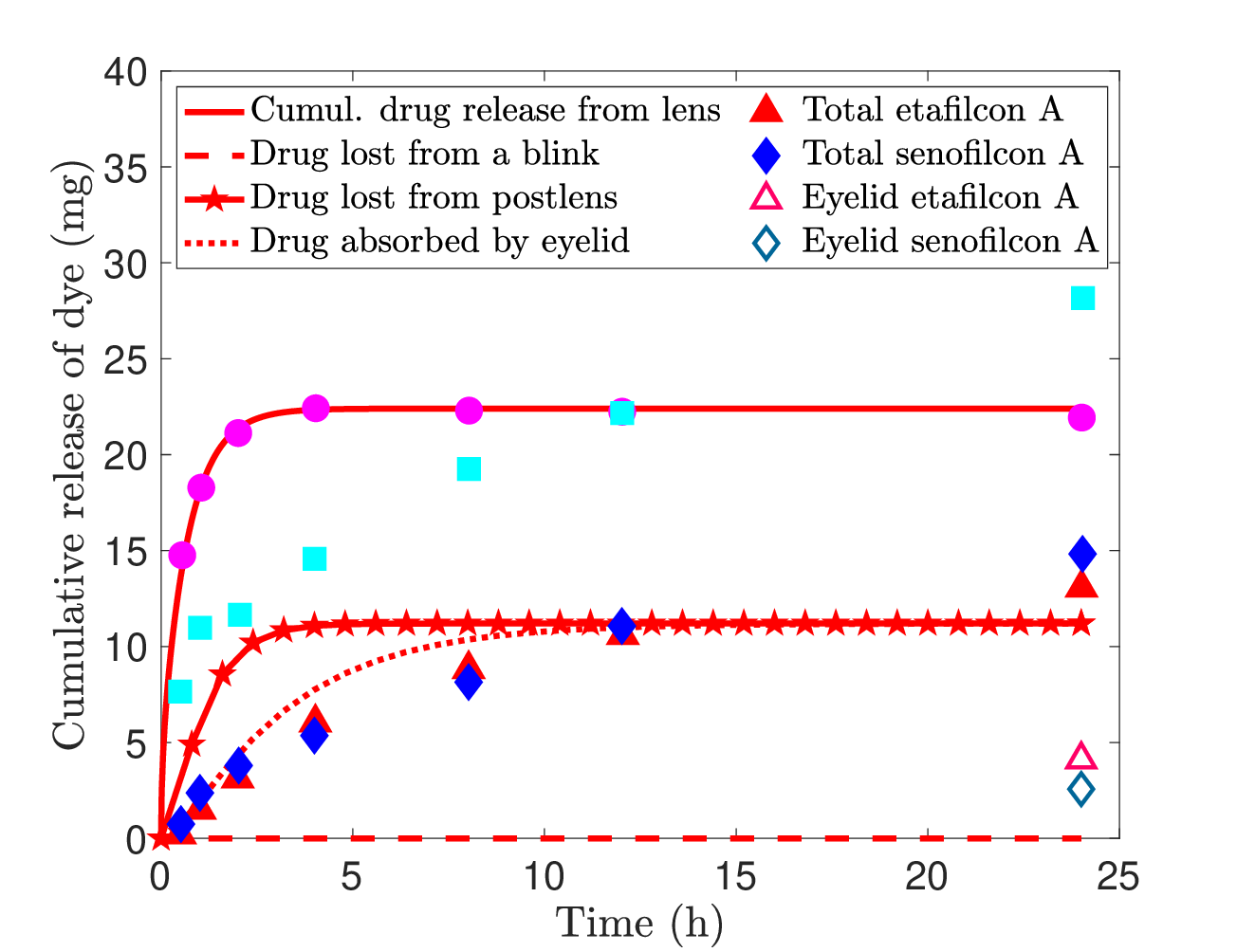}}
\subfloat[][Senofilcon A lens]{\includegraphics[scale=0.36]{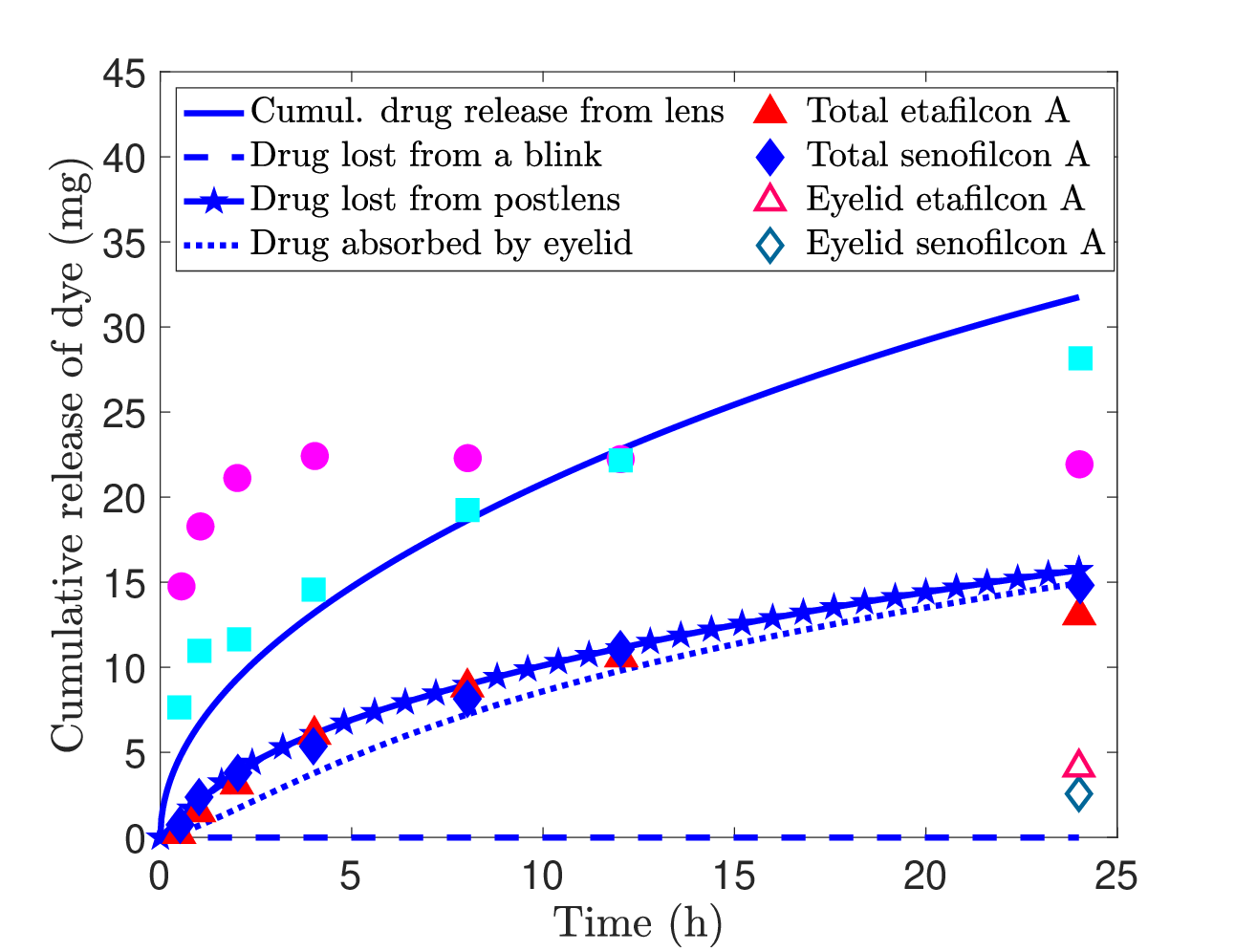}} 
\caption{Comparison of model output with Phan \textit{et al.} \cite{phan2021development} data. The model (max flux on both the pre-and post-lens sides, squeeze out option) uses $p = 0$ so that no drug is swept away by the blink. The hand-tuned parameters used are $\Delta h_{\rm post} = 0.025 \ \mu$m, and $k_{\rm lid} = 3\times 10^{-9}$ m/s.
}
\label{fig:maxflux_fit}
 \end{figure}

  For this case, the comparison to experimental measurements is shown in Figure \ref{fig:maxflux_fit}. We use $h_{\rm pre}^{\rm init} = h_{\rm post}^{\rm init} = 5 \ \mu$m, $J_E = 0$, and the diffusion coefficient is set at the vial setting value. In both the etafilcon A and senofilcon A cases, the differences between model eye and vial solutions are very small.   The cumulative drug release predicted by the eye model is  the same independent of squeeze out, slide out, or no contact lens motion options. The choice of hand-tuned model parameters has no effect on the cumulative drug release.
%Corresponding model solutions for the etafilcon A and senofilcon A lenses are shown in Figures \ref{fig:maxflux_eta} and \ref{fig:maxflux_seno}. 
Due to its larger diffusion coefficient, the etafilcon A lens releases essentially all of the drug, whereas a non-negligible concentration profile is maintained for the senofilcon A lens.
  Clearly this simple prediction for contact lens drug release for the eye model does not conform with experimental observations, as is seen from the work of Phan {\it et al.} \cite{phan2021development}.

  This observation is the basis for the models developed in the current work and our objectives have been to identify
  possible mechanisms associated with processes external to the contact lens -- in the pre-lens and/or the post-lens tear film along with consequences associated with blinking -- that
  can provide possible explanations for experimental observations.  In that direction we have identified several mechanisms in the pre-lens tear film and post-lens tear film that suggest further exploration.
 
\subsubsection{Possible pre-lens mechanisms} 

The easiest way to explore potential drug release pathways through the pre-lens tear film is to shut off substantial drug 
transport pathways into and through the post-lens tear film.  We have accomplished this in our studies by assuming
that no contact lens motion occurs, so that no drug mass is lost out of the post-lens tear
film through either the slide out or squeeze out mechanisms.  With a partition coefficient balance applied at the contact lens/post-lens boundary given by $C(z=0,t) = K C_{\rm post}$, the concentration in the post-lens is held in check by the concentration
in the contact lens.   In particular, with nonzero $K$, having found $K = {\cal O}(1)$ for both lens types, total drug mass in the post-lens is essentially limited by $C^{\rm init} V_{\rm post} /K$.  Typically $V_{\rm post} \ll V_{\rm cl}$; this suggests only a small amount of the total drug mass can be present in the post lens at any given time, and if none is removed
by a slide out or squeeze out type mechanism, this pathway is essentially turned off.  It is
important to remember that in 
clinical applications, unlike the Phan {\it et al.} \cite{phan2021development} experiments, the drug
can permeate into the cornea, thereby providing a drug pathway out of the post-lens with or without contact lens motion.  Then, under these conditions drug mass loss is dominated by pre-lens drug pathway mechanisms.
In this context we have made two key observations.
\begin{itemize}
\item First, we examine the assumption that each blink removes all pre-lens drug mass as fresh tear fluid replaces the old  after each blink ($p = 1$). \textit{In this case where $p = 1$, the senofilcon A lens cumulative drug release is reasonably 
well matched, but
%this appears fortuitous as 
the corresponding prediction for etafilcon A
is nowhere near the experimental data.}
Although it is possible that neither
  lens moves significantly as a result of a blink, there is no clear reason why a blink would completely replenish the tear fluid on one lens but not the other.  Any lens motion would only increase the drug release rate  through post-lens pathways. 
  
  \item Second,  as a consequence of a blink, it is possible that
  only a portion $0 < p < 1$ of
  the pre-lens tear fluid drug mass is washed away.  \textit{Our large
  diffusion limit results suggest that only a tiny amount of drug
  loss ($p = {\cal O}(10^{-3})$) is allowed in order to match the Phan {\it et al.} cumulative drug release data.}  
  In the full model simulations it appears that larger values of $p$ %(more drug lost due to blink motion) 
  are allowed; as noticed
  for the senofilcon A lens with
  no post-lens drug loss, this seems to include values up to $p=1$. 
  %(all pre-lens drug swept out by each blink)
  %with no post-lens drug loss appear to be allowed.
  \end{itemize}

  The application of a no-flux boundary condition closes off any drug pathways through the pre-lens tear film and necessarily  requires some type
  of post-lens pathway.  A softening of this condition is possible via a more detailed
  treatment of the tear film and evaporation effects.
  For example, evaporation could remove the tear film by some intermediate time through the interblink, rather than immediately as the no-flux condition would mimic.  This type of modulated no-flux effect could partially limit the pre-lens drug loss pathways.  We plan to explore this in future work.

\subsubsection{Possible post-lens mechanisms}

In a similar manner to our investigation of pre-lens drug transport mechanisms, 
% for our study and interpretation of post-lens mechanisms
it is easiest to imagine cases in which pre-lens mechanisms are shut off by way of the boundary condition imposed at the lens/pre-lens interface. We use either no-flux, so that no drug mass reaches the pre-lens tear film,  
%so that no drug mass reaches the pre-lens tear film 
or the partition coefficient balance  but with $p=1$, so that no drug mass is removed from the pre-lens as the result of a blink.  Under these pre-lens conditions and for reference post-lens thicknesses within a clinically-reasonable range, there appear to be both slide out and squeeze out mechanisms with feasible parameter values in terms of translational lens motion ($\Delta X_{\rm cl}$) or post-lens thickness depression ($\Delta h_{\rm post}$).  As noted earlier, in the Phan {\it et al.} experimental system there is no drug transport into the cornea; 
%(the 3D printed eye in their case) 
in clinical settings with an open drug pathway to  the cornea, the rate at which this occurs will factor into the overall drug release dynamics.

We have not included a drug pathway through the post-lens to the eyelid within the framework of our present model.  Such pathways could be included with a more detailed model of the eye and the incorporation of compartments outside of the lens periphery and under the eyelids.  This is a consideration for future work.

\subsection{Extensions}

Our models make simplifying assumptions to connect to available experimental data; future work could explore modifications of all parts of the model including the contact lens, the pre-lens tear film, the post-lens tear film, as well as other compartments (e.g.~eyelid, cornea).
%In the contact lens, 
One could focus on a more realistic characterization of diffusion, absorption, blinking, lens properties, and tear film dynamics. Our models assume linear diffusion, but could be modified to allow for specific adsorption to the contact lens. Following \cite{Liu_etal_2013}, we could modify equation~(\ref{eq:Ceq}) to differentiate between adsorption of protein to the polymer in the contact lens and the protein in the liquid-filled spaces.
In our model, 
%the pre-lens thickness remains constant during the interblink since we set $J_E = 0$.   
the post-lens thickness
remains constant during the interblink since we have assumed that all lens motion occurs during the blink only.
 One extension in the squeeze out case is to consider a restoring force after a blink that partially returns the post-lens thickness to its pre-blink height. The restorative rate could be found by knowing the elastic parameters for the lenses, or by knowing the time scale and assuming we return to $h_{\rm post}^{\rm squeeze}$ exactly after each blink. The models by Chauhan \& Radke \cite{ChauhanRadke2002} and Maki \& Ross \cite{MR2014a,MR2014b} for elastic response could serve as a starting point.

 Since eyelid absorption has not been previously modeled to the best of our knowledge, our formulation relies on simple assumptions and motivation from the experimental data in \cite{phan2021development}. Our reset condition removes a fraction $p$ of pre-lens drug concentration via the blink, and our simulations suggest that $p$ is fairly small. This seems reasonable given what is known about the residence time of drugs delivered via eye drops; the extremes $p = 0$ and $p = 1$ would not be practical assumptions in such a context. %Instead the eyelid could (i) remove none of the PreLTF with a blink or (ii) remove some fraction. One could model this second option as Couette flow, whereby the upper eyelid drags some PreLTF with a blink. 
%We could also modify the reset condition to allow the concentration of the post-lens tear film to contribute to eyelid absorption, perhaps through the introduction of another, `underlid', compartment.  
%Further, we could consider an additional compartment for fluid trapped beneath the upper eyelid in the interblink, motivated by experimental evidence from Phan et al. (2021) \cite{phan2021development}, as their video suggests that fluid pools underneath the lid between blinks. In this model option, we assume that the eyelid drags fluid upwards during a blink, and then during the interblink this fluid remains in contact with the eyelid and absorption occurs during the interblink.
 %We assume that a fresh fluid layer (zero drug) is pulled/flows in so that all of the drug in the PreLTF during the interblink is swept away into this under-eyelid compartment. We also  assume zero transfer of drug out from the eyelid (back into the PreLTF or into systemic circulation).
 %Over the course of one blink/interblink cycle, there is absorption during the blink governed by a concentration difference via
%  \begin{equation}
%\frac{d C_{\rm lid}}{dt}   =  k_{\rm lid}  [ C_{\rm pre}(t_{\rm blink}^-) - C_{\rm lid}] \times A_{\rm lid}.
% \end{equation}
%Here, $k_{\rm lid} = k_{\rm lid}' \tau/h_{\rm lid}$, where the prime denotes a dimensional quantity, and $A_{\rm lid}$ denotes the area of the eyelid.
%This assumes that the eyelid does not maintain a sink condition during the blink cycle and follows one of the options for dermal absorption by Selzer et al. \cite{selzer2015mathematical}.
An interesting alternative model may be to treat eyelid absorption as a transient mechanism and calculate the  mass absorbed at each blink. The framework of Frasch \& Barbero \cite{frasch2008transient} %note that many experiments measure the skin permeability coefficient at steady state %(constant absorption, large dose) 
%rather than during transient absorption. 
  can be adapted to model eyelid absorption as a reset condition. They assume a constant concentration is applied on the  (pre-lens/eyelid) interface for $0 < t < \Delta t_{\rm blink}$ and a max flux condition for the eyelid for $t > \Delta t_{\rm blink}$, as we do in our model.
 %two boundary conditions at the pre-lens/eyelid interface: (i) no drug/max flux% and (ii) no flux. %For now, assume this first condition holds.
%For a single event (blink),  the total drug mass accumulation is
%\begin{equation}
%$m_{\infty} = k_{\rm lid} A_{\rm lid} C_{\rm pre}(t_{\rm blink}^-) \Delta t_{\rm blink},$
%\end{equation}
%where 
%$k_p = K D_{\rm lid}/h_{\rm lid}$ is the permeability coefficient with partition coefficient $K$  (we scale $k_p$ with $h_{\rm lid}/\tau$ instead of $D_{\rm lid}/h_{\rm lid}$), diffusion coefficient $D$, and eyelid thickness $h_{\rm lid}$ (reported as 500 $\mu$m in \cite{phan2021development}); $A$ is the eyelid surface area exposed to the drug, 
%$C_{\rm pre}(t_{\rm blink}^-)$ is the constant drug concentration applied to the eyelid. 
%and $\Delta t_{\rm blink}$ is the blink duration. 
%A reasonable assumption for $A$ that connects to available data is to use the area of the contact lens $A = \pi r^2$. Perhaps more accurately, we should use $A_{\rm lid}$, which we have estimated (see Section \ref{sec:results_in_vitro}). 
For multiple transient events (blinks),  total drug mass accumulation is
%\begin{equation}
%m_{\infty} = k_p  A_{\rm lid} \sum_{i = 1}^n C_i \Delta t_{\rm blink,i}.
%\end{equation}
%Our blink durations are all the same, so the only factor that could be changing is $C_i$, the amount of drug in the pre-lens that can be absorbed by the eyelid during a blink:
%\begin{equation}
$m_{\infty} = k_{\rm lid} A_{\rm lid} \Delta t_{\rm blink} \sum_{t_i = 0}^{T_f} C_{\rm pre}(t_{\rm blink}^-),$
%\label{eq:m_infty_blink_mass_flux}
%\end{equation}
where $T_f$ is a final time. %We may extend $\Delta t_{\rm blink}$ a little bit to allow for some absorption of drug into the eyelid beyond the blink. 

 %If instead we assume that the eyelid is entirely out of the picture during the interblink (a reasonable assumption), then the second situation is most applicable since there should be no flux of drug out of the pre-lens into the air. The authors write, ``This boundary condition applies to nonvolatile chemicals that partition preferentially in skin as opposed to the surrounding air,'' and I believe we can assume that is true for our situation. Frasch and Barbero \cite{frasch2008transient} give the expression for the mass of drug absorbed at time $t = \Delta t_{\rm blink}$ (for a single exposure of time $\Delta t_{\rm blink}$) as
%\begin{equation}
%m_{\Delta t_{\rm blink}} = k_p A_{\rm lid} C_0 C_{\rm pre}(\Delta t_{\rm blink}) \left[ \Delta t_{\rm blink} + \frac{h_{\rm lid}^2}{3D_{\rm lid}} - \frac{2 h_{\rm lid}^2}{\pi^2 D_{\rm lid}} \sum_{n = 1}^{\infty} \frac{1}{n^2} \exp \left( - \frac{D_{\rm lid} n^2 \pi^2 \Delta t_{\rm blink}}{h_{\rm lid}^2} \right) \right].
%\end{equation} 
%For repeated exposures, the total mass of drug absorbed by the eyelid should be the above quantity summed over changing pre-lens concentrations:
%\begin{equation}
%m_{\infty} = k_p A_{\rm lid} C_0 \left[ \Delta t_{\rm blink} + \frac{h_{\rm lid}^2}{3D_{\rm lid}} - \frac{2 h_{\rm lid}^2}{\pi^2 D_{\rm lid}} \sum_{n = 1}^{\infty} \frac{1}{n^2} \exp \left( - \frac{D_{\rm lid} n^2 \pi^2 \Delta t_{\rm blink}}{h_{\rm lid}^2} \right) \right] \sum_{t_j = 0}^{T_f} C_{\rm pre}(t_i).
%\end{equation}

\subsection{Limitations}

It is worth noting that our numerical implementation does not include a $\Delta t_{\rm blink}$ between interblinks; the computational time domain does not take this into account. This means that the blink is occurring in the last fraction of a second of an interblink computation time period. Diffusion should be mostly independent of this $\Delta t_{\rm blink}$ period, but there is a small effect to the pre-lens that we  ignore. In fact, by this numerical implementation, there is diffusion of drug from the contact lens into the pre-lens during the blink that is perhaps at odds with the reset condition that assumes (when  $p=1$) that all drug is swept from the pre-lens by the blink. Therefore, we may be slightly \textit{overestimating} the actual amount of diffusive drug transport from the contact lens into the pre-lens.

We acknowledge the balance sought in our work between proposing a realistic model, which often requires additional complexity such as spatial variation across the cornea, with the practical identifiability of model components from comparison to experimental data.
As one example, our model does not explicitly account for friction or lubrication of the eyelid/tear film interface. As a simplification, one can view friction as rolled into the $\Delta X_{\rm cl}$ slide out parameter. As one goal of this paper was to compare with and explain a single output metric from Phan \textit{et al.} \cite{phan2021development}, we chose a simple form for this element of the model  in the context of available data.  Perhaps more to the point, discussion of lubrication is absent from the Phan \textit{ et al.} paper, and so it may be most appropriate to ignore friction here.  Future work could extend the model to consider friction and propose a lipid layer or other lubricating film on the tear film/eyelid interface.

\section{Conclusions}

In this study we have developed a mathematical model for the prediction of drug transport out of a contact lens during many hours of wear (over 8000 blinks).  We are able to make predictions about cumulative drug loss from the contact lens and assess drug lost via different pathways through the pre-lens tear film, post-lens tear film, and into adjoining regions such as the eyelid.  Our general model includes a pathway into the cornea, but as we have focused our comparison on the experimental data of Phan {\it et al.} \cite{phan2021development} for which there is no uptake of drug into the eye, we have not addressed this aspect in our simulations.

Our full model handles each blink by applying reset conditions on the various parameters.  These reset conditions encode various drug loss mechanisms such as pre-lens drug loss via fresh supply of tear fluid and post-lens drug loss due to contact lens motion that generates fluid transport via slide out
(Couette-type flow) and squeeze out (squeeze film flow) mechanisms.

We have also analyzed a simplified model that applies to a large diffusion limit in which by the end of the
interblink, the contact lens concentration is assumed to have reached an equilibrium (spatially uniform) profile.  The model in this limit can be solved analytically and has provided predictions that we  used to validate the full numerical simulations. The  large diffusion limit model offers analytical expressions that predict
the $\Delta X_{\rm cl}$ or $\Delta h_{\rm post}$ values in terms of the other system parameters and, as such, provide at least 
rough guidelines for parameter values that can be difficult to measure directly.  Although the diffusion coefficient range characteristic of this limit is not well-matched by our numerical estimates for either contact lens in the study, the simplified model predictions agree qualitatively with the full model simulations.

By setting our parameter $p=1$ to model all drug being swept out of the pre-lens by a blink, we are unable to match the cumulative drug release data from Phan \textit{et al.} \cite{phan2021development} for the etafilcon A lens, regardless of contact lens motion option (see Figure \ref{fig:noCLmot_fit_p1}). This shows that some fraction of drug must remain in the pre-lens after a blink in order to explain the observed eye model release data. 

 %We found that the experimental eyelid drug absorption could be replicated under some conditions and not others.
By using a no flux condition on the pre-lens side, we can match the cumulative release data for both lenses. However, we cannot  replicate eyelid absorption, as the no flux condition shuts off our only drug pathway to the pre-lens.   %However, the slide out and squeeze out mechanisms of drug loss out of the post-lens tear film allow the model predictions to match the total cumulative drug release data for conditions on parameters such as post-lens tear film thickness, blink-driven contact lens motion, and blink pressures that appear to be clinically feasible.

We find that the full model with either the slide out or squeeze out option can replicate both the cumulative drug release data over time and the final time eyelid absorption measurement. The hand-tuned values of $\Delta X_{\rm cl}$ in the slide out case fall near or within experimental ranges \cite{Hayashi1977, Gilman1982,Cui_etal2012,Wolffsohn_etal2009}, and the values of post-lens thickness depression $\Delta h_{\rm post}$ in the squeeze out case correspond to reasonable blink pressures 
primarily in the range ${\cal O}(0.1)$ kPa -- ${\cal O}(1)$ kPa
(see Talbott {\it et al.}~\cite{Talbott_etal2014} Section 3 and references therein).   Further, our hand-tuned eyelid permeability constant values fall within experimental ranges for rat models \cite{see2017eyelid}. The results from the no lens motion version of the full model suggest that this option may not be physically observed, as the final time eyelid value and cumulative drug release data cannot be simultaneously replicated by the model output.

We expect that the model and its several variations that we have presented here should serve as a valuable framework under which further mathematical and computational models can be developed and tested.  Further predictions of this and future models should be able to guide understanding of transport from
drug-eluting contact lenses to various target tissues for treatment of different ophthalmic conditions.

\section*{Conflict of Interest} On behalf of all authors, the corresponding author states that there is no conflict of interest.

\section*{Acknowledgments} DMA and RAL note that our initial exploration into contact lens drug delivery modeling came during the week-long Graduate Student Mathematical Modeling Camp held at the University of Delaware in June 2019.  We would particularly like to acknowledge the organizers of that camp as well as the other members of the contact lens modeling group.

 \bibliographystyle{unsrt}
 
 \bibliography{CL_bib.bib}

\setcounter{equation}{0}
\setcounter{figure}{0}
\setcounter{table}{0}
\appendix
\renewcommand{\theequation}{A\arabic{equation}}
\renewcommand{\thefigure}{A\arabic{figure}}
\renewcommand{\thetable}{A\arabic{table}}

\end{document}